\documentclass[useAMS,usenatbib]{mn2e}
\usepackage{graphicx}
\usepackage{latexsym}
\usepackage{psfig}
\usepackage{amssymb,amsmath}
\usepackage{subfig}
\usepackage{caption}
\usepackage[outercaption]{sidecap}

\title{Ionization--induced star formation V: Triggering in partially unbound clusters}

\author[J. E. Dale, B. Ercolano, I.A. Bonnell]{J. E. Dale$^{1}$\thanks{E-mail: dale@usm.lmu.de (JED)}, B. Ercolano$^{1}$, I. A. Bonnell$^{2}$\\
$^{1}$Excellence Cluster `Universe', Boltzmannstr. 2, 85748 Garching, Germany.\\
$^{2}$Department of Physics and Astronomy, University of St Andrews, North Haugh, St Andrews, Fife KY16 9SS}

\begin{document}

\pagerange{\pageref{firstpage}--\pageref{lastpage}} \pubyear{2006}

\maketitle

\label{firstpage}

\def\mnras{MNRAS}
\def\apj{ApJ}
\def\aj{AJ}
\def\aap{A\&A}
\def\apjl{ApJL}
\def\apjs{ApJS}
\def\araa{ARA\&A}
\def\pasj{PASJ}
 
\begin{abstract}
We present the fourth in a series of papers detailing our SPH study of the effects of ionizing feedback from O--type stars on turbulent star forming clouds. Here, we study the effects of photoionization on a series of initially partially unbound clouds with masses ranging from $10^{4}$--$10^{6}$M$_{\odot}$ and initial sizes from 2.5--45pc. We find that ionizing feedback profoundly affects the structure of the gas in most of our model clouds, creating large and often well--cleared bubble structures and pillars. However, changes in the structures of the embedded clusters produced are much weaker and not well correlated to the evolution of the gas. We find that in all cases, star formation efficiencies and rates are reduced by feedback and numbers of objects increased, relative to control simulations. We find that local triggered star formation does occur and that there is a good correlation between triggered objects and pillars or bubble walls, but that triggered objects are often spatially--mixed with those formed spontaneously. Some triggered objects acquire large enough masses to become ionizing sources themselves, lending support to the concept of propagating star formation. We find scant evidence for spatial age gradients in most simulations, and where we do see them, they are not a good indicator of triggering, as they apply equally to spontaneously--formed objects as triggered ones. Overall, we conclude that inferring the global or local effects of feedback on stellar populations from observing a system at a single epoch is very problematic.
\end{abstract}

 
\section{Introduction}
We do not understand what initiates the process of star formation in giant molecular clouds (GMCs). If GMCs form in a gravitationally unstable state, so that they inevitably and immediately begin star formation, the distinction between cloud formation and star formation would disappear. They would essentially both be part of the same continuous process. However, it is clear from both the observational \citep[e.g][]{2009ApJ...699.1092H} and theoretical \citep[e.g.][]{2011MNRAS.413.2935D} perspectives that not all GMCs are born gravitationally bound. Several authors have studied star formation occurring in such clouds as a possible means of circumventing the problem of unbinding embedded stellar clusters or explaining slow star formation rates \citep[e.g][]{2005MNRAS.359..809C,2010ApJ...715.1302V,2008MNRAS.386....3C,2011MNRAS.410.2339B}.\\
\indent There are also GMCs which do not appear to be forming stars, or are doing so very slowly \citep[e.g.][]{2009AJ....137.4072M,2012ApJ...746..117L}. Such clouds could be induced to form stars, for example by collisions with other clouds \citep[e.g.][]{2011ApJ...738...46T} or by a sudden increase in the pressure of the intercloud medium due to a galactic collision or merger \citep[e.g.][]{2004ApJS..154..193W}, or by the action of spiral arms (Bonnell, Dobbs \& Smith 2012, submitted). These are examples of \emph{external} triggering.\\
\indent However, there are also \emph{internal} processes which, once star formation has already begun, are thought to have the ability to accelerate it, to increase its efficiency, or to spread it over larger volumes and the most commonly invoked are the HII regions and winds generated by massive stars. Supernovae may also have a triggering effect provided that the host clouds live long enough for at least some of their O--stars to reach the ends of their live while still enveloped in the clouds (see \cite{2011EAS....51...45E} for a recent review of this topic).\\
\begin{table*}
\begin{tabular}{|l|l|l|l|l|l|l|l|l|l|}
Run&Mass (M$_{\odot}$)&R$_{0}$(pc)&$\langle n(H_{2})\rangle$ (cm$^{-3}$) & v$_{\rm RMS,0}$(km s$^{-1}$)&v$_{\rm RMS,i}$(km s$^{-1}$)&v$_{\rm esc,i}$(km s$^{-1}$)&$t_{\rm i}$ (Myr) &t$_{\rm ff,0}$ (Myr)\\
\hline
UZ&$10^{6}$&45&149&18.2&9.4&13.8&4.33&2.9\\
\hline
UB&$3\times10^{5}$&45&45&10.0&4.6&7.6&9.40&6.0\\
\hline
UC&$3\times10^{5}$&21&443&14.6&6.0&11.1&4.03&1.9\\
\hline
UV&$10^{5}$&21&148&12.2&3.9&6.4&10.44&3.3\\
\hline
UU&$10^{5}$&10&1371&8.4&5.8&9.3&3.73&1.1\\
\hline
UF&$3\times10^{4}$&10&410&6.7&3.5&5.1&3.28&2.0\\
\hline
UP&$10^{4}$&2.5&9096&7.6&3.6&5.9&1.83&0.4\\
\hline
UQ&$10^{4}$&5.0&1137&5.4&2.6&4.1&3.13&1.2\\
\hline
\hline
I&$10^{4}$&10&136&2.1&1.2&2.9&5.38&2.56\\
\hline
J&$10^{4}$&5.0&1135&3.0&1.7&4.1&2.09&0.90\\
\end{tabular}
\caption{Initial properties of clouds listed in descending order by mass, including all unbound clouds and the bound clouds Run I and J. Columns are the run name, cloud mass, initial radius, initial RMS turbulent velocity, RMS turbulent velocity at the time ionization becomes active, the escape velocity at the same epoch, the time at which ionization begins, and the initial cloud freefall time.}
\label{tab:init}
\end{table*}
\indent There is an obvious tension here, because these same processes are also often blamed for the destruction of molecular clouds and their low global star formation efficiencies. \citep[e,g,][]{1979MNRAS.186...59W,1988ARA&A..26..145T,1994ApJ...436..795F,2002ApJ...566..302M}. Stellar feedback can obviously be destructive as well as constructive. The questions which must be answered are: is the net effect of feedback positive or negative?; under what circumstances can its global effects be opposite to its local effects?; more generally, in what ways are GMCs and clusters different because of the action of feedback?\\
\indent Observational studies of feedback abound but there is as yet a lack of consensus in three main areas:\\
\indent (i) What is actually meant by triggering; it could be construed as an increase in the star formation rate (SFR), an increase in the star formation efficiency (SFE), or the formation of stars that would not otherwise have been born. Neither an increase in the SFR nor the formation of different stars necessarily entails an increase in the SFE, since they can both occur without altering the total final stellar mass. Conversely, an increase in SFE can occur without a change in SFR if star formation continues for a longer period. Additionally, any of these outcomes could be local or global. Many studies do not make these distinctions, probably because they are very hard to make without knowledge of the history of the system under study or the availability of a counterfactual model to serve as a control.\\
\indent (ii) What is the best way to detect triggering; studies have variously invoked interaction between HII regions or wind bubbles and molecular gas \citep[e.g][]{2009ApJ...706...83K}, proximity of young stars to ionization fronts, bubbles or swept--up shells \citep[e.g][]{2006ApJ...649..759C,2008ApJ...688.1050G,2009ApJ...700..506S}, or to the tips of pillars or bright--rimmed clouds \citep[e.g][]{1996AJ....111.2349H,2007ApJ...654..316G,2012arXiv1208.1471G}, dynamical ages or spatial age--gradients from massive stars \citep[e.g][]{1985ApJ...297..599D,2005ApJ...624..808L,2007ApJ...654..316G,2009MNRAS.396..964C}.\\
\indent (iii) Which stellar feedback mechanism is the most important; some authors invoke HII regions and some invoke winds, occasionally when referring to the same object \citep[e.g.][]{2008ApJ...681.1341W,2012ApJ...756..151D}.\\
\indent This paper forms part of a series \citep[e.g][]{2011MNRAS.414..321D,2012MNRAS.424..377D,2012arXiv1208.4486D,deb3} (the latter three are hereafter referred to as Papers I, II and III respectively) in which we study the effects of internal feedback from HII regions on GMCs. Paper II and this work in particular are devoted to examining the ability of HII regions to trigger star formation in bound and unbound clouds respectively.\\
\indent We studied in detail a parameter space of unbound clouds in Paper III to determine whether the clouds' dynamical state helped or hindered photoionization in destroying the clouds. Here, we concentrate instead on the details of the star formation process in the same clouds presented in Paper III. We briefly outline our numerical methods in Section 2 and present our results in Section 3. We discuss and draw conclusions in Section 4 and 5.\\
\section{Numerical methods}
Our numerical methods are identical to those detailed in Paper III and we will describe them only very briefly here. We use a hybrid N--body/SPH code based on that described by Benz \citep{1990nmns.work..269B}, updated to model star formation using the sink particle technique \citep{1995MNRAS.277..362B} and with an algorithm to simulate photoionization from multiple point sources \citep{2007MNRAS.382.1759D}, Dale and Ercolano (2012), in prep.\\
\indent The cold neutral gas is treated using a piecewise barotropic equation of state from \cite{2005MNRAS.359..211L}, defined so that $P = k \rho^{\gamma}$, where
\begin{eqnarray}
\begin{array}{rlrl}
\gamma  &=  0.75  ; & \hfill &\rho \le \rho_1 \\
\gamma  &=  1.0  ; & \rho_1 \le & \rho  \le \rho_2 \\
\gamma  &=  1.4  ; & \hfill \rho_2 \le &\rho \le \rho_3 \\
\gamma  &=  1.0  ; & \hfill &\rho \ge \rho_3, \\
\end{array}
\label{eqn:eos}
\end{eqnarray}
and $\rho_1= 5.5 \times 10^{-19} {\rm g\ cm}^{-3} , \rho_2=5.5 \times10^{-15} {\rm g cm}^{-3} , \rho_3=2 \times 10^{-13} {\rm g\ cm}^{-3}$. The thermodynamics are taken to be dominated by line cooling at low densities, dust cooling at intermediate densities optically--thick heating at high densities, with a final isothermal phase to permit sink particle formation. This choice of equation of state is extensively discussed justified and tested in Paper I.\\
\indent Sink particles represent stars or clusters depending on mass resolution and are given ionizing luminosities if they are sufficiently massive as detailed in Paper I. In the simulations where sink particles represent clusters, sinks approaching each other to within their accretion radii are merged if they are bound to one another. All simulations were run for as close to 3Myr as was numerically practicable.\\
\indent Our initial conditions are a set of turbulent clouds with masses ranging from $10^{4}$ to $10^{6}$M$_{\odot}$, radii from a few to $\sim100$ pc and with undriven turbulent fields such that their initial virial ratios of 2.3. For convenience, we reproduce Table 1 from Paper III summarizing the principal characteristics of each cloud as Table \ref{tab:init} here.\\
\section{Results}
\subsection{Changes in gaseous and stellar structures due to feedback}
\begin{figure*}
    \centering
     \subfloat[Run UB]{\includegraphics[width=0.45\textwidth]{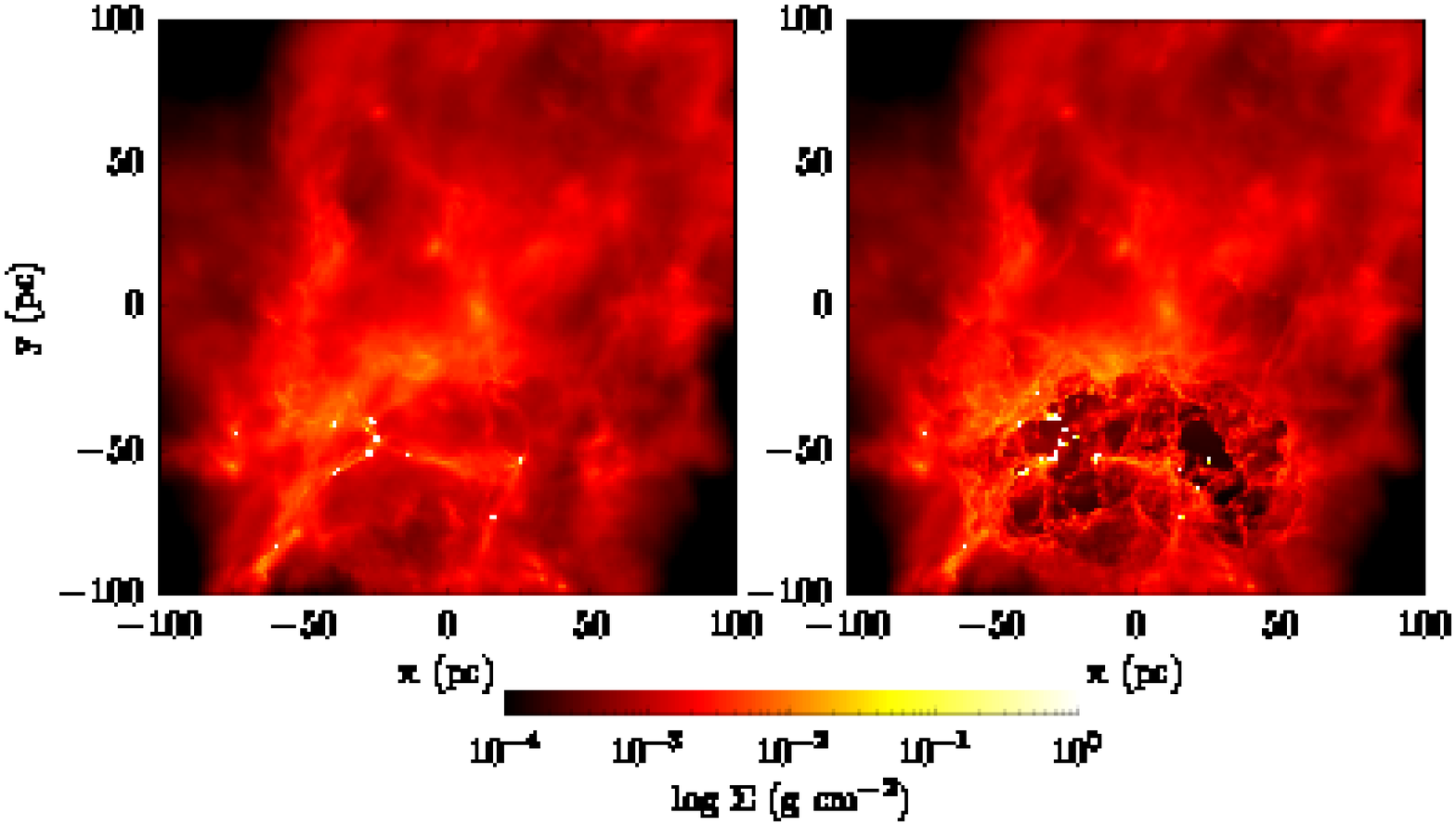}}     
     \hspace{.1in}
     \subfloat[Run UZ]{\includegraphics[width=0.47\textwidth]{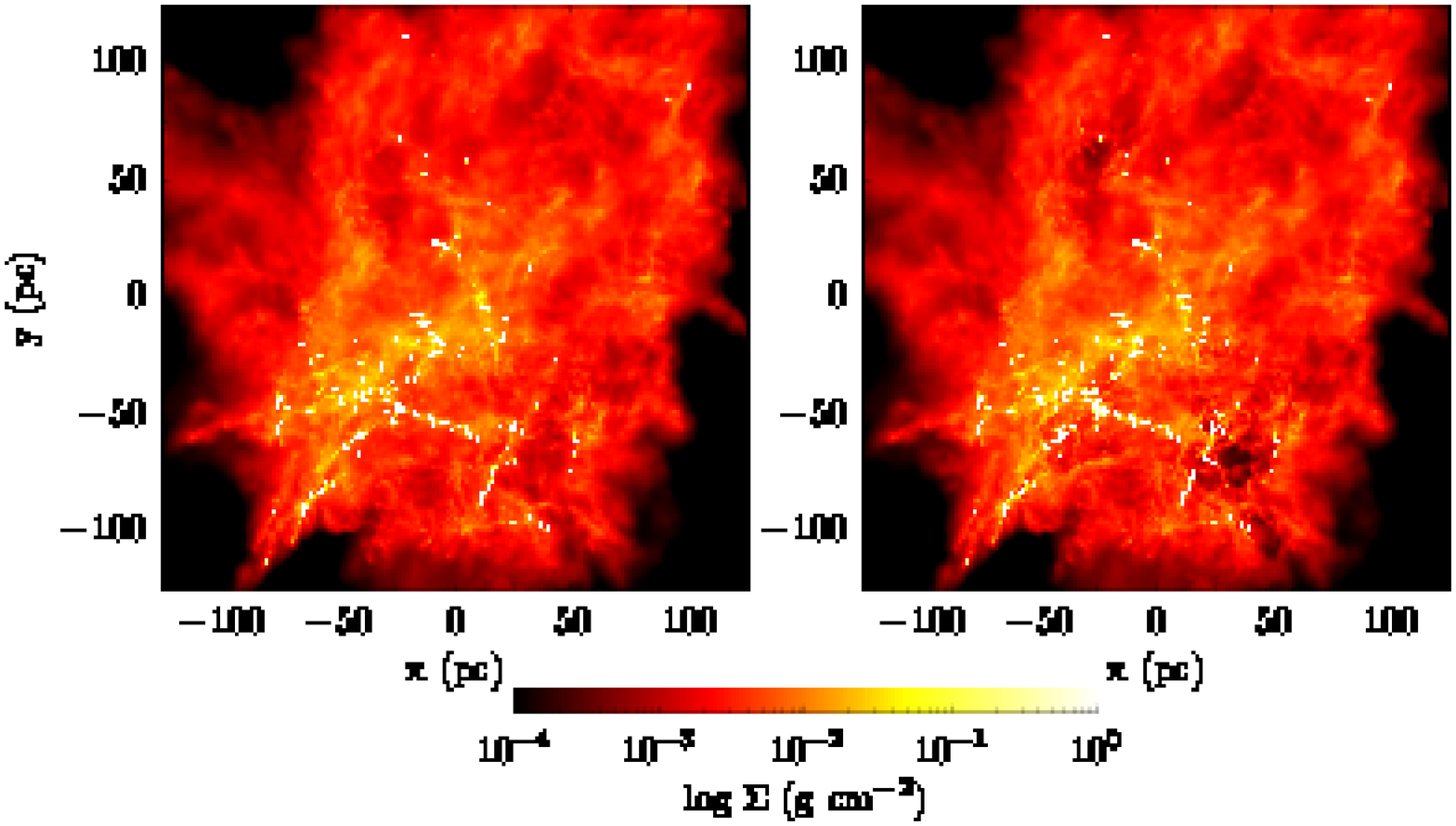}}     
     \vspace{.1in}
      \subfloat[Run UC]{\includegraphics[width=0.47\textwidth]{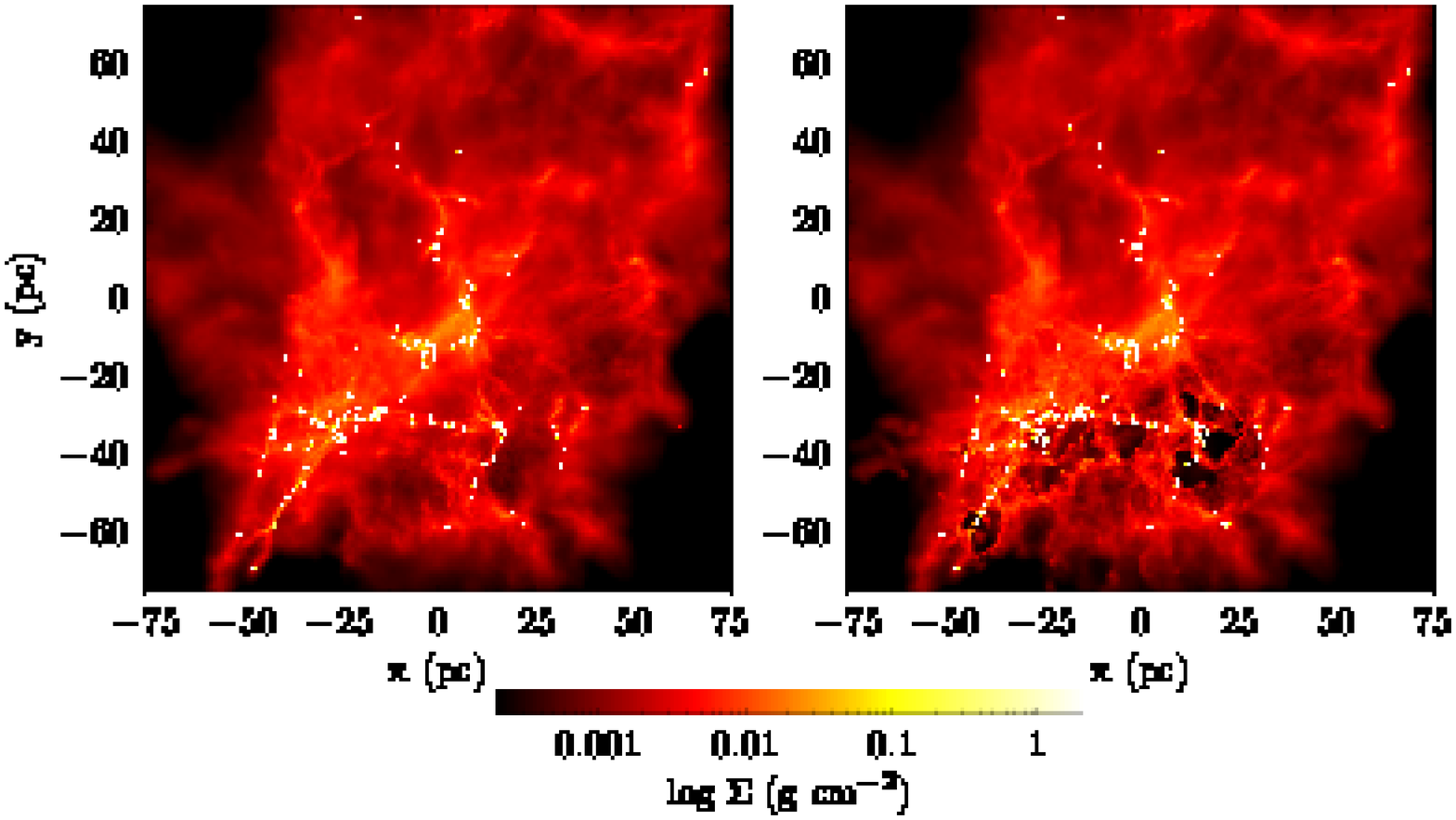}}
     \hspace{.1in}
     \subfloat[Run UU]{\includegraphics[width=0.47\textwidth]{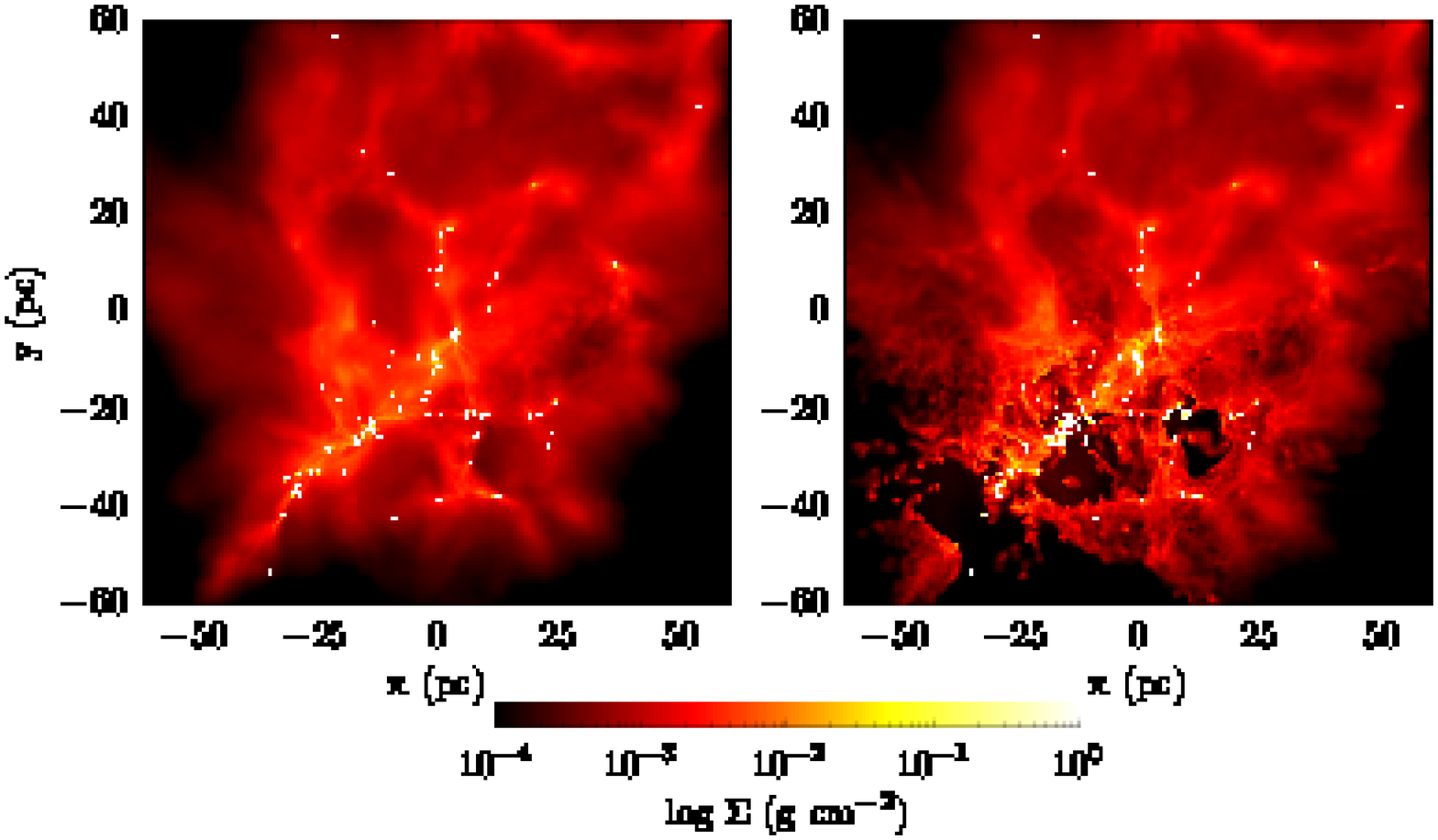}}     
     \vspace{.1in}
     \subfloat[Run UV]{\includegraphics[width=0.47\textwidth]{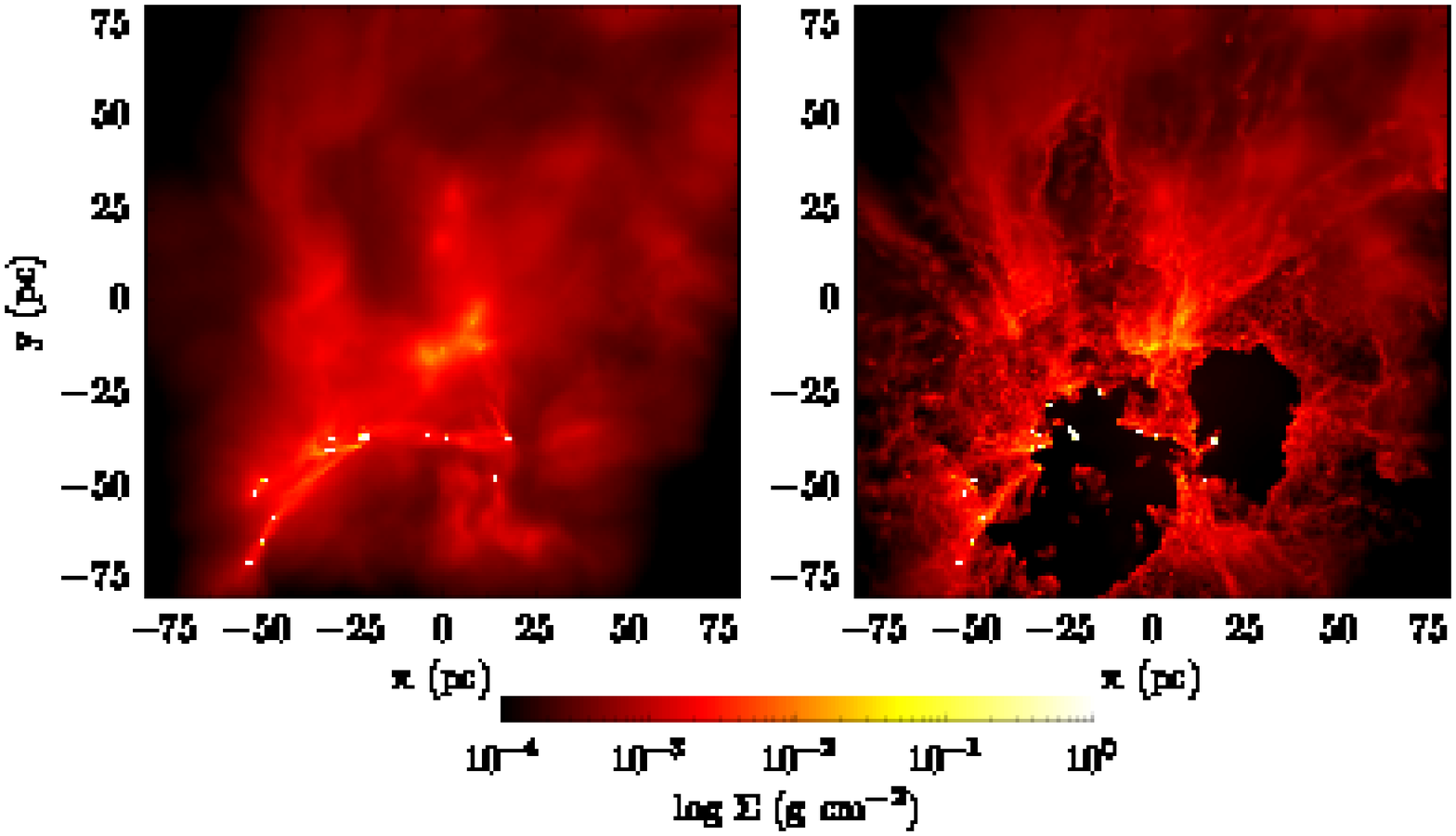}}
     \hspace{.1in}
     \subfloat[Run UF]{\includegraphics[width=0.47\textwidth]{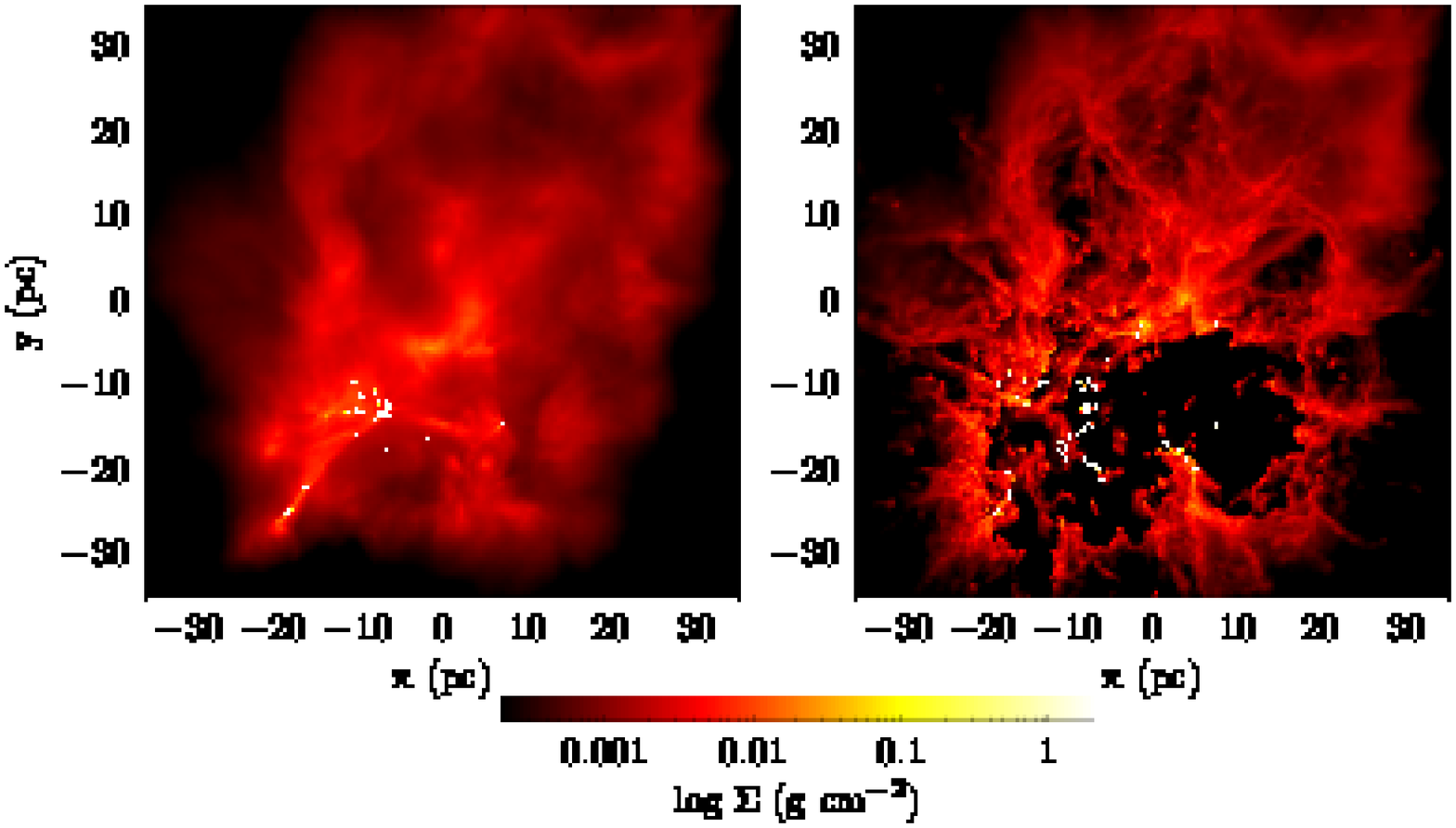}}     
     \vspace{.1in}
     \subfloat[Run UP]{\includegraphics[width=0.47\textwidth]{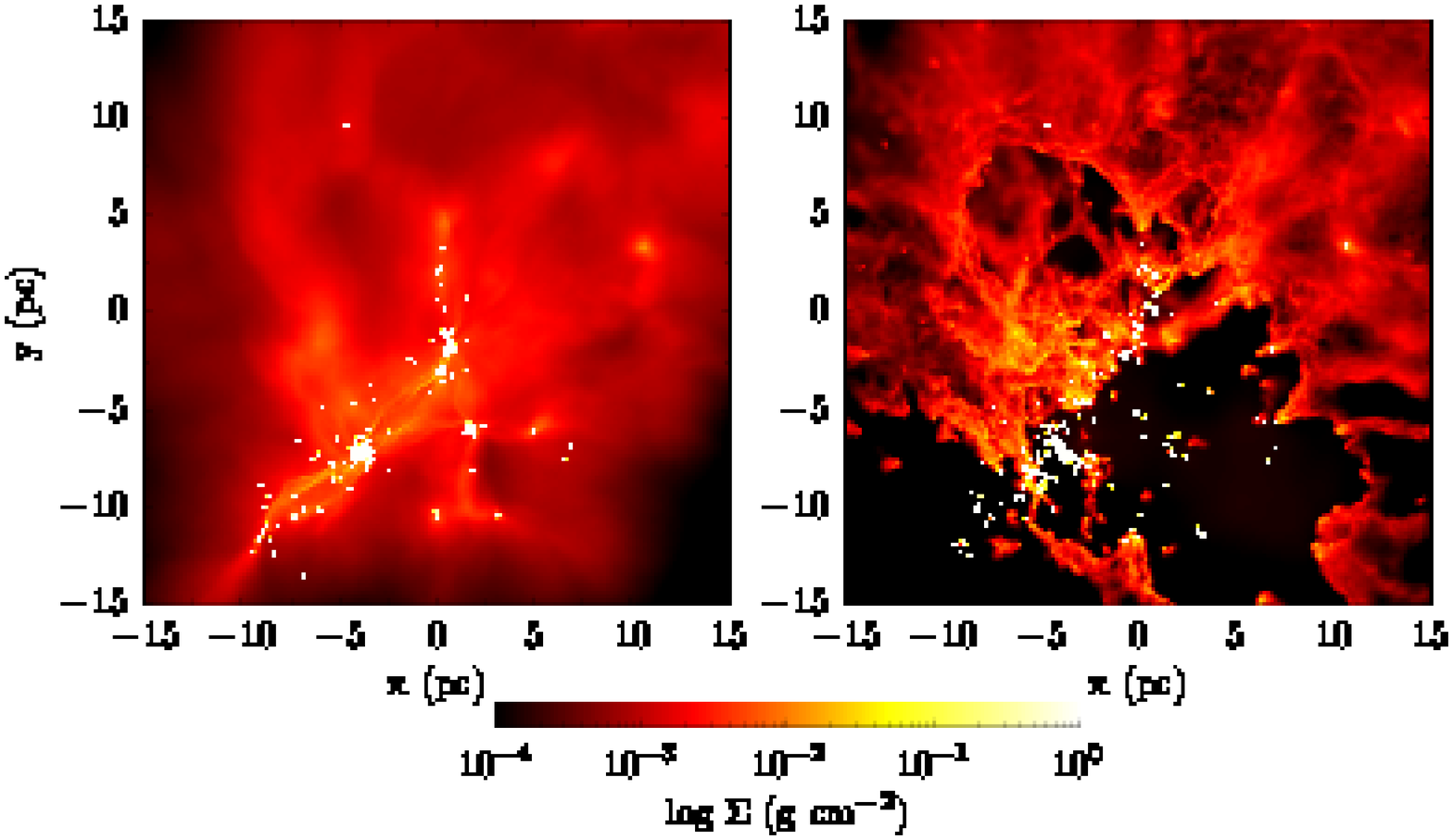}}     
     \hspace{.1in}
     \subfloat[Run UQ]{\includegraphics[width=0.47\textwidth]{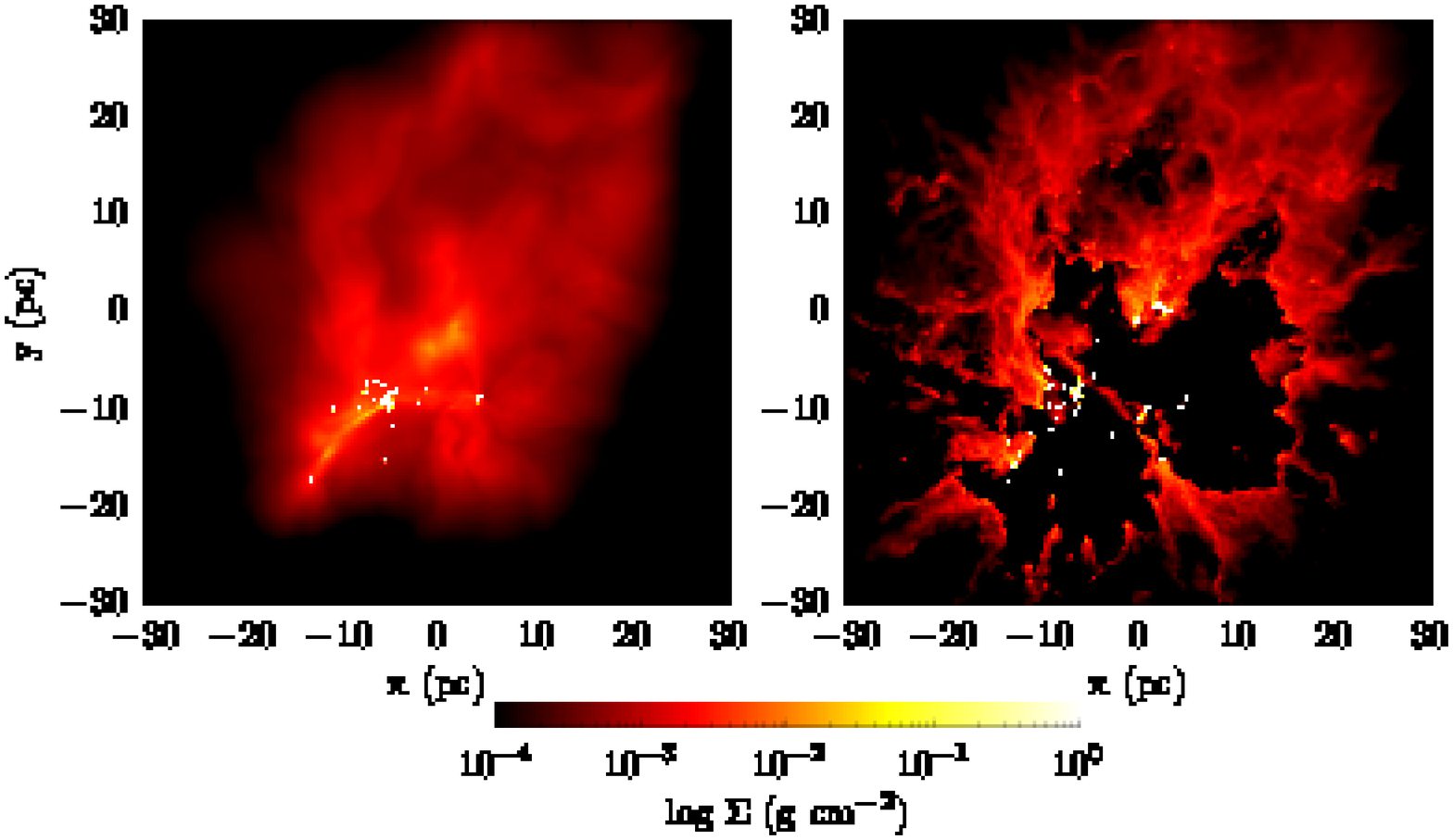}}
     \hspace{.1in}
    \caption{Comparison of column--density maps viewed down the z--axis of the end states of the control (left panels) and ionized (right panels) runs. Simulation parameters are given in Table 1.}
   \label{fig:compare_end}
\end{figure*}

\begin{figure*}
     \centering
     \subfloat[Run UB]{\includegraphics[width=0.32\textwidth]{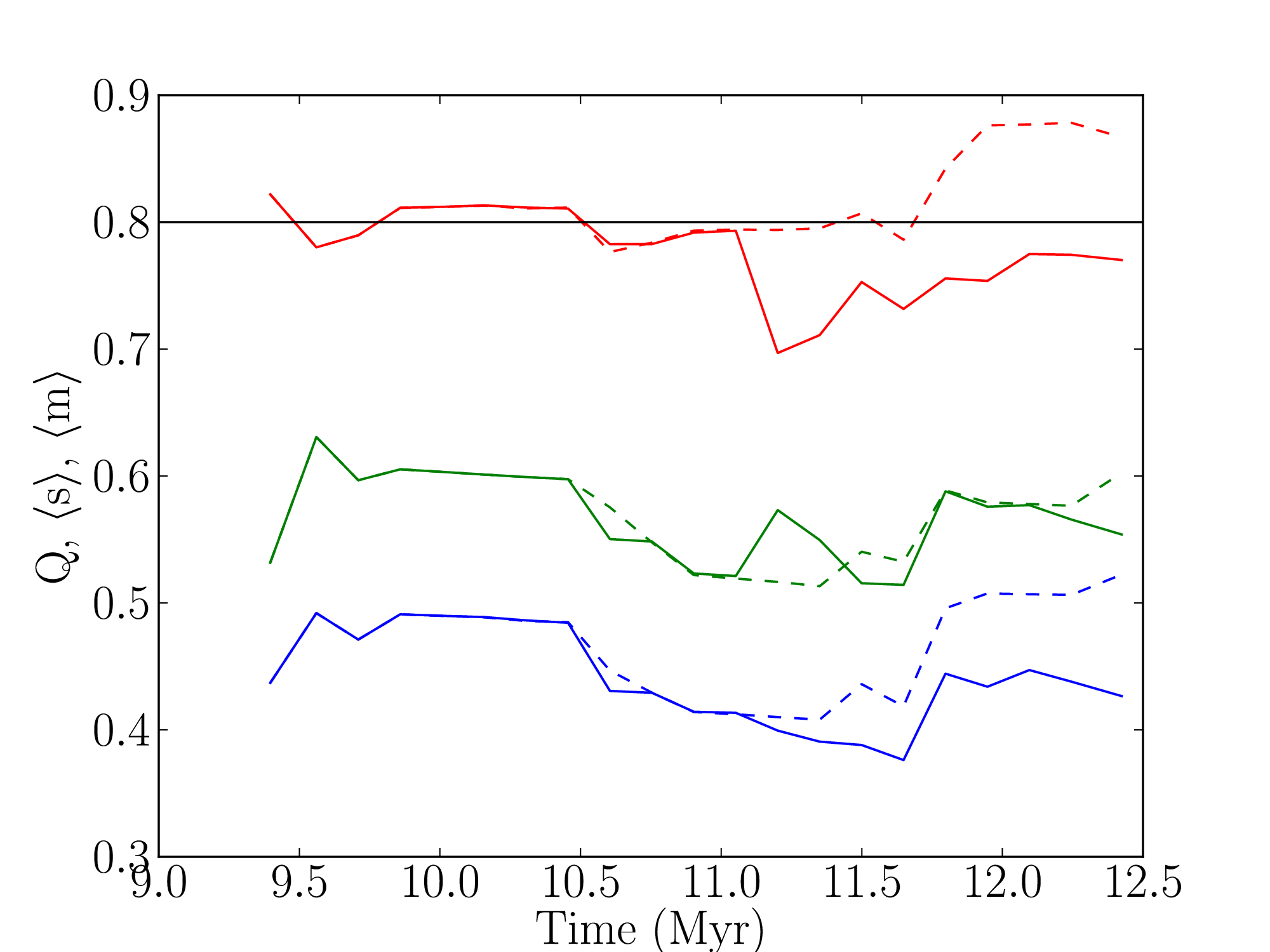}}     
     \hspace{.1in}
     \subfloat[Run UC]{\includegraphics[width=0.32\textwidth]{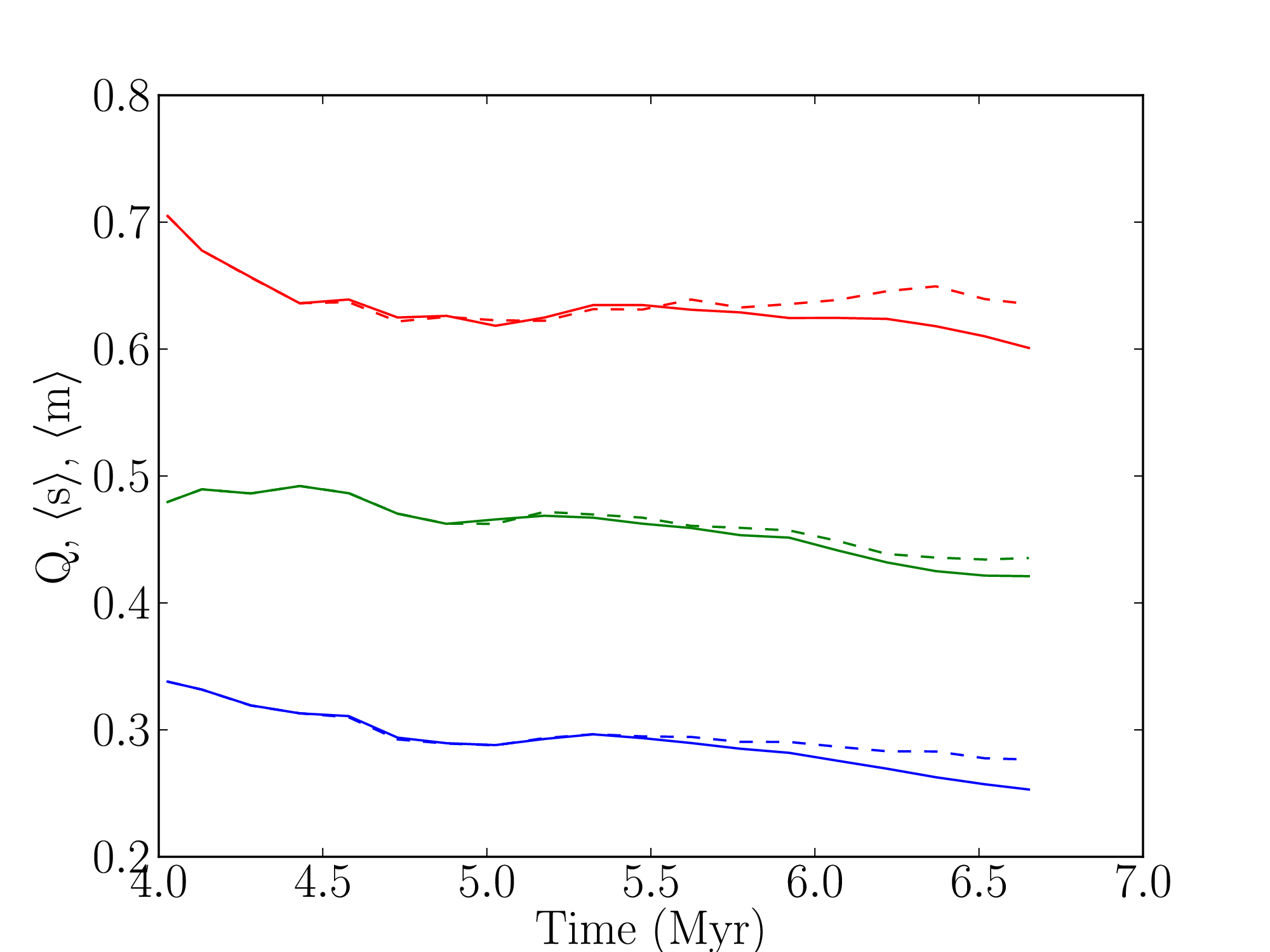}}
     \hspace{.1in}
     \subfloat[Run UZ]{\includegraphics[width=0.32\textwidth]{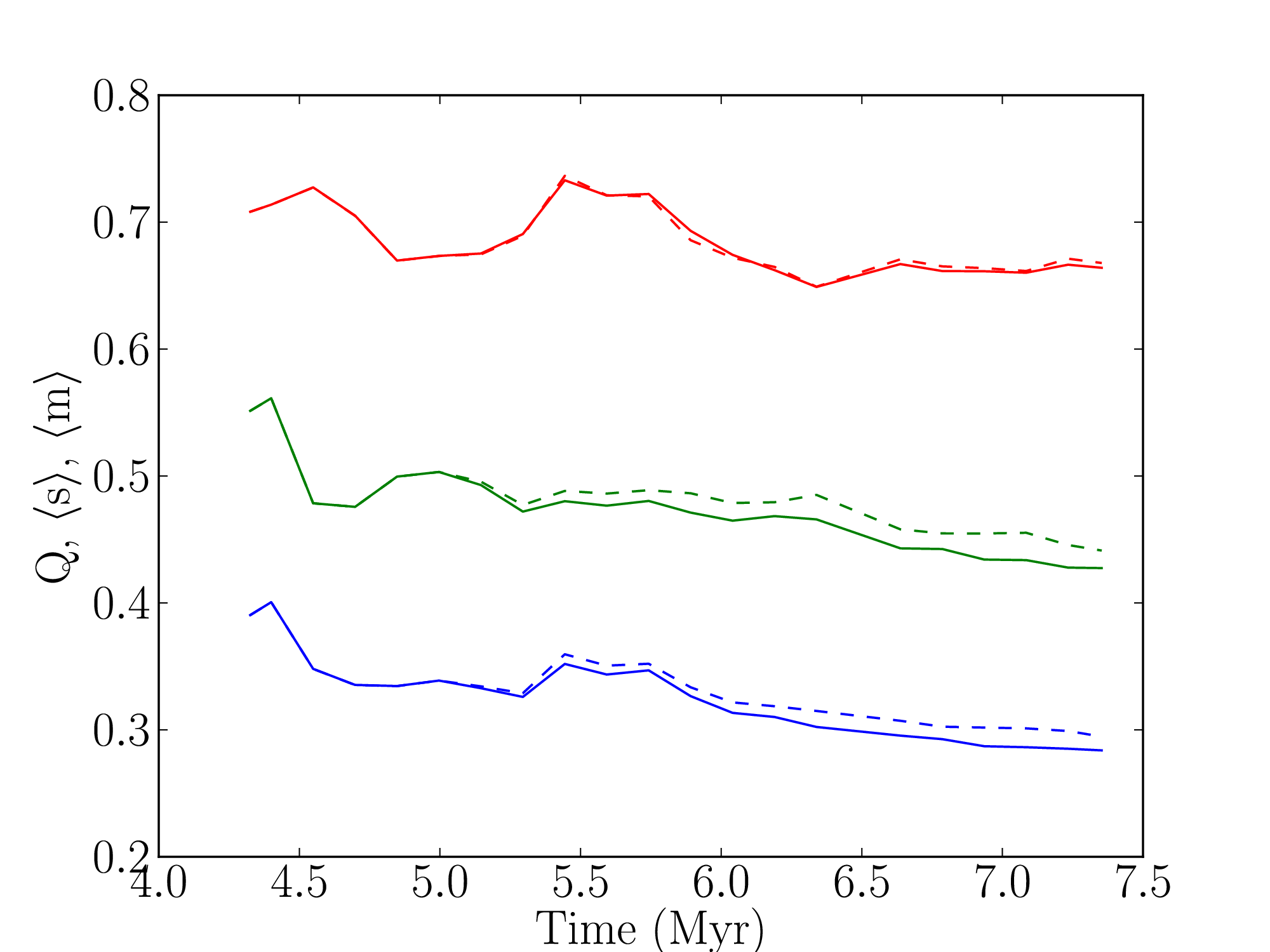}}     
     \vspace{.1in}
     \subfloat[Run UU]{\includegraphics[width=0.32\textwidth]{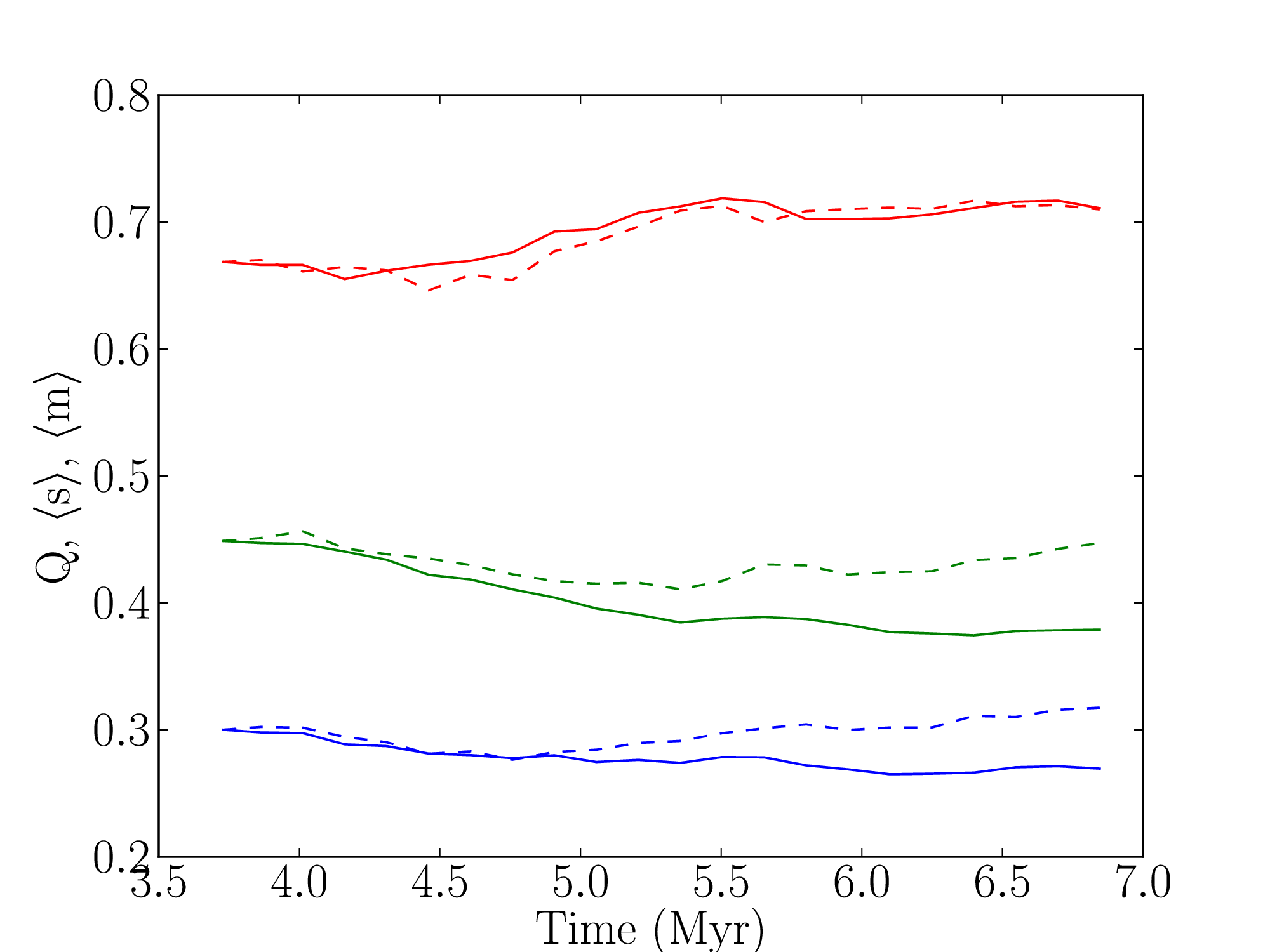}}
     \hspace{.1in}
     \subfloat[Run UV]{\includegraphics[width=0.32\textwidth]{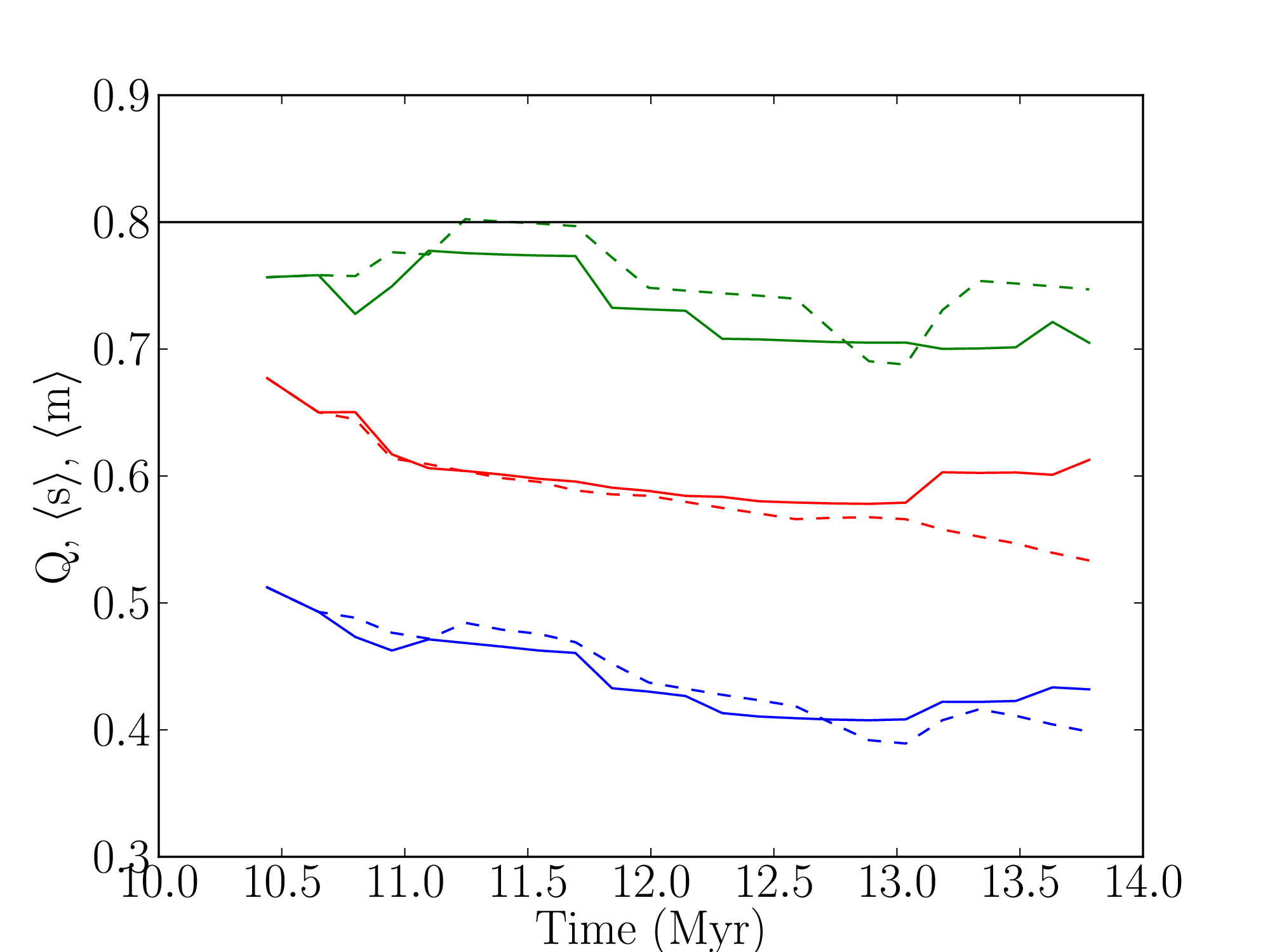}}
     \hspace{.1in}
     \subfloat[Run UF]{\includegraphics[width=0.32\textwidth]{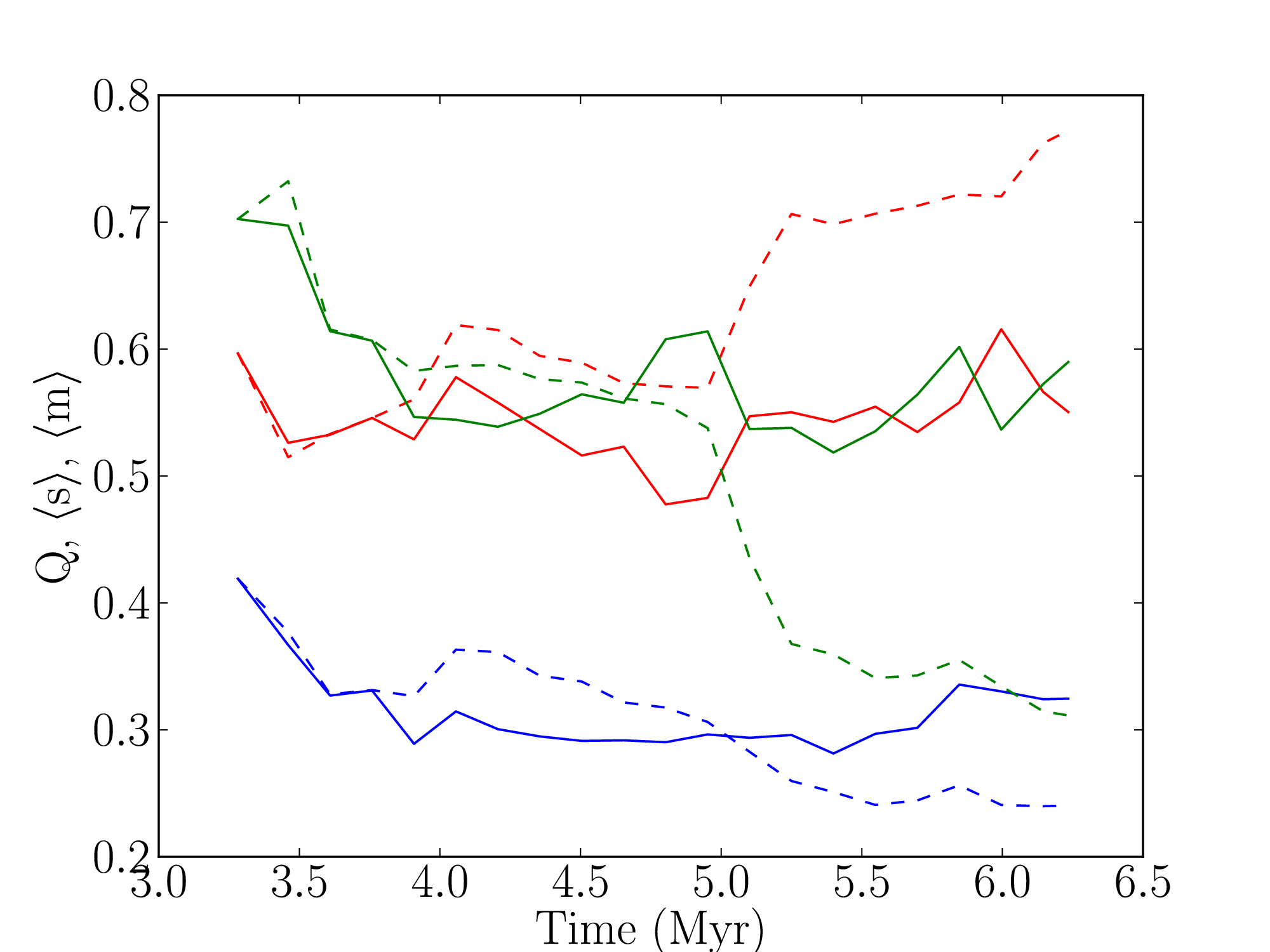}}
     \vspace{.1in}
     \subfloat[Run UP]{\includegraphics[width=0.32\textwidth]{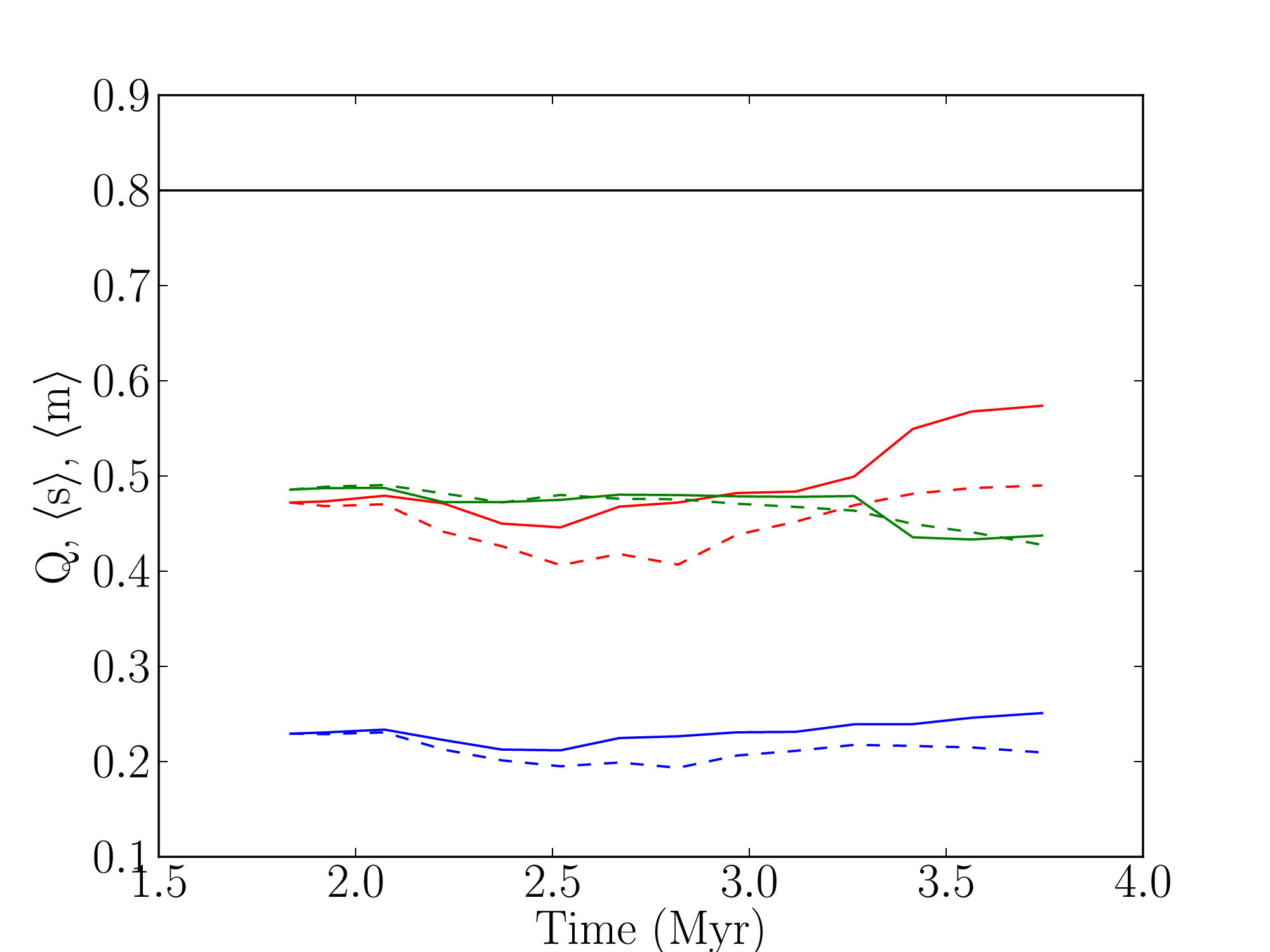}}
     \hspace{.1in}
     \subfloat[Run UQ]{\includegraphics[width=0.32\textwidth]{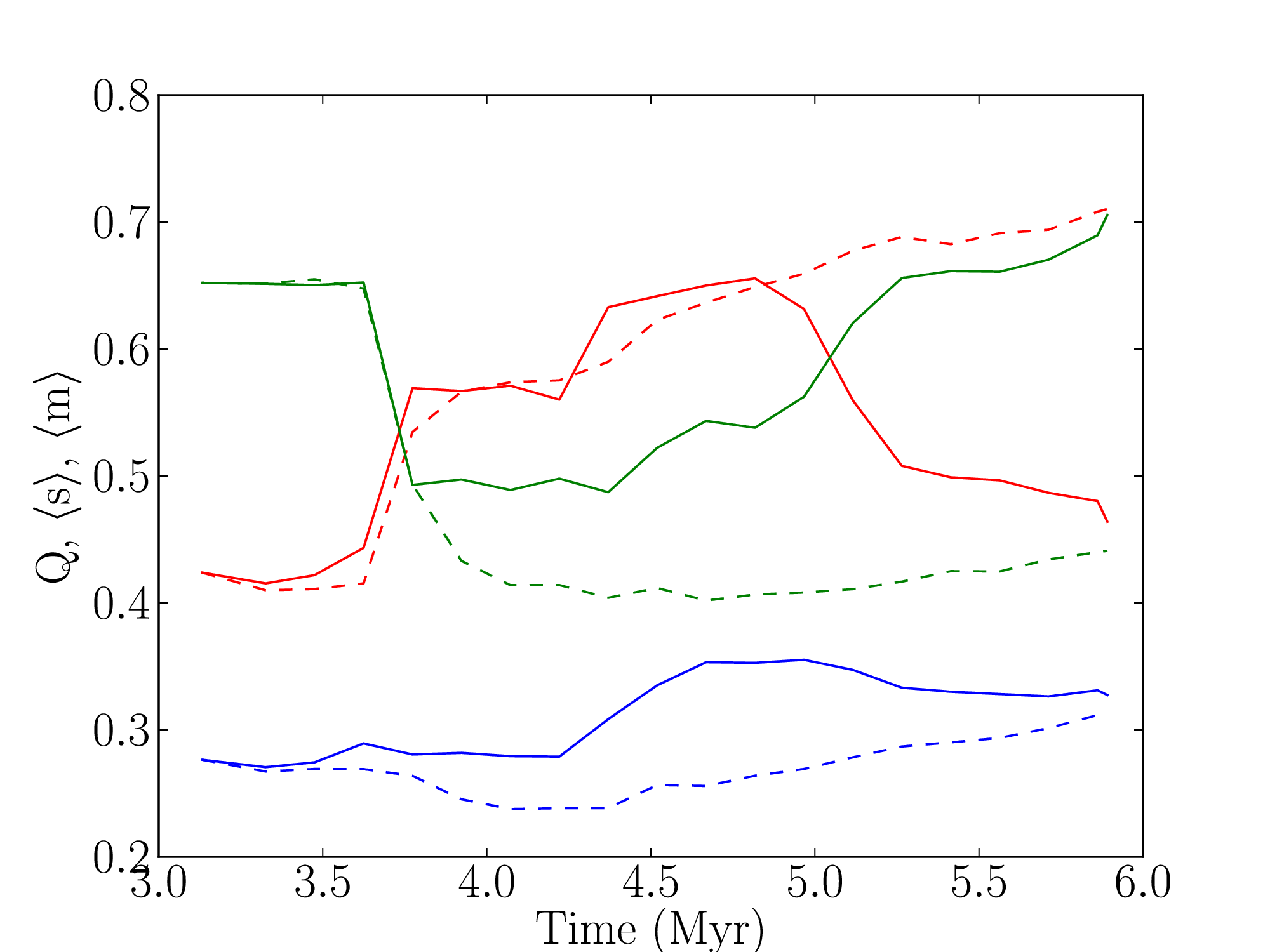}}
     \hspace{.1in}
     \caption{Comparison of the time--evolution of the Q--parameter ($Q=\langle l \rangle/\langle s \rangle$, red lines), where $\langle s \rangle$ (green lines) denotes the mean separation between pairs of clusters and $\langle l \rangle$ denotes the mean edge--length of the minimum spanning tree connecting all objects (blue lines), in the ionized (solid lines) and control (dashed lines) in al runs. Q$<$0.8 indicates a fractally--substructured system, Q$>0.8$ indicates a smooth distribution of clusters with a global density gradient and Q=0.8 indicates a uniform distribution.}
   \label{fig:compare_Q}
\end{figure*}

In Figure \ref{fig:compare_end} we plot the endpoints of each pair of control/ionized simulations after as close as possible to 3Myr of evolution from the epoch at which ionization is enabled in the relevant ionized run. These images give a direct indication of how the structure of the systems has been affected by the feedback and the ways in which they are different as a consequence.\\
\indent In common with the bound clouds, the control runs exhibit a filamentary structure, with most of the star formation occurring along the filaments, and with the more populous clusters appearing at filament junctions. The filaments serve as conduits for gas and sometimes stars which feed these clusters. Some clusters gain enough stars in this way that dynamical interactions eject members, and this process is largely responsible for the more distributed objects which do not appear to be associated with significant quantities of dense gas.\\
\indent Turning to the ionized runs, as we noted in Paper III, all simulations save Run UZ exhibit bubble--like structures created by expanding HII regions, although they are not always well--cleared. Careful examination of Figure \ref{fig:compare_end} reveals that, in virtually all cases, the bubbles in the ionized simulations are at locations where the gas in the corresponding control run is of lower than average density. In Runs UB, UC and UU, the effect of feedback in fact appears mostly to be to increase the gas density contrast in low--density voids which were already present in the turbulent velocity fields of the clouds. This is similar to the phenomenon we reported in \cite{2011MNRAS.414..321D} where HII preferentially seeks out pre--existing holes in the density field. In the case where the escape velocity of the cluster is low, the expansion of this hot gas may be sufficient to do substantial damage to the cloud. Otherwise, the HII simply quiescently fills the bubbles. The mere presence of bubbles does not therefore signify that a cloud has been strongly dynamically influenced by feedback.\\
\indent It also appears from Figure \ref{fig:compare_end} that there is an anticorrellation in the ionized runs between prominent, well--evacuated bubbles and surviving filamentary structure from the pre--feedback cloud. Taking extremes, the filamentary structure of Run UZ is virtually untouched by feedback and there are no well--defined bubbles in this simulation, whereas in Runs UF and UQ, the appearances of the clouds are completely dominated by very large bubbles and almost no vestiges of the original gaseous filaments remain.\\
\indent In most simulations, it appears that the geometrical distribution of the stars or subclusters is rather similar between corresponding control and ionized calculations. To place this observation on a more objective footing, we utilize the Q--parameter devised by \cite{2004MNRAS.348..589C} as a scale--independent means of characterizing the geometrical structure of the stellar components of each pair of control/ionized simulations. The Q--parameter of a distribution of points is defined as the ratio of the mean edge--length $l$ of the unique minimum spanning tree connecting the points to the mean distance $s$ between two points. \cite{2004MNRAS.348..589C} showed empirically that Q has a value of 0.8 for a uniform distribution. Values less than 0.8 indicate a fractal or subclusterd distribution, with smaller values denoting smaller fractal dimensions. Values in excess of 0.8 indicate a smooth distribution but with an overall density gradient, larger values implying steeper gradients.\\
\indent Q can be defined in any number of dimensions, and here we use the two--dimensional projections of our clusters along the z--axis. We found in Paper II that the results of this analysis were immune to projection effects.  We plot $\langle l\rangle$, $\langle s\rangle$ and Q as functions of time for all simulations in Figure \ref{fig:compare_Q}. The plot confirms the impression giving by simply looking at the clouds and clusters in that, in most cases, in common with the bound clouds from Paper I, the structure of the embedded clusters formed in these simulations is little altered by feedback. In the control runs and ionized runs alike, the Q--values in most of the simulations do not change very much in the $\sim$3Myr timespans of interest here. The exceptions are Runs UF and UQ. The control runs evolve generally towards smoother systems with Q--values somewhat in excess of 0.8, indicating clusters which may be characterized by radial density profiles. The ionized runs, by contrast, retain Q--values less 0.8 which, diagnostic of significant subclustering.\\
\indent That, in most cases, the structure of the clusters is largely unaffected is counterintuitive, given the profound changes suffered by the gas and clearly visible in the form of vast evacuated bubbles. This is particularly striking in the case of Run UP, where the gas structures in the control and ionized calculations are incomparably different, but the stellar distribution, whether quantified using the Q--parameter, or inferred from simple observation of the clouds, is almost the same. However, as we noted in Paper III, strong alterations of the gas structure, e.g. the presence of bubbles, is not a good indicator that the cloud has been strongly dynamically influenced by feedback.
\subsection{Star formation efficiency and numbers of stars}
\begin{figure*}
     \centering
     \subfloat[Run UB]{\includegraphics[width=0.32\textwidth]{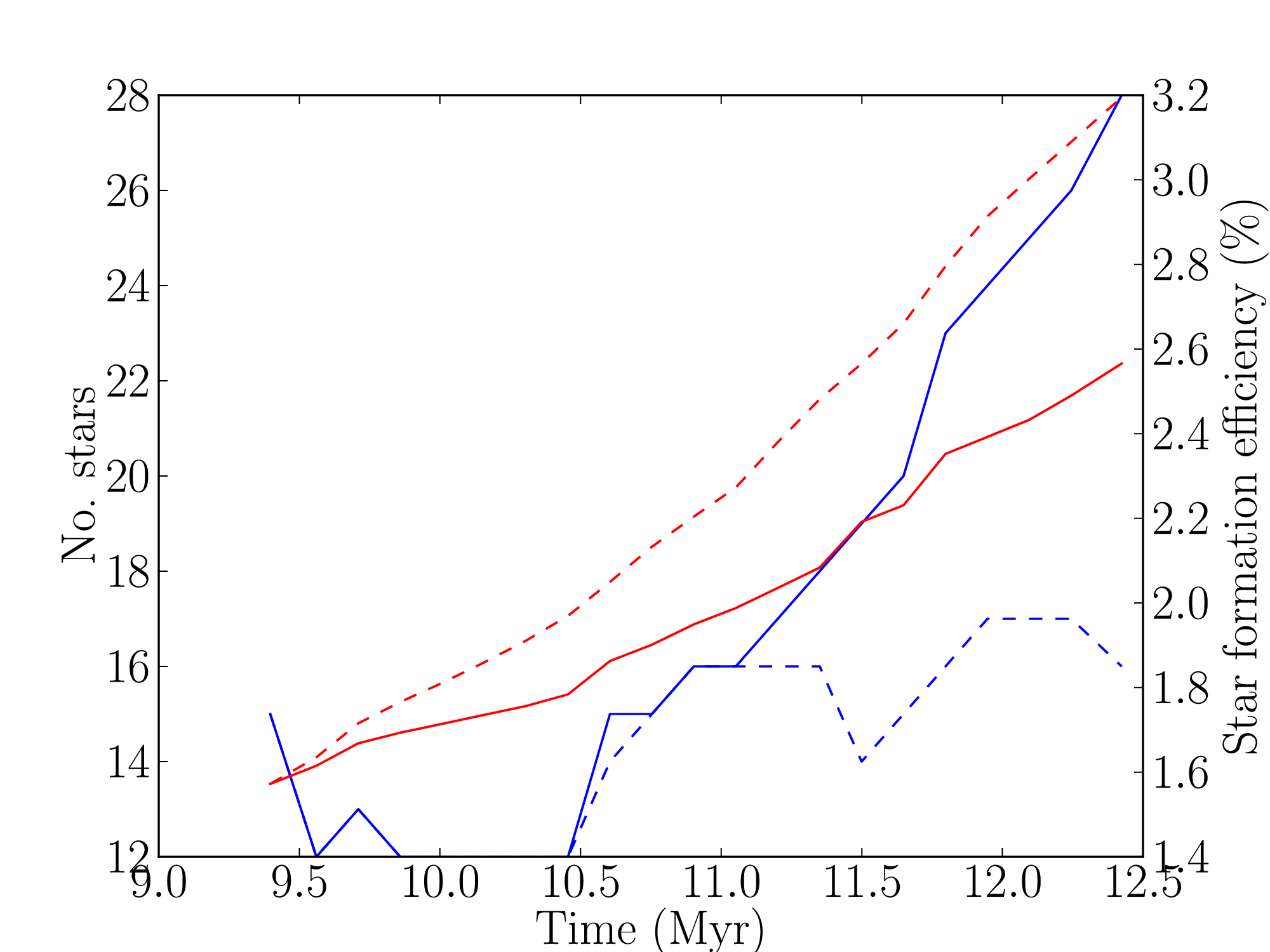}}     
     \hspace{.1in}
     \subfloat[Run UC]{\includegraphics[width=0.32\textwidth]{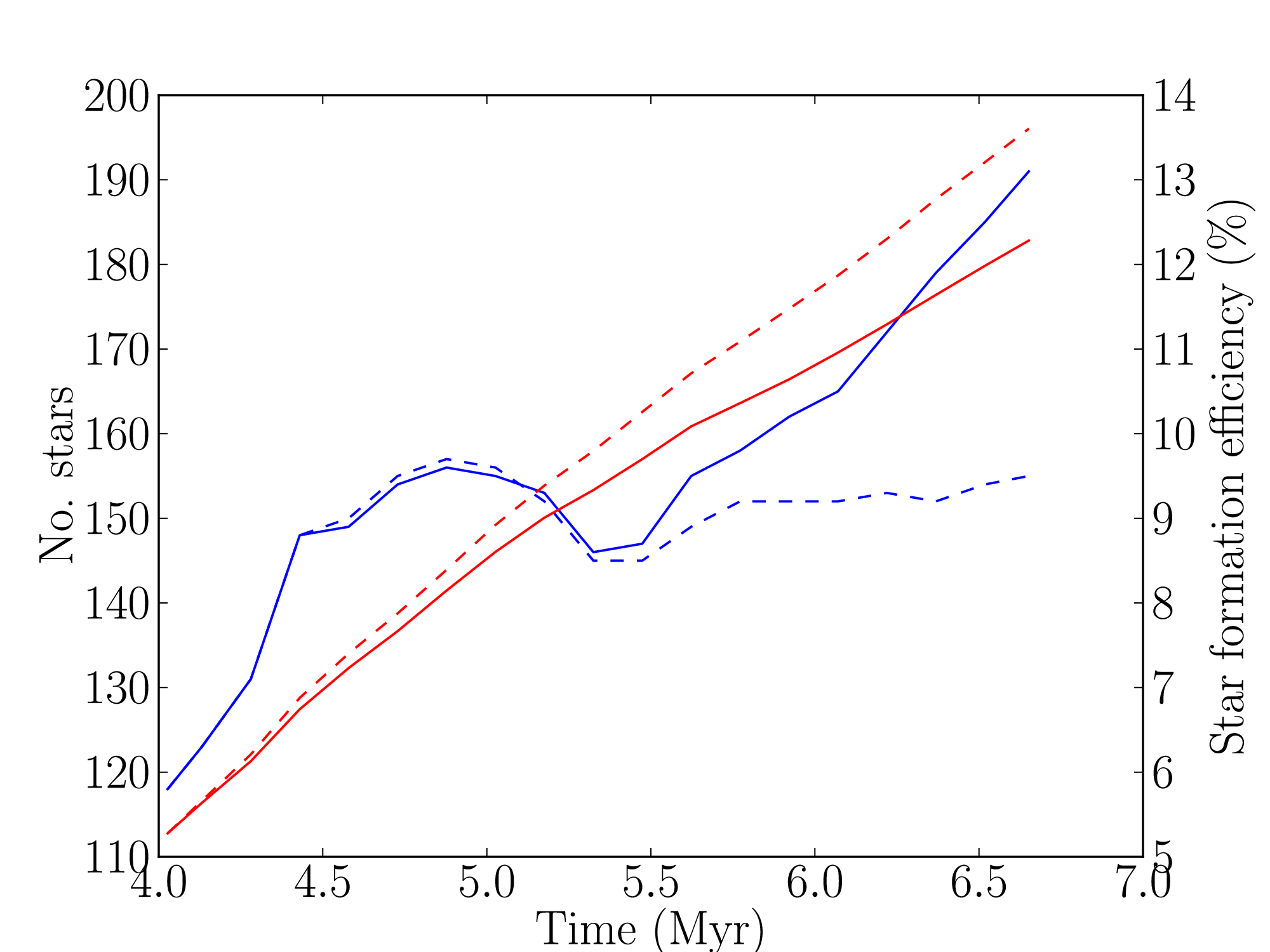}}
     \hspace{.1in}
     \subfloat[Run UZ]{\includegraphics[width=0.32\textwidth]{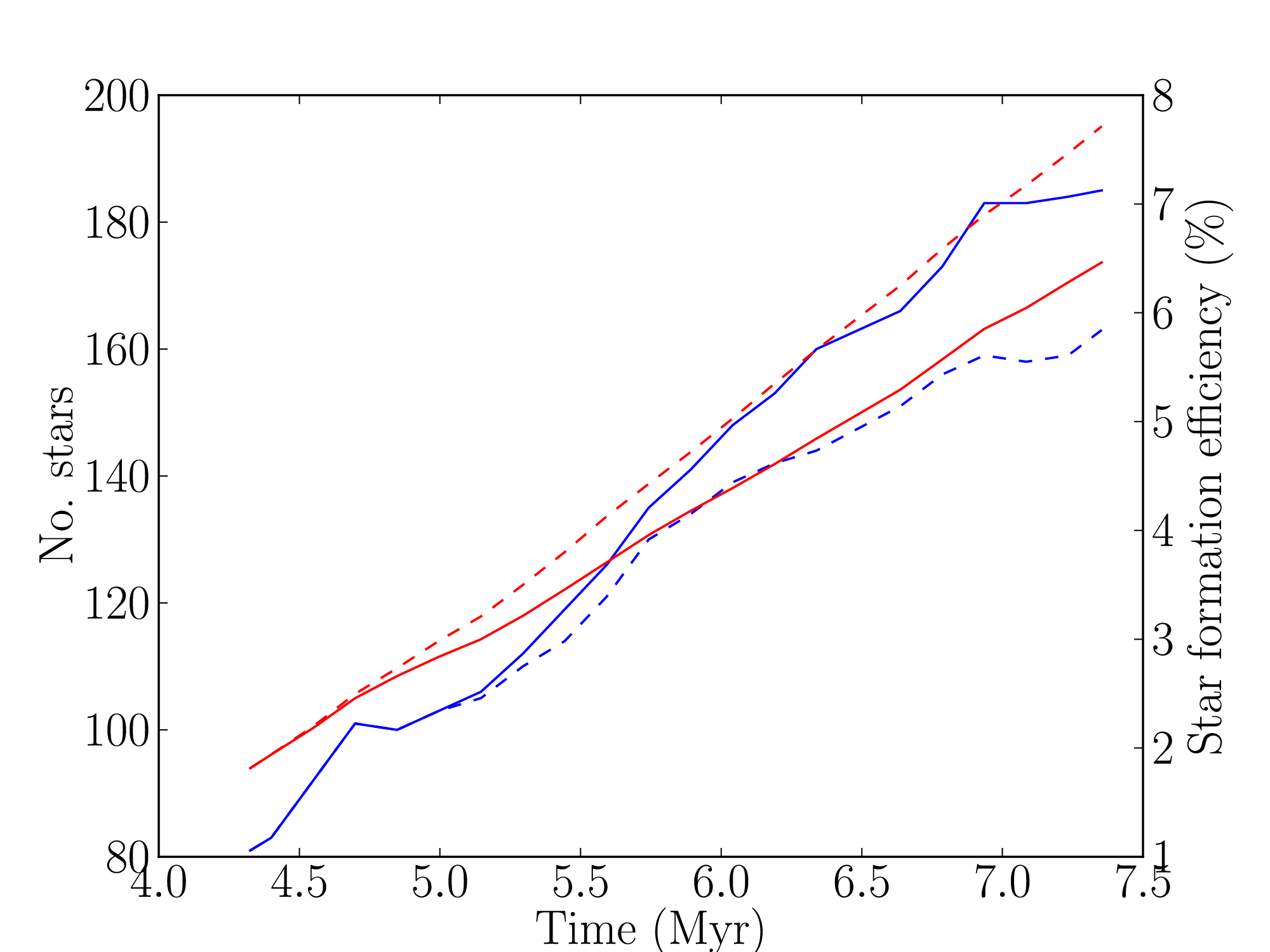}}     
     \vspace{.1in}
     \subfloat[Run UU]{\includegraphics[width=0.32\textwidth]{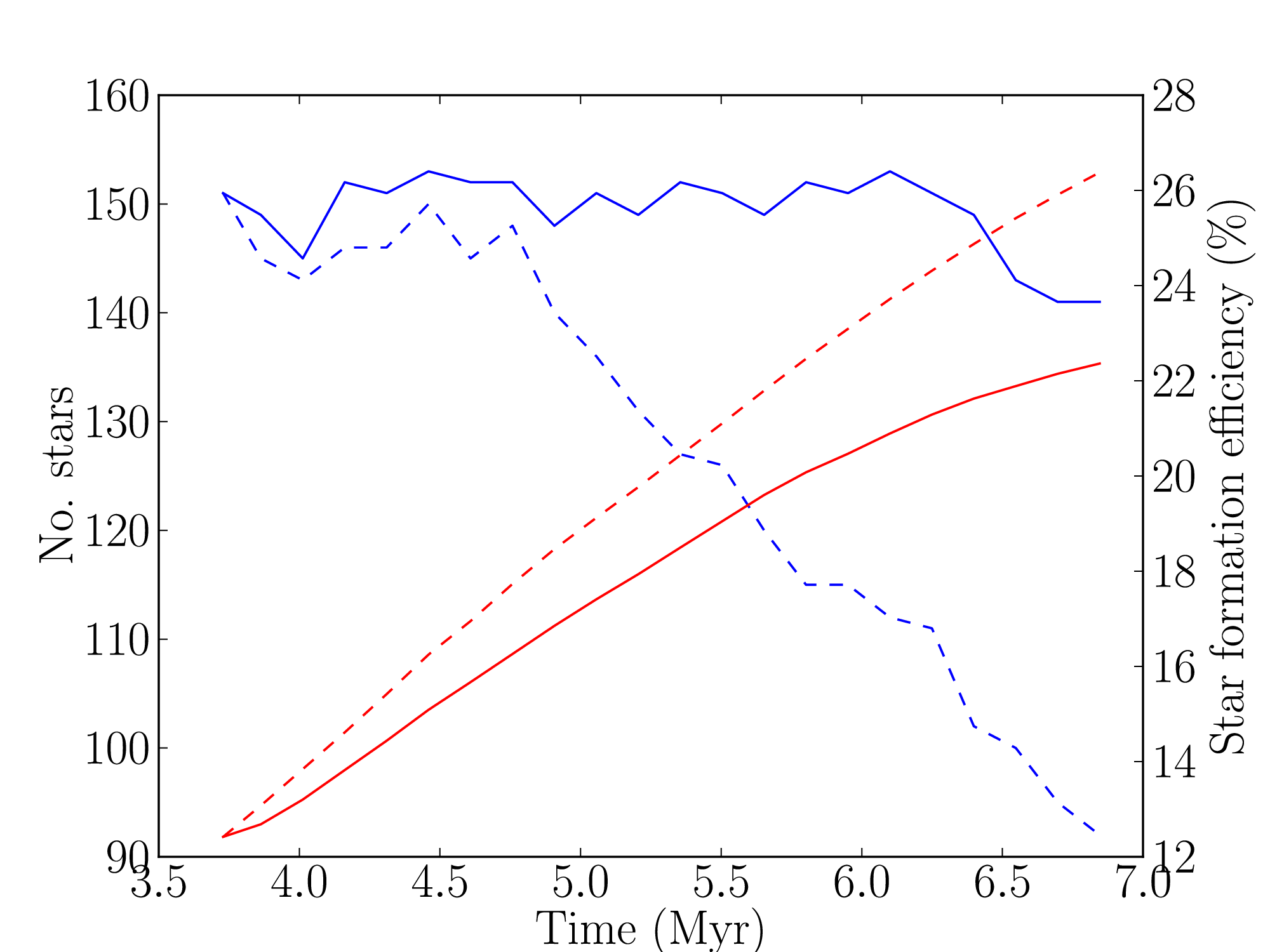}}
     \hspace{.1in}
     \subfloat[Run UV]{\includegraphics[width=0.32\textwidth]{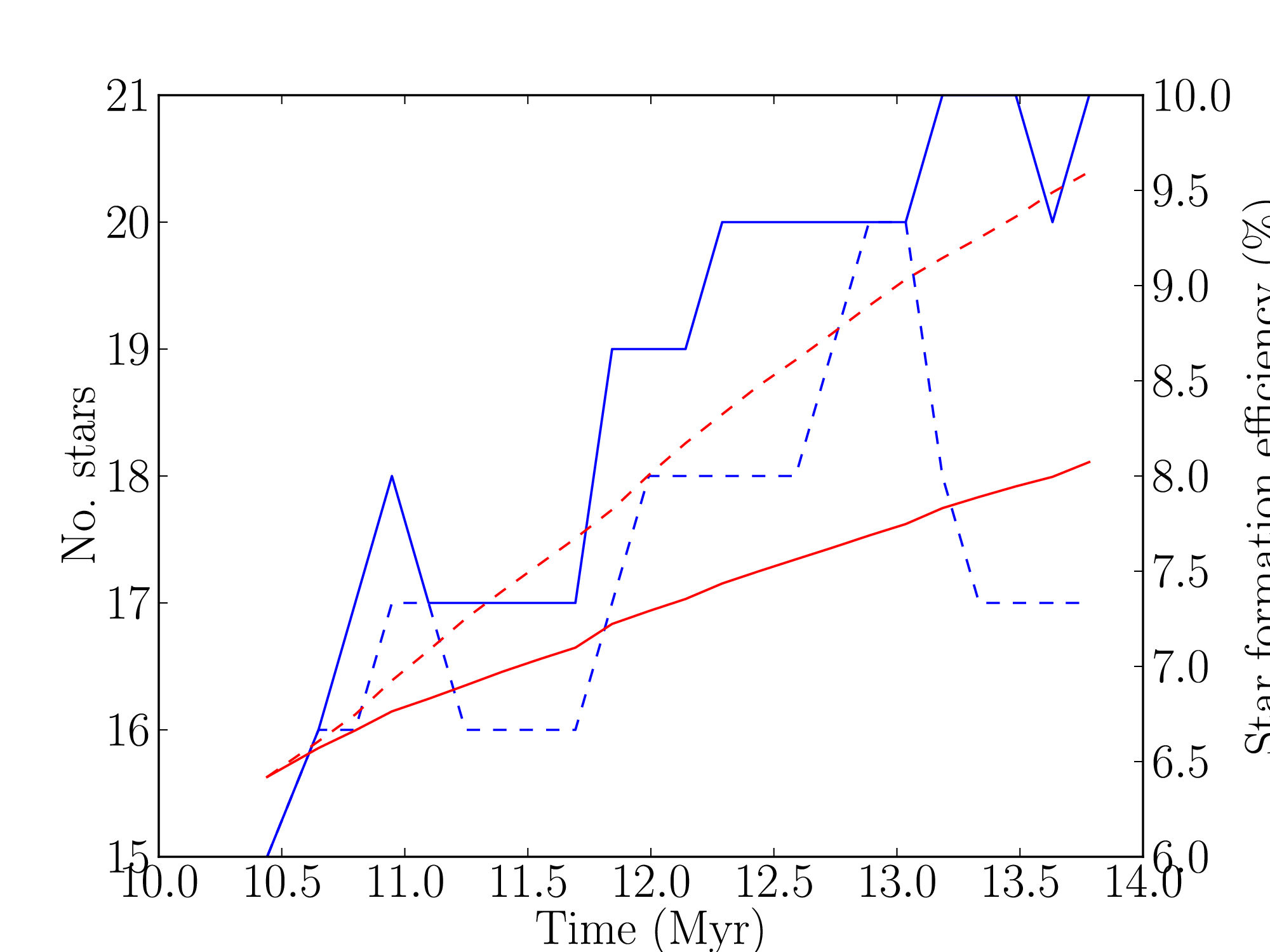}}
     \hspace{.1in}
     \subfloat[Run UF]{\includegraphics[width=0.32\textwidth]{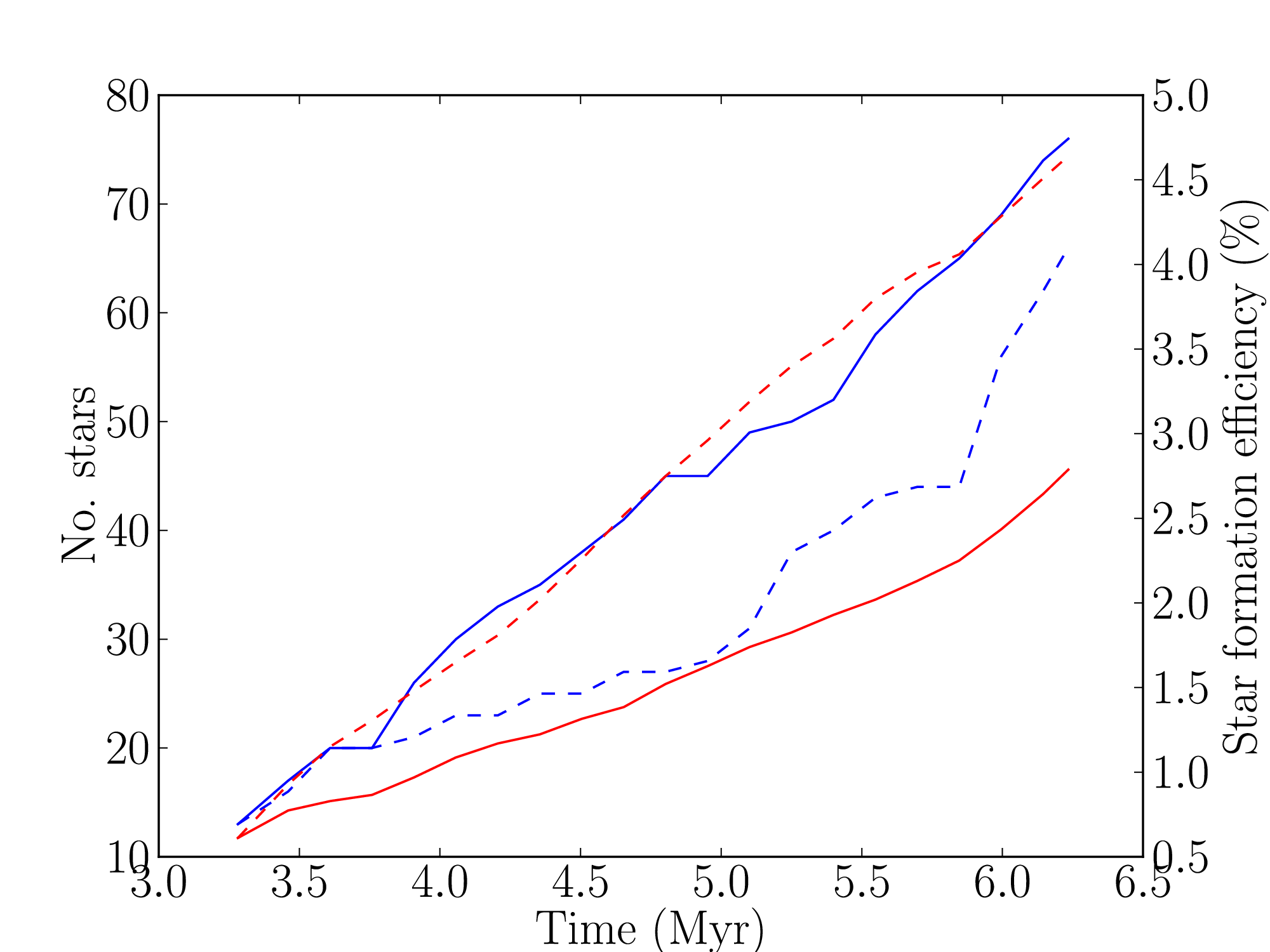}}
     \vspace{.1in}
     \subfloat[Run UP]{\includegraphics[width=0.32\textwidth]{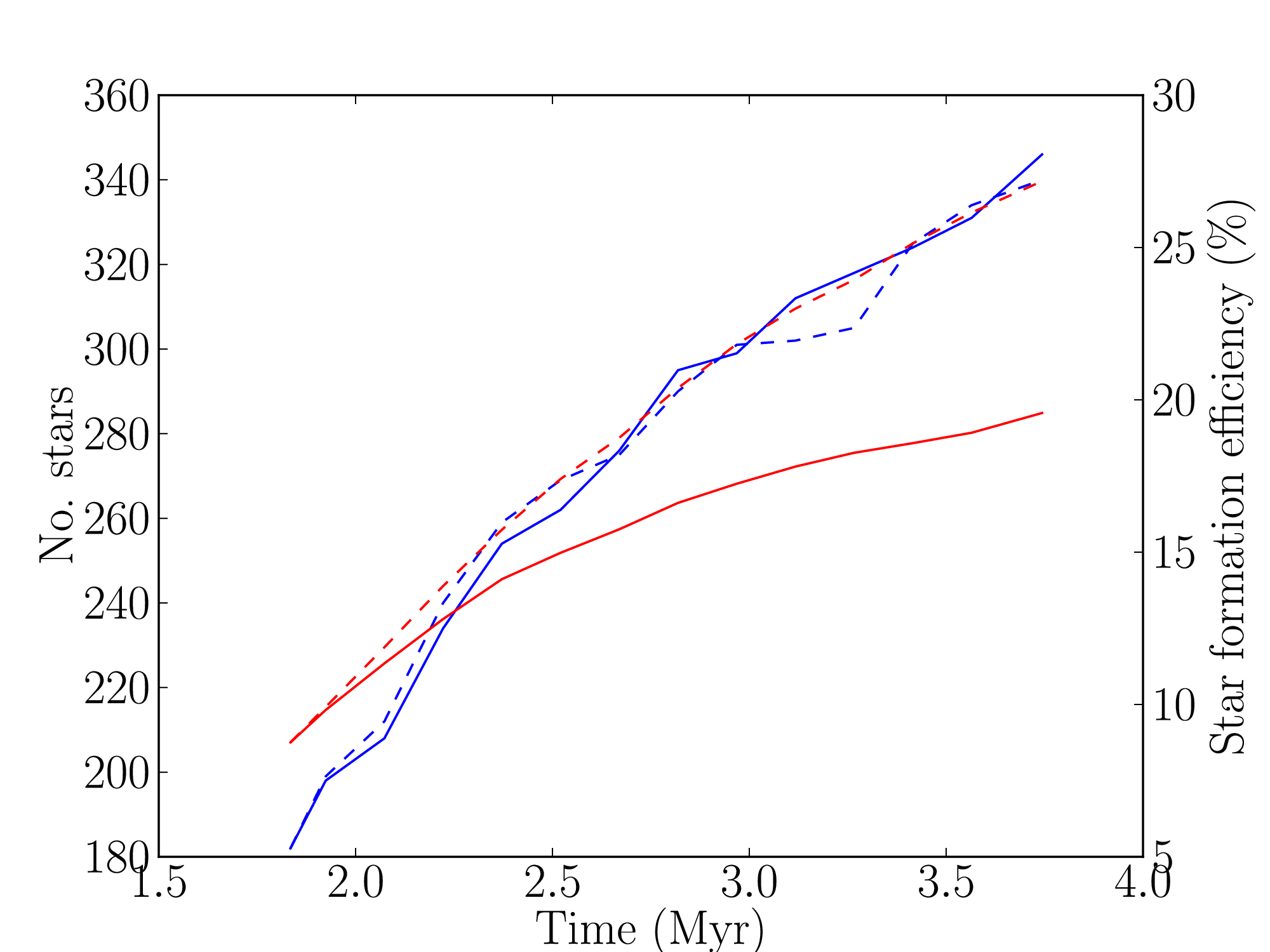}}
     \hspace{.1in}
     \subfloat[Run UQ]{\includegraphics[width=0.32\textwidth]{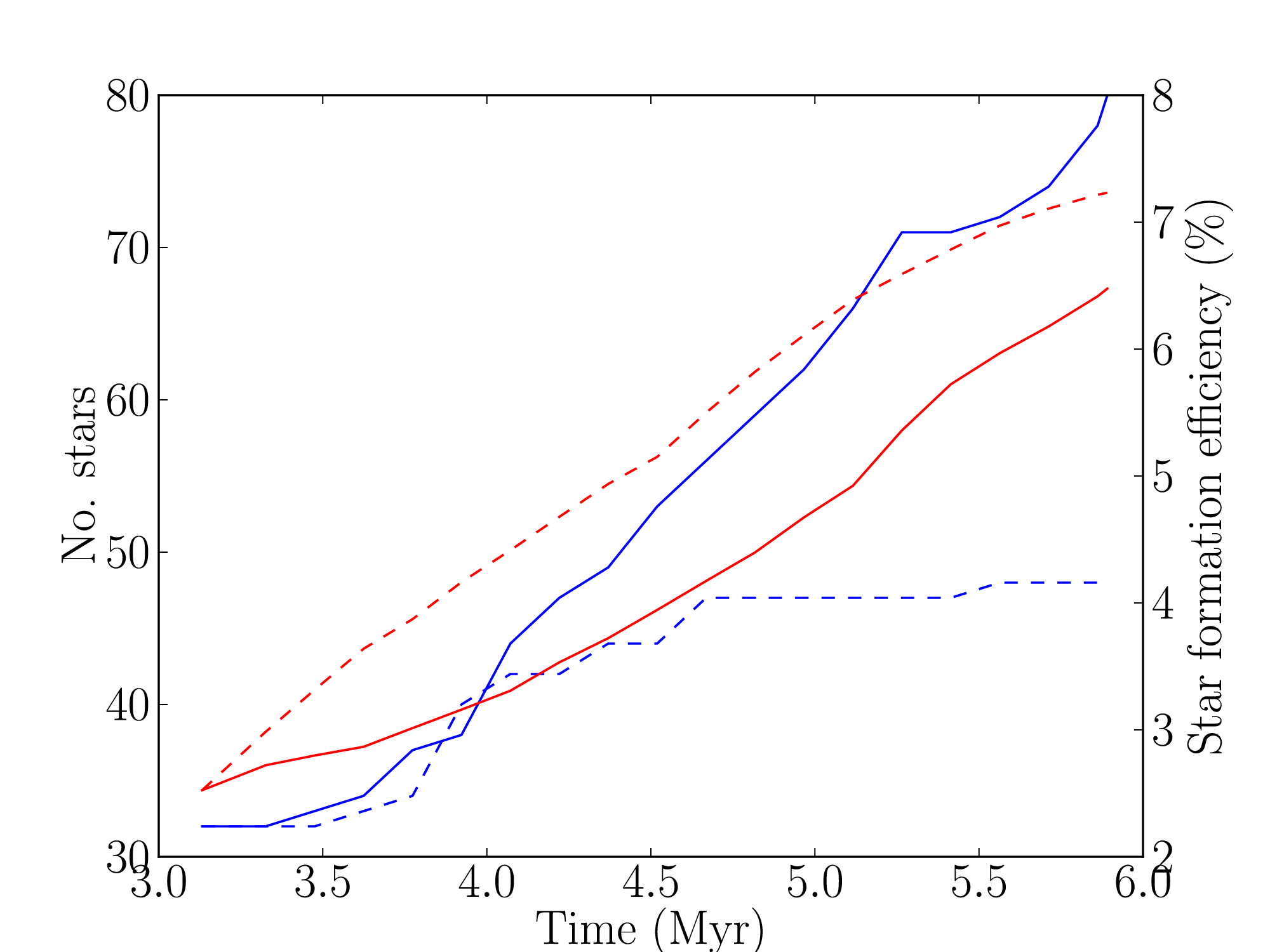}}
     \hspace{.1in}
     \caption{Comparison of the star formation efficiency (red lines) and numbers of stars (blue lines) in the ionized (solid lines) and control (dashed lines) runs.}
   \label{fig:compare_sfe}
\end{figure*}
\indent In Figure \ref{fig:compare_sfe}, we plot the time evolution of the star formation efficiency (red lines) and the numbers of objects (subclusters in Runs UB, UC, UZ, UU and UV, stars otherwise, blue lines), comparing the control runs (dashed lines) to the ionized runs (solid lines). In agreement with the simulations of the bound clouds, the star formation efficiency at every epoch is in every case lower in the ionized simulations, by factors ranging from less than 10 percent in Run UC to more than 40 percent in Run UF. Even in Run UZ, where the apparent effects of feedback are small, the star formation efficiency is smaller in the ionized run. Figure \ref{fig:compare_sfe} also shows that in most runs, most of the time, the instantaneous star formation rates (i.e. the gradients of the red lines) are smaller in the feedback runs. If the global star formation efficiencies or instantaneous star formation rates were used as arbitrators therefore, \emph{none of the clouds studied here or in Papers I and II show evidence of triggering}. One might expect feedback to assist an unbound cloud that is not vigorously forming stars in doing so but we see that even clouds such as UV where star formation is very sparse, the overall effect on the SFE and SFR is negative. The principal outcome is once again the disruption of the high-density gas from which most stars/subclusters are forming and accreting. This result is in accordance with the work of \cite{2010ApJ...725..134P}, who simulated the action of HII regions in a 10$^{3}$ M$_{\odot}$ gaseous disk. They found, again by comparison with control runs in which feedback was absent, that the effect of ionizing feedback was also to reduce both the SFE and SFR and that there was no evidence for triggering in their simulations.\\
\indent From the point of view of the numbers of objects formed, the picture is again somewhat complicated in Runs UB, UC, UZ, UU and UV by the fact that the sink particles in these runs represent small subclusters and are permitted to merge if they approach each other within small separations (0.1--0.5pc) and are mutually bound. That such mergers take place can be seen from the non--monotonic nature of the plots for these systems, particularly in the case of Run UU in which a series of mergers at a filament junction in the control simulation substantially reduces the number of separate objects and results in the formation of a massive cluster. In the corresponding ionized simulation, the partial destruction of the filaments and local gas expulsion reduces the merger frequency. This is true in all the simulations where sinks are permitted to merge, so that their ionized runs exhibit larger numbers of objects. Contrary to what was seen in the case of the bound clouds, here the number of stars formed in the calculations where sinks represented individual stars and mergers are forbidden (UF, UP, UQ) is also always larger in the ionized simulations. \cite{2010ApJ...725..134P} found that ionizing feedback in their simulations reduced the number of objects forming. While it seems safe to conclude that feedback decreases star--formation rates and efficiencies, it is difficult to make any general inferences about its effect on the total numbers of objects formed.\\
\subsection{Triggered star formation}
\begin{figure*}
\includegraphics[width=0.73\textwidth]{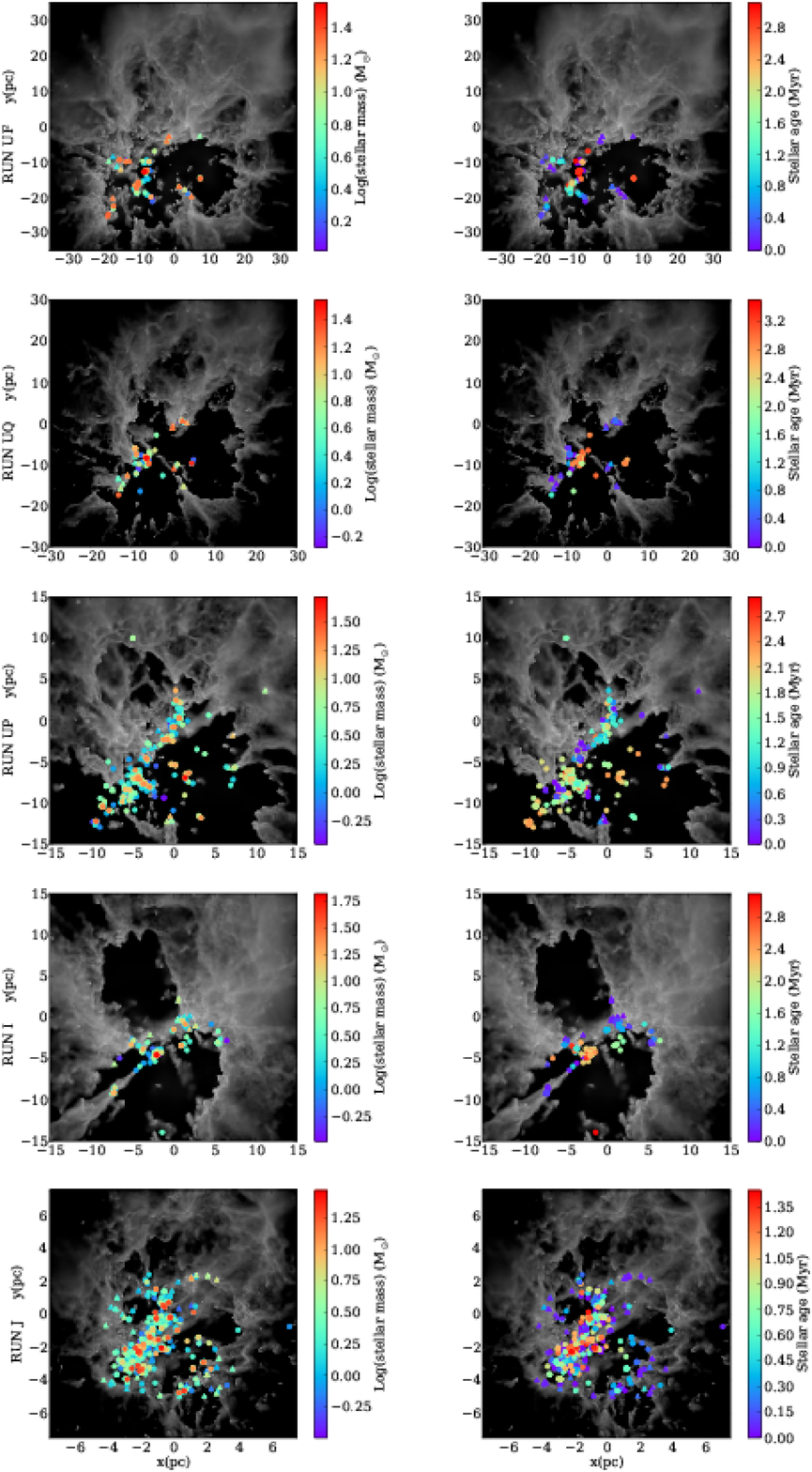}
\caption{Locations of triggered (triangles) and spontaneously--formed (circles) objects colour--coded by mass (left column) and age computed from the time at which 95 $\%$ mass acheived (right column) in the ionized Runs UF, UQ, UP, (unbound clouds), and I and J (bound clouds).}
\label{fig:trig_mass_age}
\end{figure*}

\begin{figure}
\includegraphics[width=0.45\textwidth]{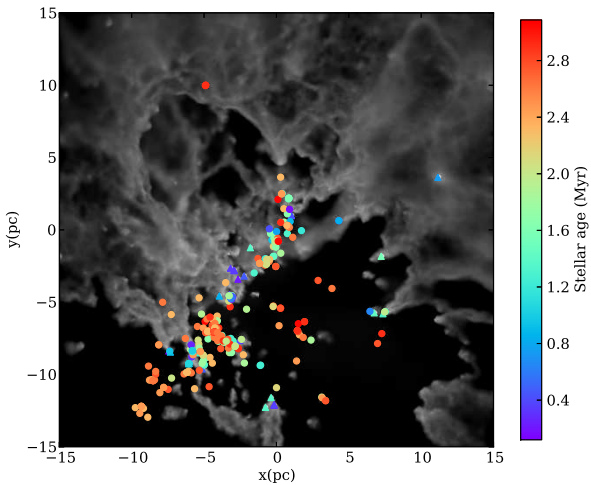}
\caption{Locations of triggered (triangles) and spontaneously--formed (circles) objects colour--coded by age computed from time of sink formation in the ionized Run UP.}
\label{fig:upcl_age2}
\end{figure}

\begin{figure*}
     \centering
     \subfloat[Run UP (this work)]{\includegraphics[width=0.32\textwidth]{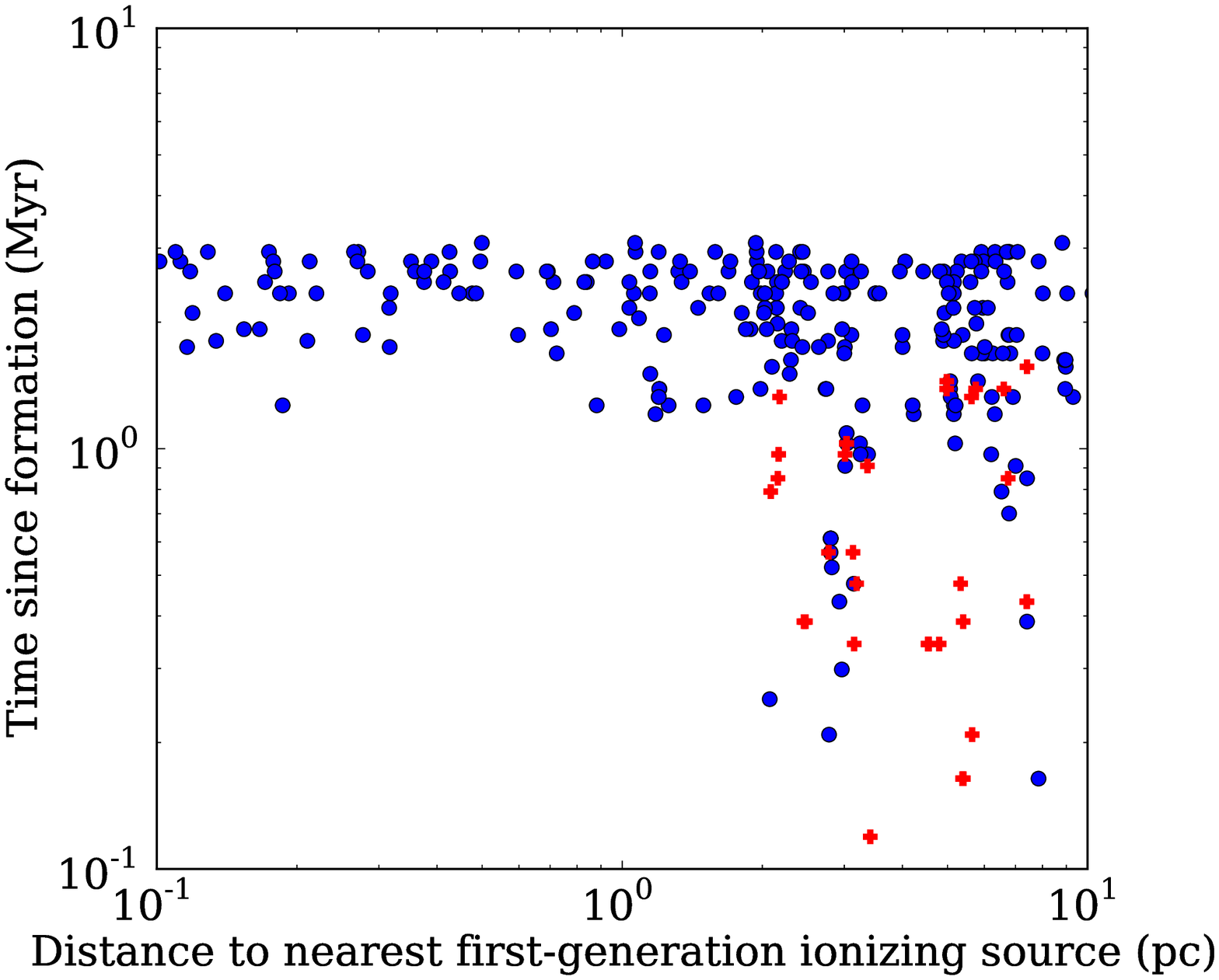}}     
     \hspace{.01in}
     \subfloat[Run UQ (this work)]{\includegraphics[width=0.32\textwidth]{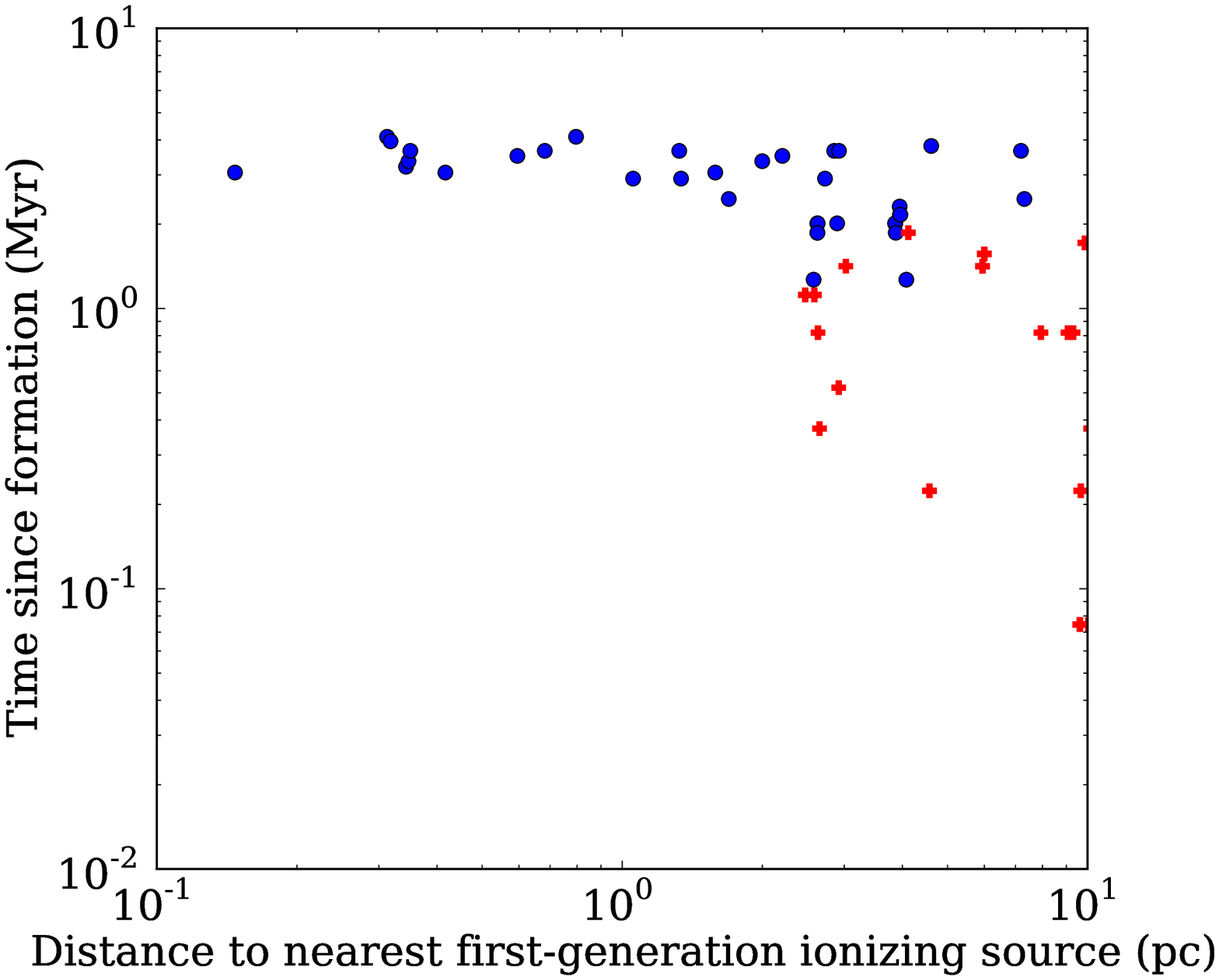}}
     \hspace{.01in}
     \subfloat[Run UF (this work)]{\includegraphics[width=0.32\textwidth]{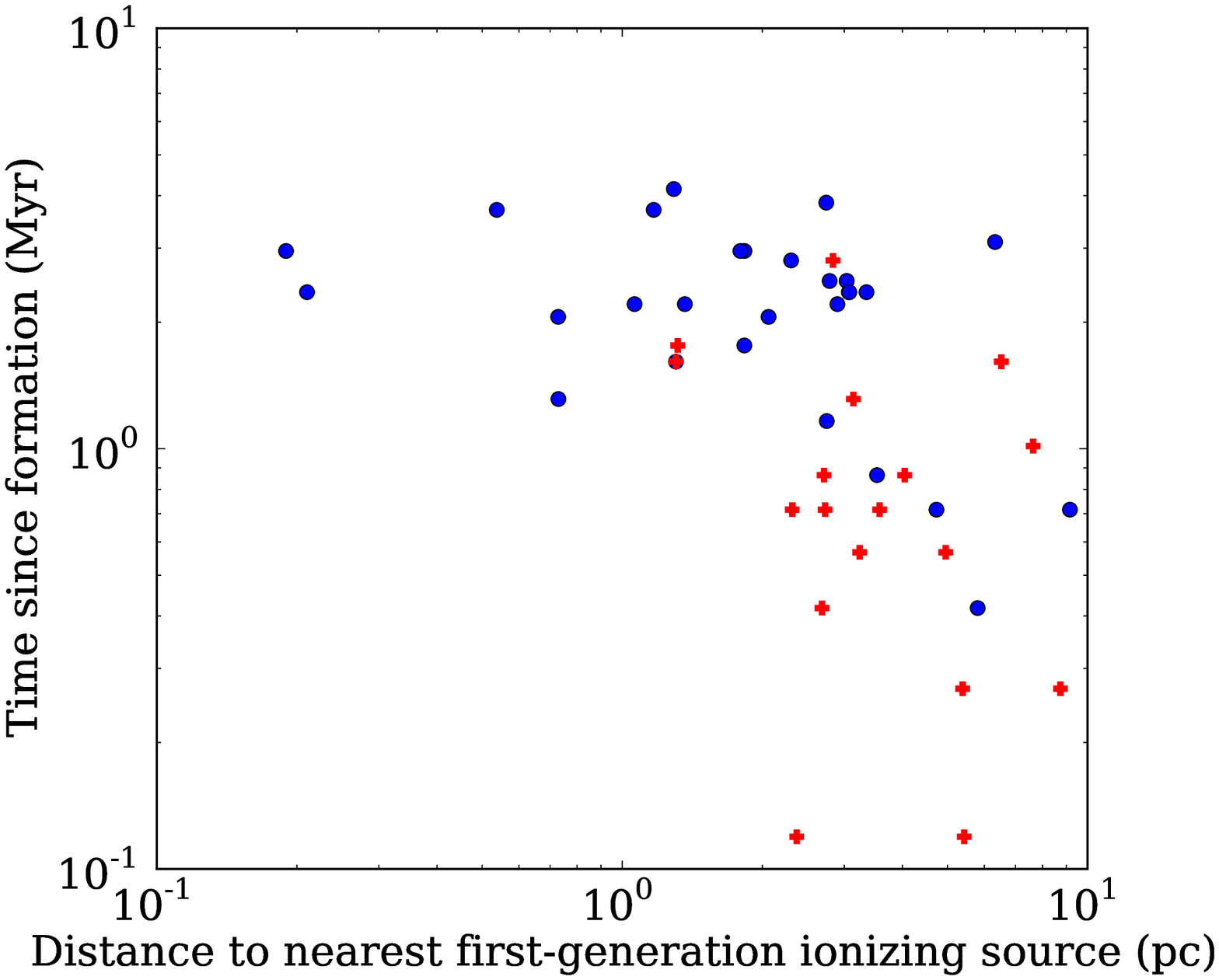}}     
     \vspace{.01in}
     \subfloat[Run I (Papers I \& II)]{\includegraphics[width=0.32\textwidth]{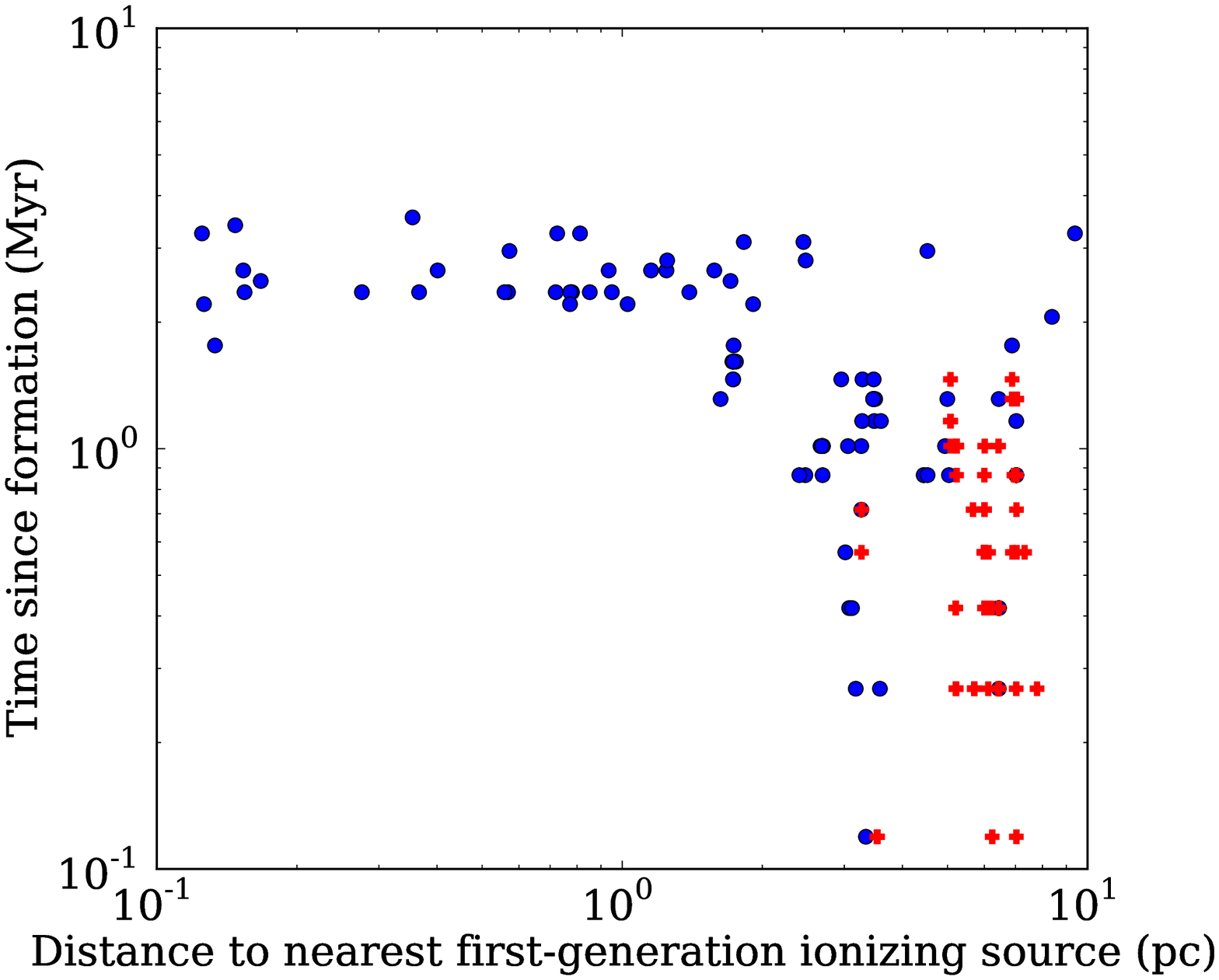}}     
     \hspace{.01in}
     \subfloat[Run J (Papers I \& II)]{\includegraphics[width=0.32\textwidth]{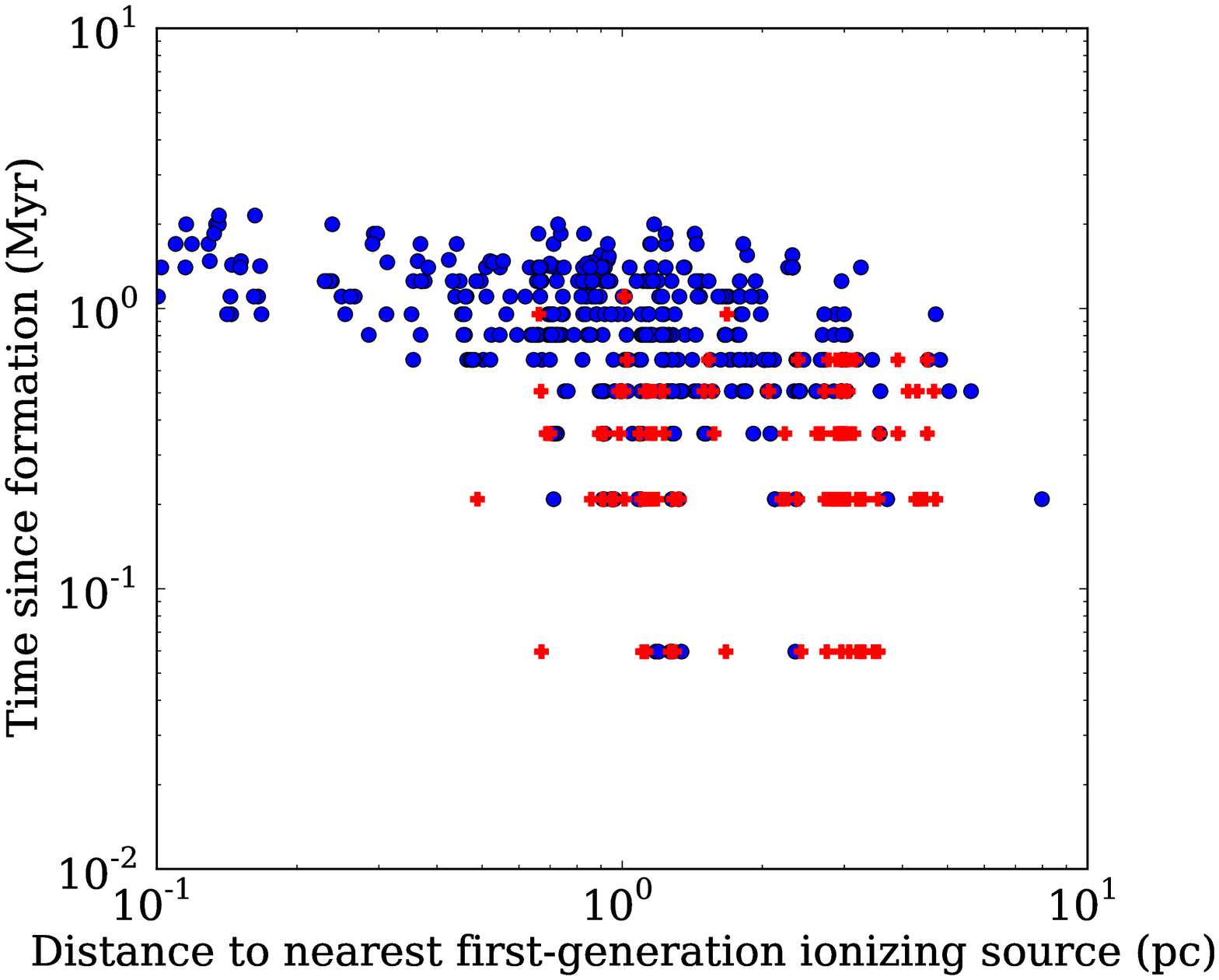}}
     \hspace{.01in}
     \caption{Plots of ages (as defined by the time each sink formed) of triggered (red crosses) and spontaneously--formed (blue circles) objects as functions of distance from the nearest \emph{first--generation} ionizing source at the ends of the ionized Runs UP, UQ and UF (this work) and I and J (Papers I and II). The ionizing sources themselves are not included in the plots.}
   \label{fig:dist_old_age1}
\end{figure*}

\begin{figure*}
     \centering
     \subfloat[Run UP (this work)]{\includegraphics[width=0.32\textwidth]{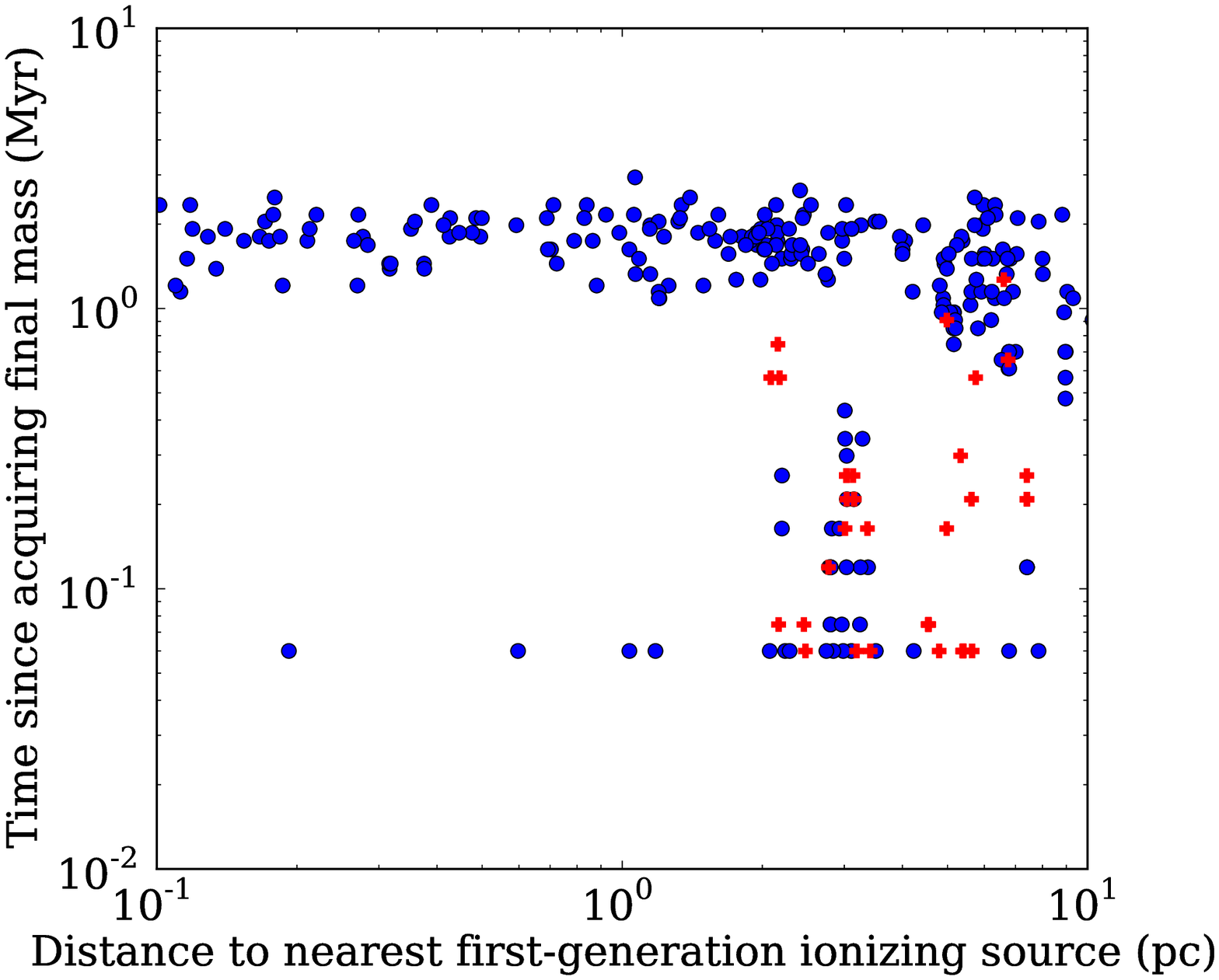}}     
     \hspace{.01in}
     \subfloat[Run UQ (this work)]{\includegraphics[width=0.32\textwidth]{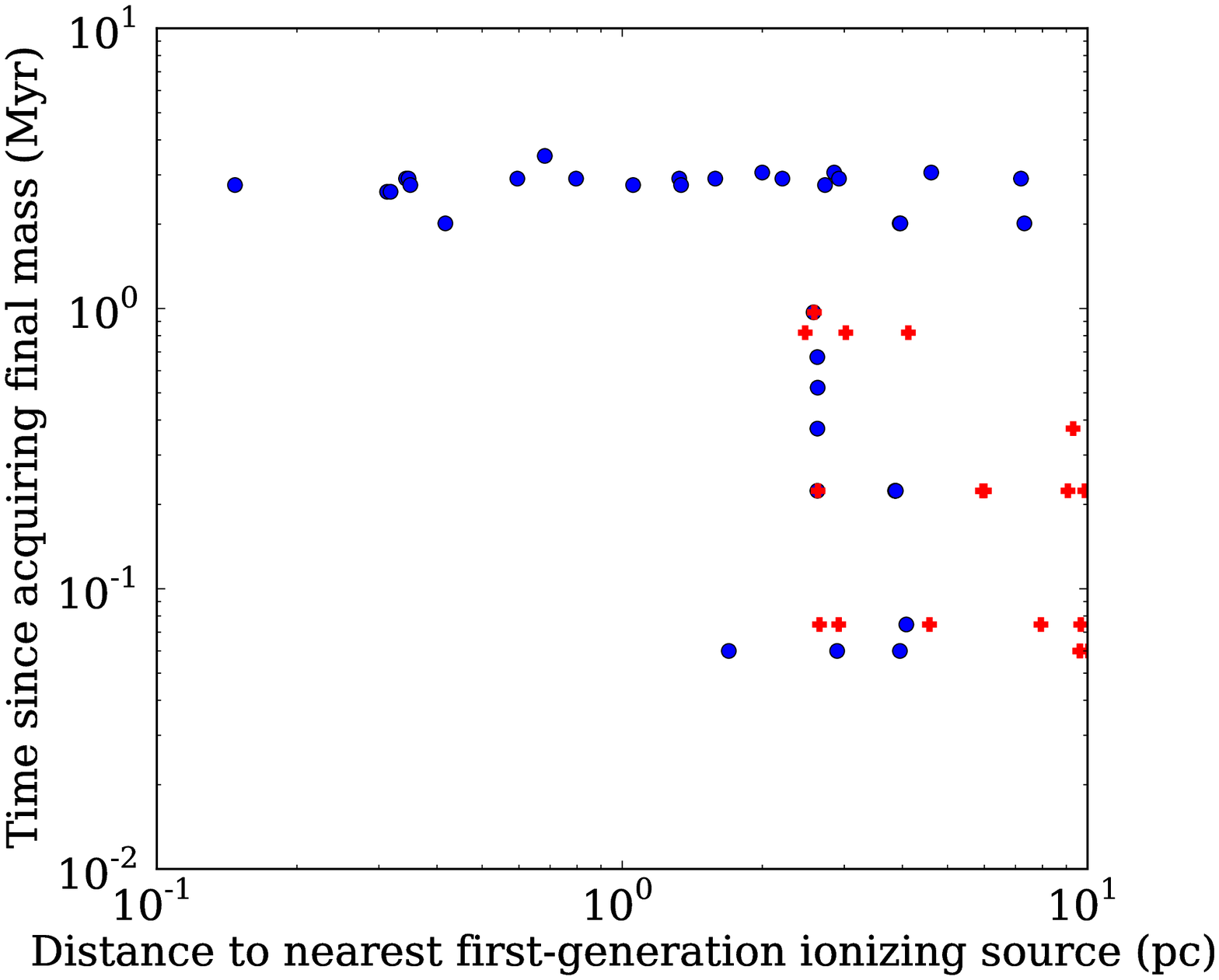}}
     \hspace{.01in}
     \subfloat[Run UF (this work)]{\includegraphics[width=0.32\textwidth]{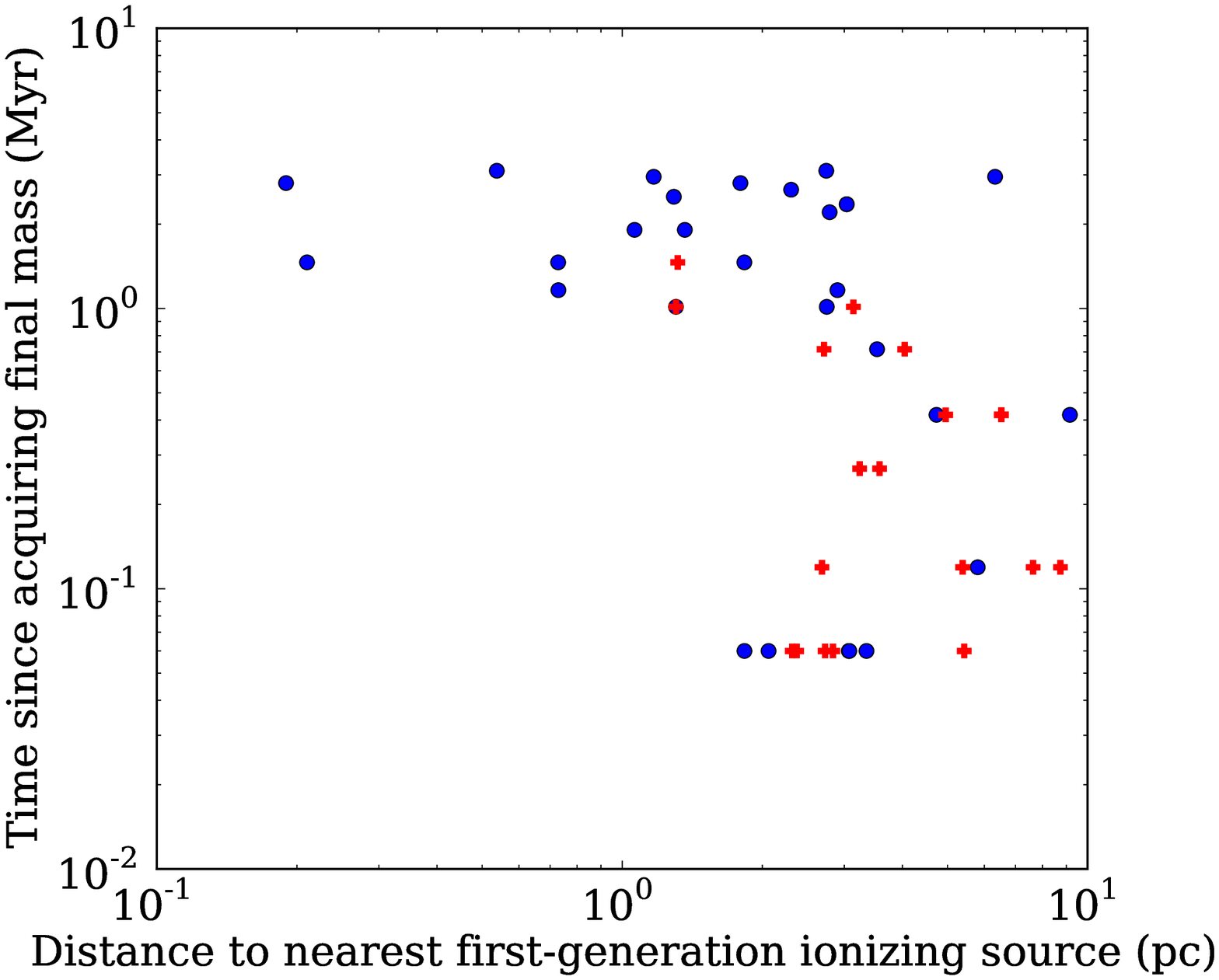}}     
     \vspace{.01in}
     \subfloat[Run I (Papers I \& II)]{\includegraphics[width=0.32\textwidth]{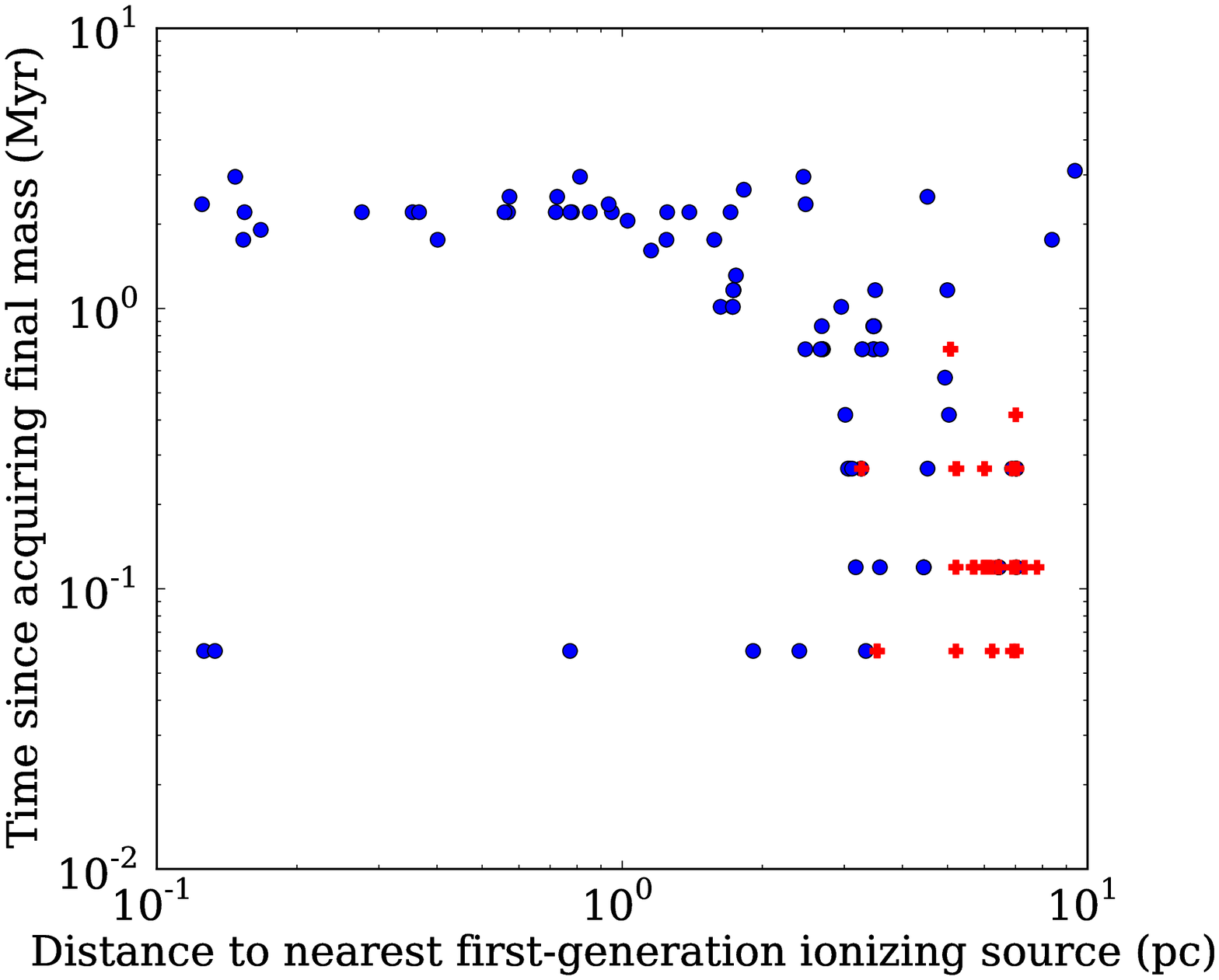}}     
     \hspace{.01in}
     \subfloat[Run J (Papers I \& II)]{\includegraphics[width=0.32\textwidth]{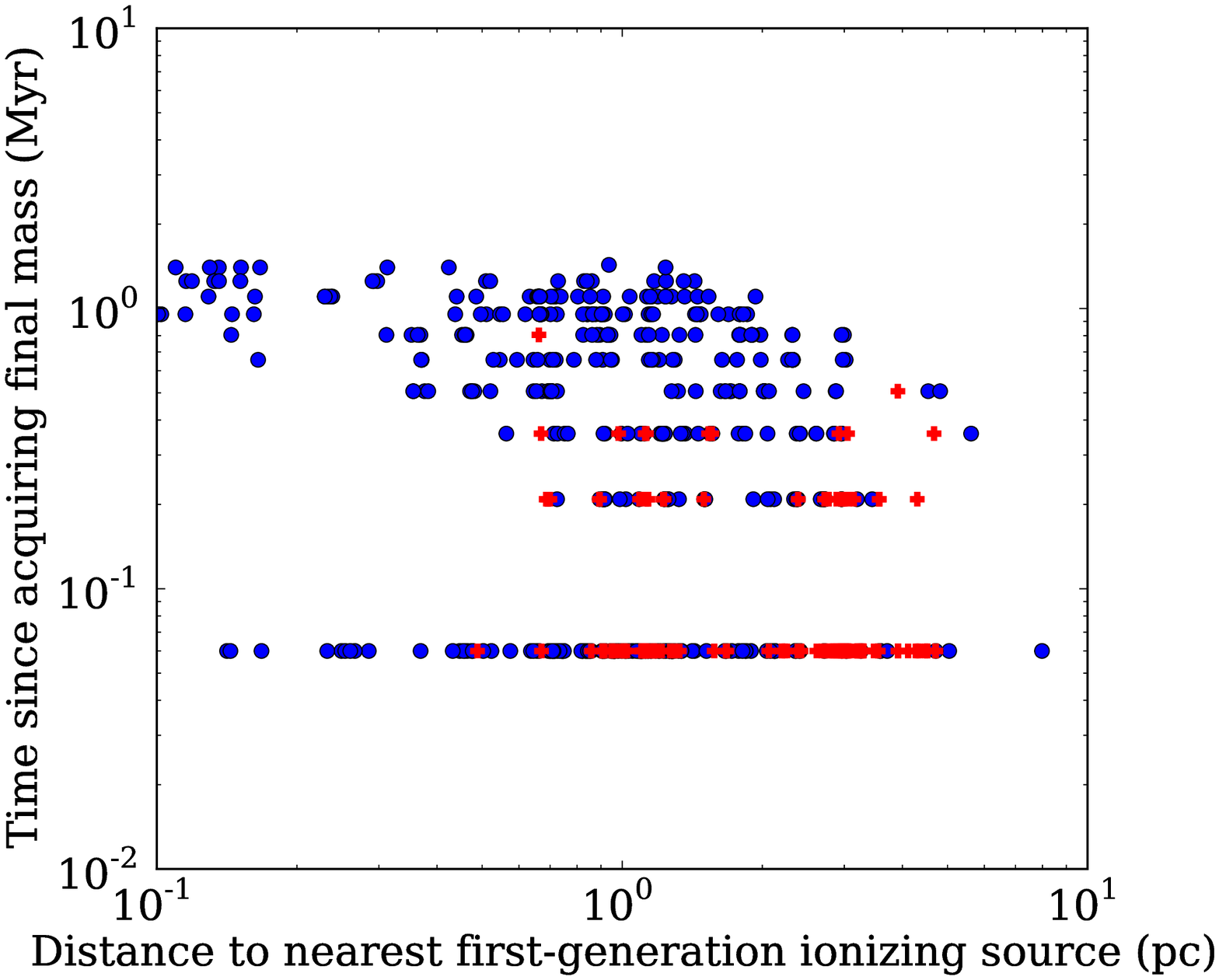}}
     \hspace{.01in}
     \caption{Plots of ages (as defined by the time final mass achieved) of triggered (red crosses) and spontaneously--formed (blue circles) objects as functions of distance from the nearest \emph{first--generation} ionizing source at the ends of the ionized Runs UP, UQ and UF (this work) and I and J (Papers I and II). The ionizing sources themselves are not included in the plots.}
   \label{fig:dist_old_age2}
\end{figure*}

We use the control runs and the same techniques detailed in \cite{2007MNRAS.377..535D,2012MNRAS.tmp.2723D} and in Paper II to determine objectively which stars or clusters had their formation triggered by ionization. We use the same three methods to cross--identify stars between simulations: (A) same seed method -- this traces the $\sim$100 seed particles from which each sink initially forms (and excludes those which it accretes later) to see if $\>50\%$ those particles also seed a single sink in the companion run; (B) same star method -- here we trace all particles from which a sink formed Ð seed particles as well as those subsequently accreted -- if $\>50\%$ of these particles form a single sink in the companion run, it can be said that the same object forms in both simulations; (C) involved method -- this traces all particles from which each sink forms and asks only whether $\>50\%$ of them are involved in star formation in the companion run, regardless of which objects they become part of. Method (A) traces starÐformation events Ð the initial collapse of a core to form a protostar -- while method (B) follows the whole process determining of an objectÕs final mass. Method (C) is less restrictive and concerns the fate of the starÐforming gas, determining only whether stars in the two compared runs are forming from the same pool of material.\\
\begin{table*}
\begin{tabular}{|l|l|l|l|l|l|l|l|l|l|l|l|}
Run&N$_{\rm TOT}^{0}$&N$_{\rm TOT}^{i}$& N$_{\rm TOT}^{c}$ &N$_{\rm untrig}^{\rm A}$&N$_{\rm untrig}^{\rm B}$&N$_{\rm untrig}^{\rm C}$&Common gas&M$_{\rm TOT}^{c}$ (M$_{\odot}$)&M$_{\rm TOT}^{i}$ (M$_{\odot}$)&M$_{\rm untrig}^{\rm C}$ (M$_{\odot}$)\\
&&&&&&&fraction&&&\\
\hline
UB&15&28&16&20&22&23&0.90&9584.7&7697.4&7428.0\\
\hline
UC&119&192&153&179&177&189&0.95&40930&36850&34090\\
\hline
UZ&80&164&185&156&163&174&0.92&77140&64630&64436\\
\hline
UU&151&141&88&120&122&138&0.92&26380&22360&20125\\
\hline
UV&15&17&21&18&19&19&0.97&9629.5&8088.1&8069.5\\
\hline
UF&12&66&76&20&25&48&0.63&1392.2&836.3&577.0\\
\hline
UP&183&342&354&212&198&317&0.83&2784.6&1994.5&1832.9\\
\hline
UQ&32&48&80&33&33&48&0.61&723.1&665.5&419.3\\
\end{tabular}
\caption{Total numbers of stars/clusters in the ionized runs before ionization (N$_{\rm TOT}^{0}$), total numbers of stars at the ends of the ionized (N$_{\rm TOT}^{i}$) and control (N$_{\rm TOT}^{c}$) runs, numbers of untriggered objects in runs derived using the three methods described in the text (N$_{\rm untrig}^{\rm A}$, N$_{\rm untrig}^{\rm B}$, N$_{\rm untrig}^{\rm C}$), the fraction of all gas involved in star formation in the ionized runs which was also involved in star formation in the corresponding control run, and the total final stellar masses in the control and ionized runs (M$_{\rm TOT}^{c}$ (M$_{\odot}$) and M$_{\rm TOT}^{i}$) and the untriggered mass identified by method C (M$_{\rm untrig}^{\rm C}$).}
\label{tab:trig_analysis}
\end{table*}
\begin{figure}
\includegraphics[width=0.45\textwidth]{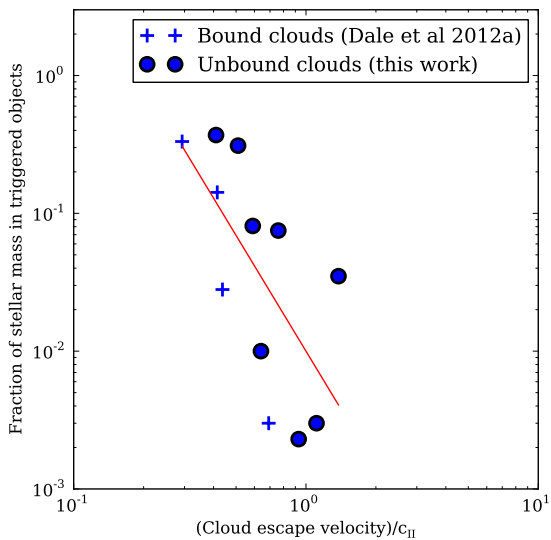}
\caption{Mass fraction of triggered stars (i.e. the total mass of triggered objects as fractions of total stellar mass at the ends of simulations) as a function of escape velocity from simulations in this work (circles) and Runs A, D, I and J from Paper I and II (crosses).}
\label{fig:ftrig_vesc}
\end{figure}
\indent In Table \ref{tab:trig_analysis} we present the results of analysing the four simulations using the same--seed method (A), the same--star method (B) and the involved--fraction (C). The columns in the table are: total numbers of stars/clusters in the ionized runs before ionization (N$_{\rm TOT}^{0}$), total numbers of stars at the ends of the ionized (N$_{\rm TOT}^{i}$) and control (N$_{\rm TOT}^{c}$) runs, numbers of untriggered objects in runs derived using the three methods (N$_{\rm untrig}^{\rm A}$, N$_{\rm untrig}^{\rm B}$, N$_{\rm untrig}^{\rm C}$), the fraction of all gas involved in star formation in the ionized runs which was also involved in star formation in the corresponding control run, the total stellar masses in the control and ionized simulations, and the total mass of untriggered objects in the ionized runs identified by method C.\\
\indent We see that there is relatively little triggering in most of the calculations. We noted in Paper II that there was a correlation between the fraction of the total stellar mass consisting of triggered objects and the cloud escape velocity. In Figure \ref{fig:ftrig_vesc}, we plot this fraction for all calculations presented here, along with Runs A, D. I and J from Papers I and II. We see that the correlation persists for the unbound clouds considered here. The red fit line is a power law with a logarithmic slope of -2.8, rather similar to the relation found in Paper III between the fractions of mass unbound and the cloud escape velocity. We find that therefore that both positive and negative effects of photoionizing feedback are most strongly correlated with the escape velocity of the host cloud. Since the sound speed in the ionized gas is fixed, the escape velocity controls what fraction of the cloud volume the HII regions are able to explore and how effective they are at sweeping up and compressing the cold gas.\\
\indent In Figure \ref{fig:trig_mass_age}, we plot the locations of the triggered (triangles) and spontaneously--formed (circles) objects with respect to the gas column density as viewed along the z--axis (greyscale) in the three runs from this work and the two from Papers I and II in which the most triggering was found. The symbols for the sinks are coloured by mass and age respectively. The ages plotted here are computed from the times when each object reaches 95$\%$ of its final mass. Another way of measuring the age of a sink is simply from the time of its formation, which we discuss later.\\
\indent It is clear that many of the triggered objects are spatially well mixed with their spontaneous counterparts and are therefore difficult to tell apart. Where bubbles are present, the triggered stars appear to be reasonably well correlated with the bubble walls, at least in projection, but many spontaneously--formed objects are also located near bubble walls. However, stellar objects are not uniformly or smoothly distributed around the bubbles. Instead, they tend to be clustered around pillar--like structures projecting from the bubble walls. It appears that the correlation of triggered stars with pillar--like structures is stronger in these unbound cloud simulations than in the bound clouds presented in Papers I and II.\\
\indent Triggered objects tend to be of low mass, but this is certainly not always the case. In Runs UQ and UF in particular, there are several intermediate/high--mass objects triggered objects embedded in the bubble walls. Some of these objects are of sufficiently high mass to be ionizing sources themselves. The existence of these objects lends support to the idea of propagating star formation \citep[e.g.][]{1981ApJ...249...93S,1983ApJ...265..202S}, since their HII regions could in principle drive another round of star formation further out in these clouds. However, as we showed in Paper III, the supernova explosions of the first--generation massive stars may disrupt remaining molecular material before second--generation O--stars begin triggering starbirth themselves.\\
\indent The triggered stars are always among the youngest objects in each system, and many are still accreting, so have ages close to zero as measured in this way. In clouds UQ, UF, I and J, there are substantial numbers of stars in or near the bubble walls which are both young and triggered. This is particularly clear in Run J, where there is a pronounced halo of young, triggered stars surrounding the main cluster. However, we stress that there are also young, spontaneously--formed objects geometrically mixed with the triggered ones, so that \emph{neither young age, nor location in or near bubble walls is a foolproof indicator that a given object has been triggered}.\\
\indent There is an abundant literature on age--gradients in star forming regions as signposts of triggering \citep[e.g][]{1995ApJ...455L..39S,2005AJ....129..776N,2011MNRAS.415.1202C}. The idea is simply that feedback--driven bubbles proceed outward from the massive stars, triggering star formation as they go. The oldest low--/intermediate--mass objects are then to be found nearest the O--stars, with such objects getting progressively younger with distance, so that there is a gradient of age with respect to distance from the massive stars. Such age gradients are rather difficult to discern in the above images, but several observations can be made that call this picture into question.\\
\indent In Run UP, there does appear to be an age gradient along the linear central cluster, with oldest objects at the bottom left and youngest at the top right, save for a small group of very young stars halfway along. Since there is a champagne flow at the bottom left of the image and a large bubble at the bottom right, this gradient could be interpreted as evidence of triggering. However, it clearly is not, since the vast majority of stars along the central filament are spontaneously--formed. Their ages, as determined from when they stopped accreting, are connected to feedback nonetheless. The action of the champagne flow and bubble have strongly influenced when the gas supply for each object was cut off, but has merely altered the masses and apparent ages of stars that were forming anyway. In Run I (using the same definition for the ages of sinks) there is a central oldest cluster, a second cluster of intermediate age a few pc away, and a younger population embedded in the nearby bubble wall. Again, most of these objects are not triggered and the age gradient is a result of when these objects were overrun by the ionization front emanating from the older and more massive cluster. The same is true for the line of four sinks at (-8,-18) in Run UF, which show a perfect age gradient pointing towards the nearest ionizing source. All these objects are spontaneously--formed, and the clump from which they were accreting was progressively destroyed in the radial direction from the ionizing sources. \cite{2009A&A...504...97B} suggest this mechanism as a possible cause for the age gradients observed in IC1396N.\\
\indent An alternative means of measuring the age of objects represented by sink particles is simply to record the time at which each sink particle initially forms. If this definition is used in place of that used above, we see no evidence of age gradients of any kind across the faces of the clouds. We illustrate this in Figure \ref{fig:upcl_age2} where we show again the gas in greyscale and the sink particles as circles for spontaneously--formed objects and triangles for triggered objects, but this time colour--coding the sinks according to the time when they formed. There is evidently no clear age gradient at all, with sinks of all ages being spatially well mixed. Since the vast majority of stars in Run UP first formed in the dense gas before there was any interaction between this gas and the expanding HII regions, their ages as measured in this way are entirely uncorrelated with the evolution of the ionized gas. These uncorrelated ages are of course preserved when the HII regions later wash over this part of the cloud, resulting in a large ionized volume containing many stars, none of whose formation times were determined by feedback.\\
\indent To reinforce this point, we plot in Figures \ref{fig:dist_old_age1} and \ref{fig:dist_old_age2} the age (measured as the time since sink formation, and the time since 95$\%$ final mass achieved, respectively) against the 3D distance to the nearest \emph{first--generation} ionizing source for all sinks in Runs UP, UQ, UF, I and J. Spontaneously--formed sinks are shown in blue, triggered sinks in red. Which definition of the objects' ages is used is evidently not very important in these plots. There is very little evidence of any kind of age--gradient \emph{with respect to distance from the massive stars} in these plots in either the triggered or spontaneously--formed stars. If all such distances are computed with respect to the nearest ionzing source regardless of whether it is a first--generation source or not, there is even less structure present. This is not surprising, since many of the second--generation sources are spatially mixed with the triggered and spontaneously--formed stars.\\
\indent These plots do show some structure, in that they show that first--generation ionizing sources tend to be surrounded by relatively compact (except in the case of Run UP) clusters of coeval and spontaneously--formed stars -- the horizontal features in the plots -- with a mixed population of triggered and spontaneously--formed objects of various ages distributed at characteristic distances of a few to ten pc -- the vertical features. These latter are the populations within the walls of the prominent bubbles found in most simulations, although less prevalent in Run J. We again see a general tendency for triggered objects to be younger than their untriggered counterparts, but they also tend to be spatially mixed and there is scant evidence of age gradients in either population.\\
\subsection{Suppressed star formation}
\begin{figure*}
     \centering
     \subfloat[Run UF]{\includegraphics[width=0.32\textwidth]{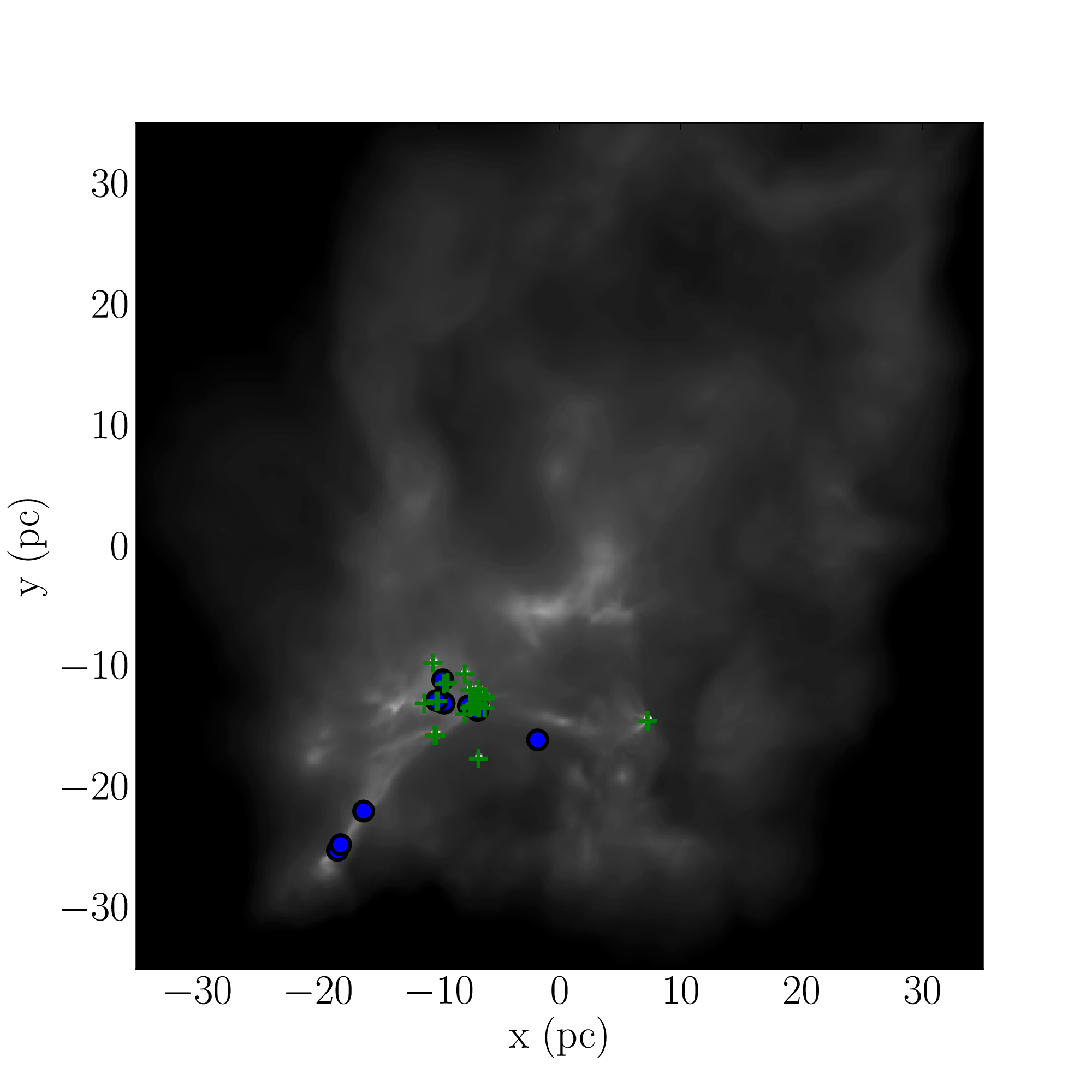}}     
     \hspace{.1in}
     \subfloat[Run UP]{\includegraphics[width=0.32\textwidth]{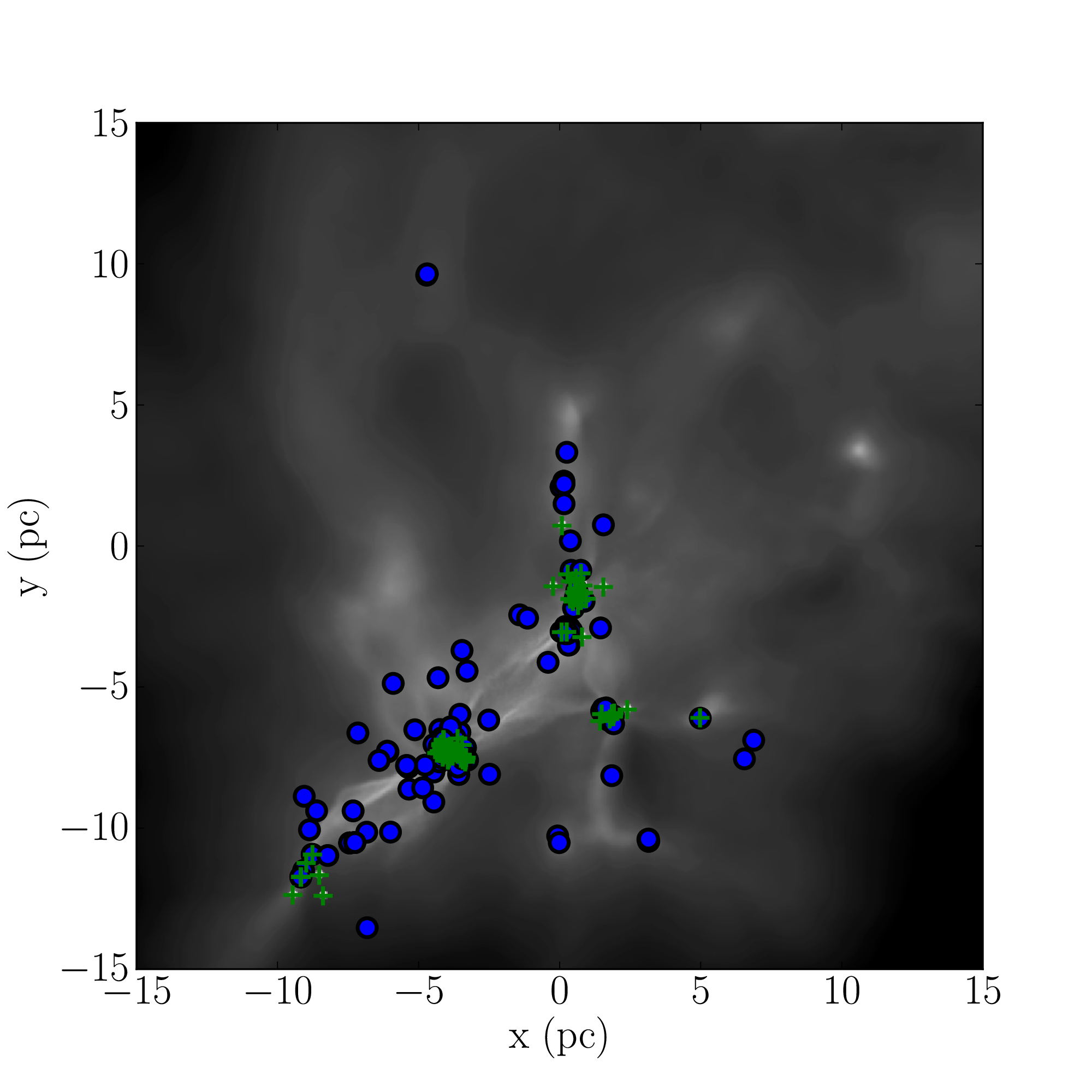}}     
     \hspace{.1in}
     \subfloat[Run UQ]{\includegraphics[width=0.32\textwidth]{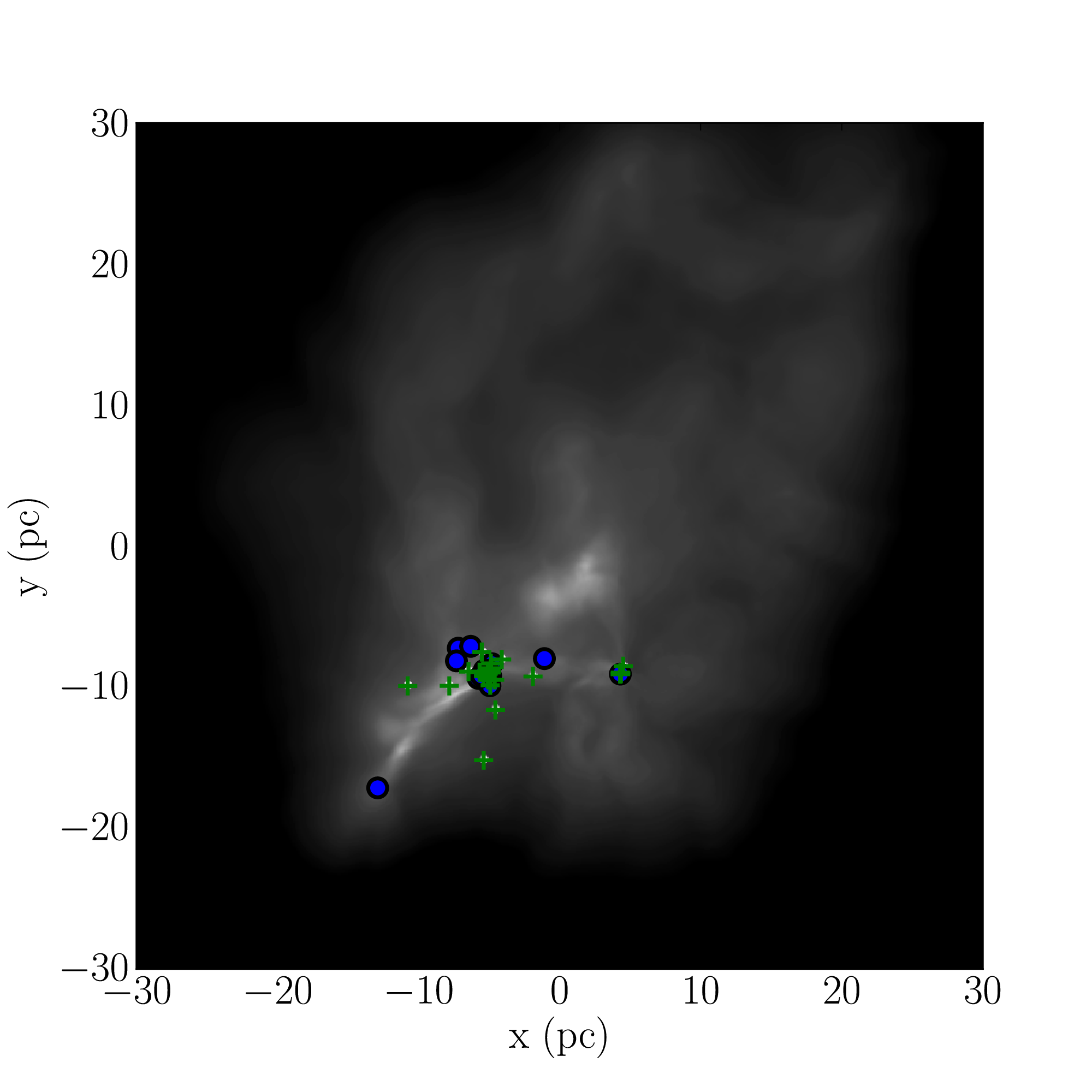}}     
     \vspace{.1in}
     \subfloat[Run I]{\includegraphics[width=0.32\textwidth]{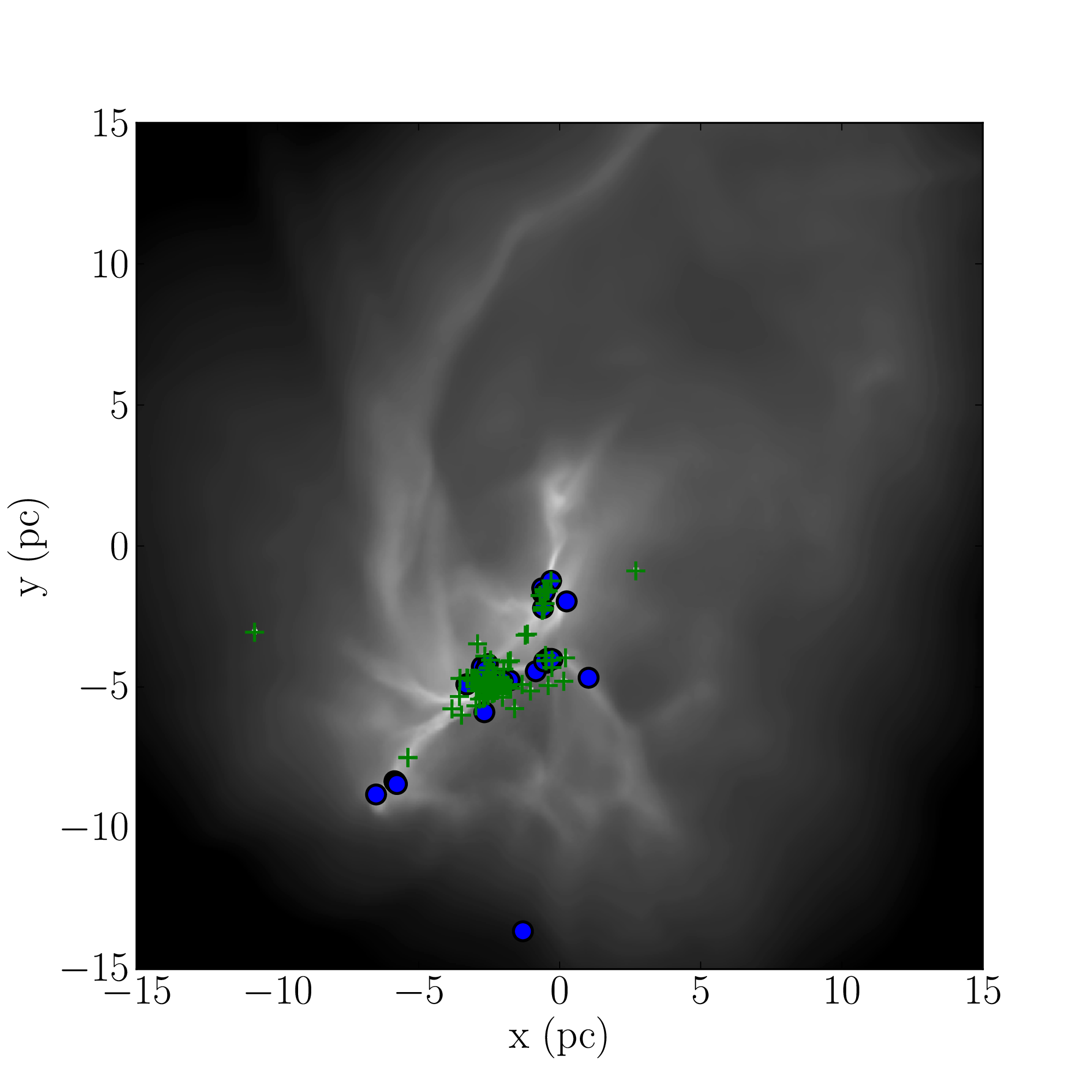}}     
     \hspace{.1in}
     \subfloat[Run J]{\includegraphics[width=0.32\textwidth]{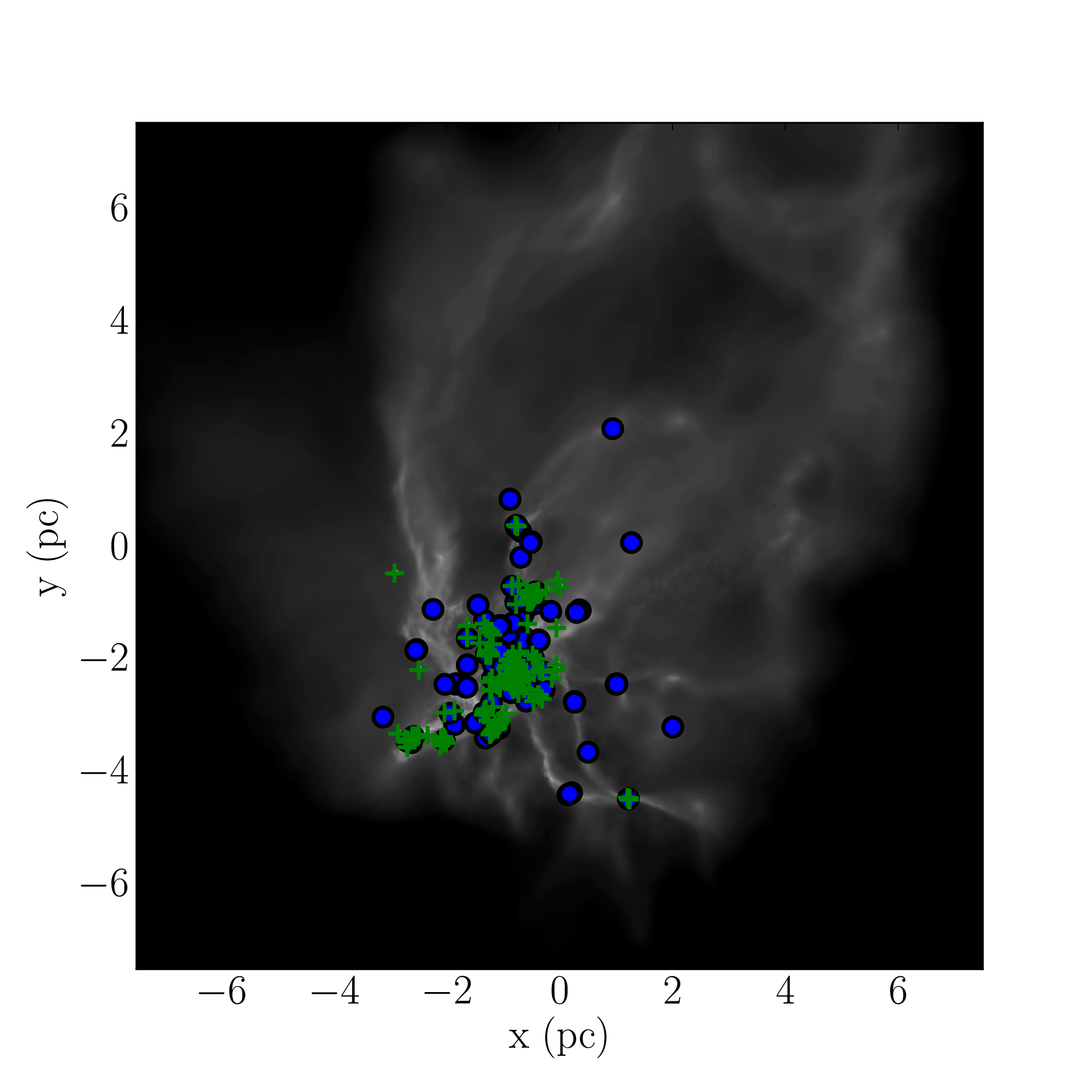}}     
     \caption{Locations of objects in the control runs which are either also found in the corresponding ionized runs (blue), or are aborted by photoionisation (green), in Runs UP, UQ and UF (this work, top row) and I and J (Papers I and II, bottom row).}
   \label{fig:abort}
\end{figure*}

The control runs can also be used to establish which stars are prevented from forming in the ionized runs. The numbers of aborted objects correlate reasonably well with the numbers of triggered objects, in the sense that there are considerably more aborted objects in simulations where there is also more triggering. Both processes can be used as a proxy for how much influence feedback has on a given cloud. We show in Figure \ref{fig:abort} locations in Runs UF, UP and UQ from this work and in Runs I and J from Papers I and II of from the control simulations which also form in the corresponding feedback run (blue), or which are aborted by ionization (green). We do not attempt to colour--code the sinks by age or mass in these images for reasons of confusion.\\
\indent In all cases, the majority of aborted stars are members of the clouds' central clusters. These are the clusters where the ionizing sources reside and the first outcome of feedback in strongly--affected clouds is, by definition, the destruction of the accretion flows (or alternatively, filaments) which are delivering gas to these clusters. This has the dual effect of curtailing accretion onto the stars already in the clusters, and of preventing the clusters from forming new members.\\
\subsection{Redistributed star formation}
\begin{figure*}
     \centering
     \subfloat[Run UF]{\includegraphics[width=0.32\textwidth]{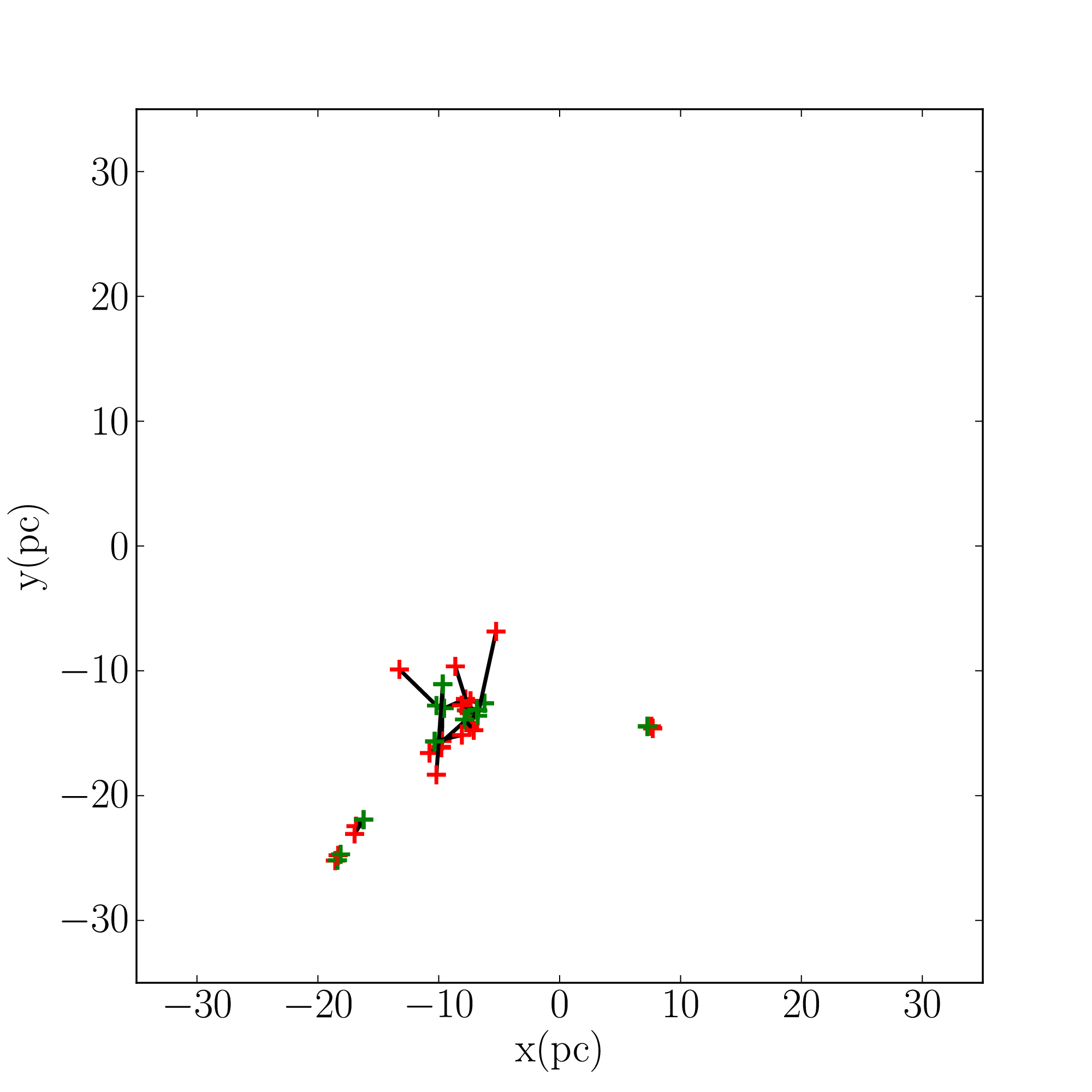}}     
     \hspace{.1in}
     \subfloat[Run UP]{\includegraphics[width=0.32\textwidth]{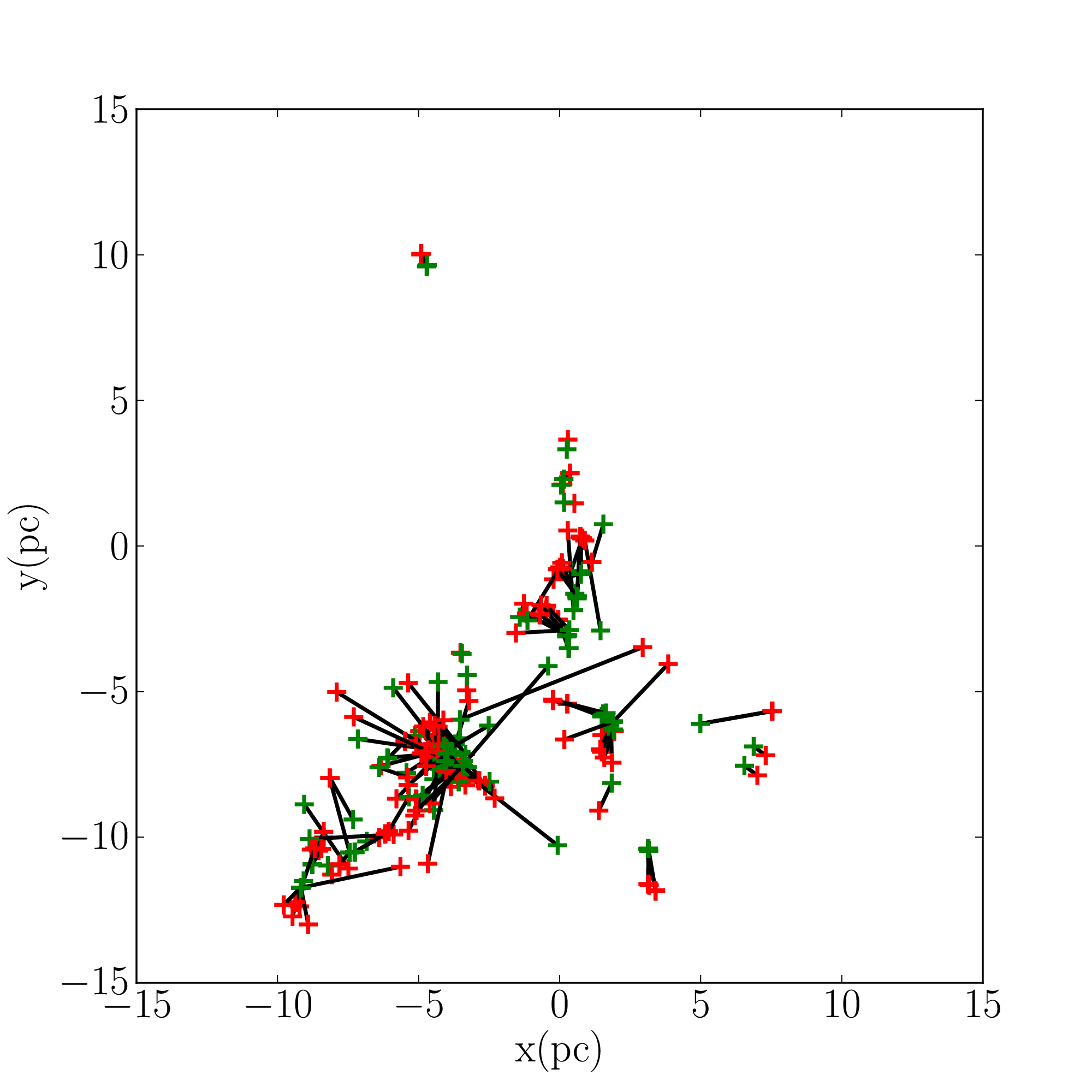}}     
     \hspace{.1in}
     \subfloat[Run UQ]{\includegraphics[width=0.32\textwidth]{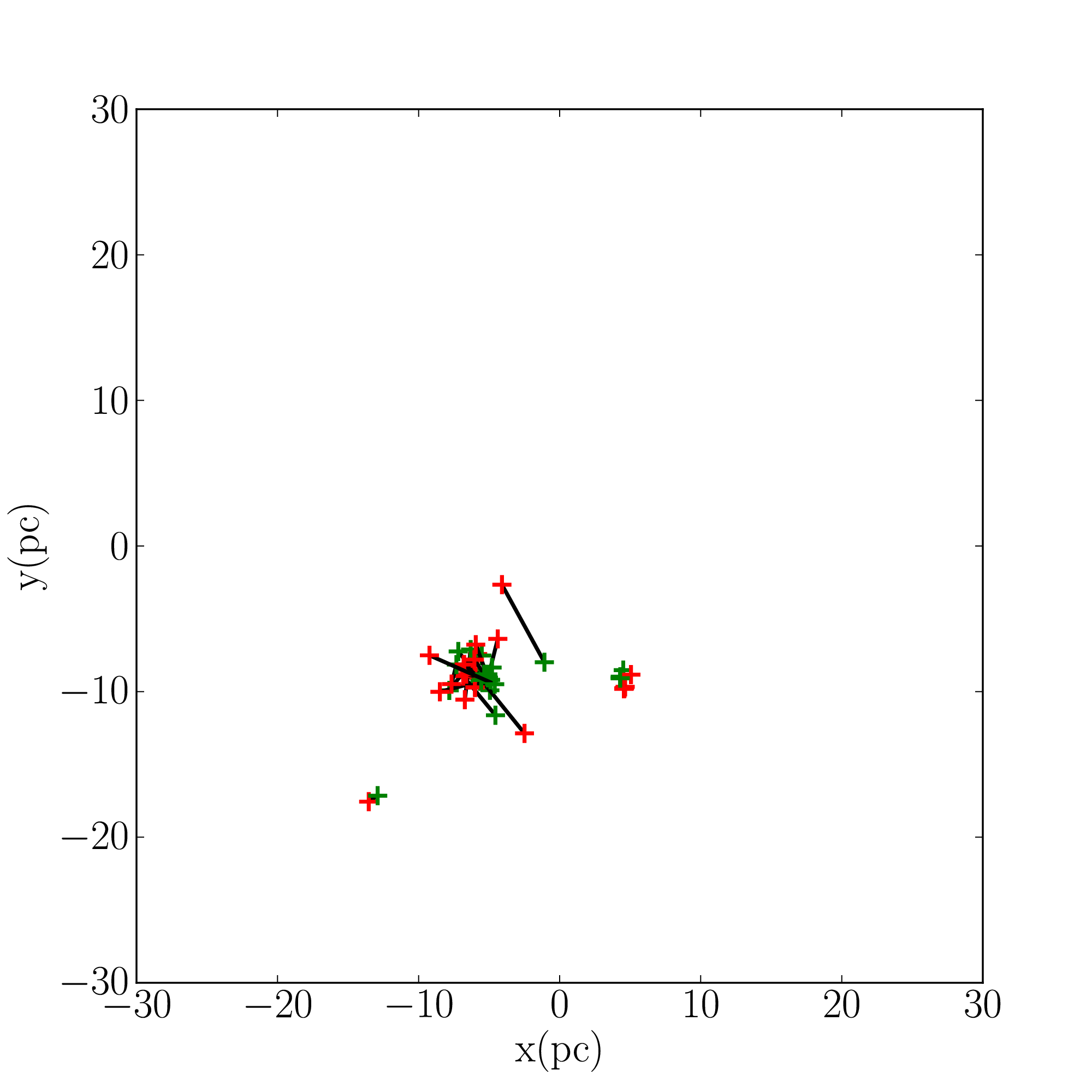}}     
     \vspace{.1in}
     \subfloat[Run I]{\includegraphics[width=0.32\textwidth]{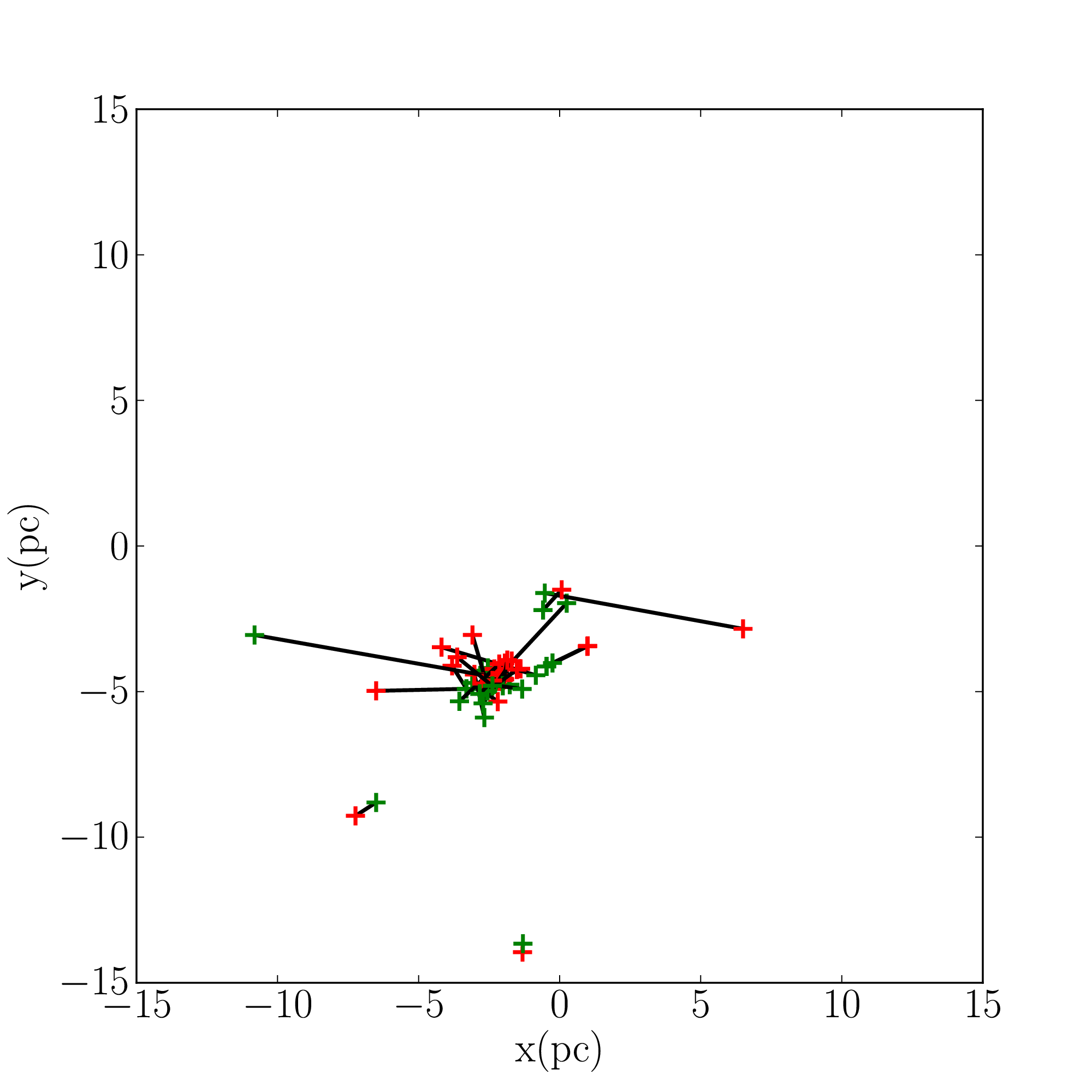}}     
     \hspace{.1in}
     \subfloat[Run J]{\includegraphics[width=0.32\textwidth]{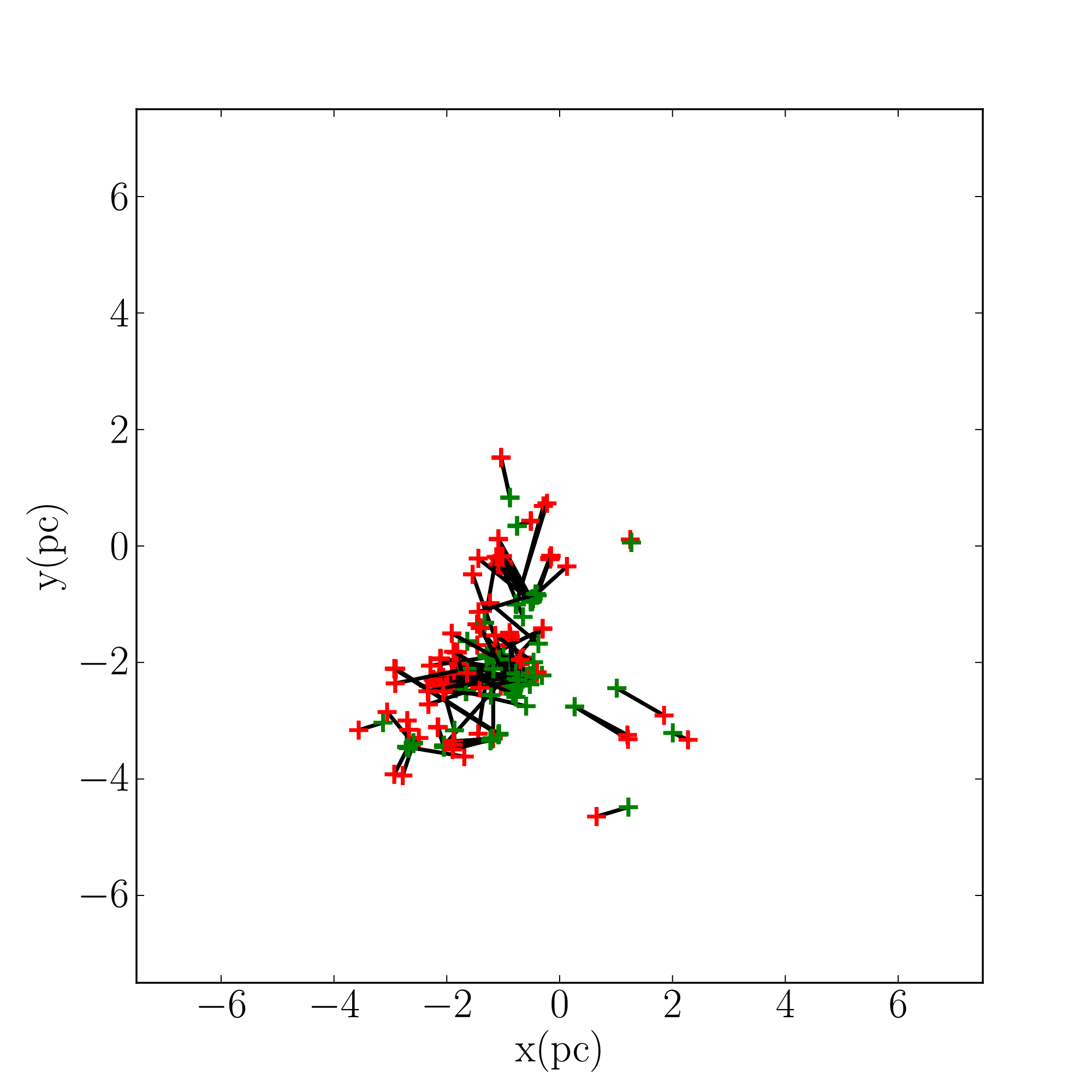}}     
     \caption{Locations of stars common to control and ionized runs. Location in the control run is shown in green, location in the ionized run in red and identical stars are joined by black lines.}
   \label{fig:redist}
\end{figure*}

We are using the most conservative possible definition to identify triggered stars. Unless more than half of the gas from which an object forms in a given feedback calculation is not involved in star formation at all in the corresponding control run, the object is taken to be spontaneously--formed (the involved method). However, there are cases when almost the same group of gas particles form a single sink particle in both halves of a feedback--control pair, so that the same sink forms in both calculations (these are identified by the same--star method. These are a subset of the objects we generally refer to as spontaneously--formed\\
\indent However, this is not to say that these objects are unaffected by feedback. Since the HII regions are expanding from inside the dense star--forming gas in the cloud cores, some of the gas from which stars form in the control run may not be prevented from doing so in the feedback run, but merely be moved to another location. If the distance between this alternative location and the position of the same object in the control run is a substantial fraction of the size of the embedded cluster, this may be termed redistributed star formation.\\
\indent The systems in which redistribution is most prevalent are also the systems which suffer the most local triggering and suppression. If the disturbances to the gas density and velocity field caused by feedback are sufficient to noticeably influence the star formation process in the clouds, it always does so in all three ways simultaneously. In Figure \ref{fig:redist}, we show the locations in the control (green) and ionized (red) simulations of all objects identified by the same--star method in Runs UF, UP and UQ from the simulations presented here, and from Runs I and J from Papers I and II. In all cases, there are stars in the feedback runs which are to be found several parsecs away from their equivalent positions in the control simulations, generally in directions radially outward from the main sites of star formation.\\
\subsection{Mass functions}
\begin{figure*}
     \centering
     \subfloat[Run UU]{\includegraphics[width=0.39\textwidth]{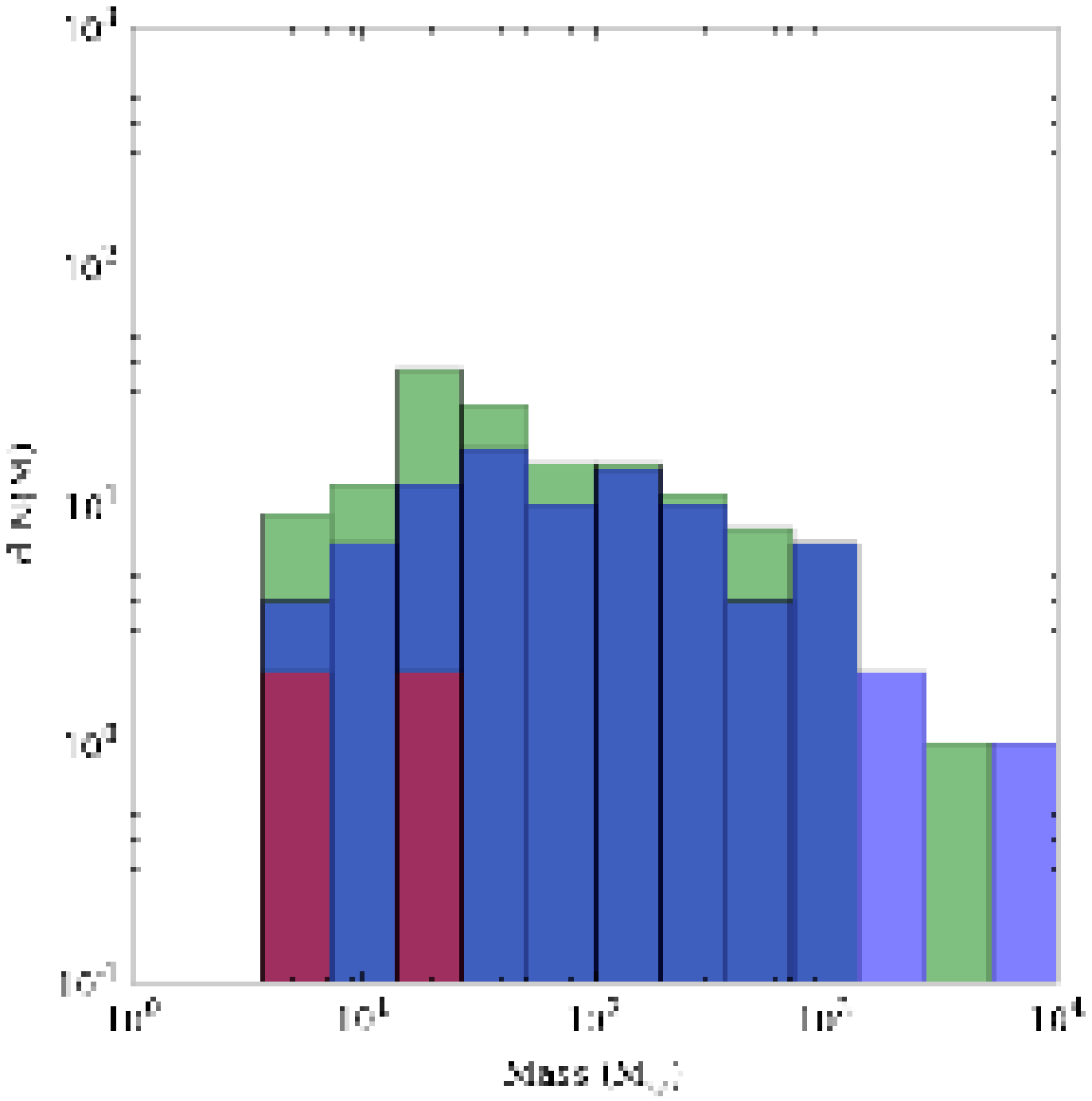}}     
     \hspace{.1in}
     \subfloat[Run UC]{\includegraphics[width=0.39\textwidth]{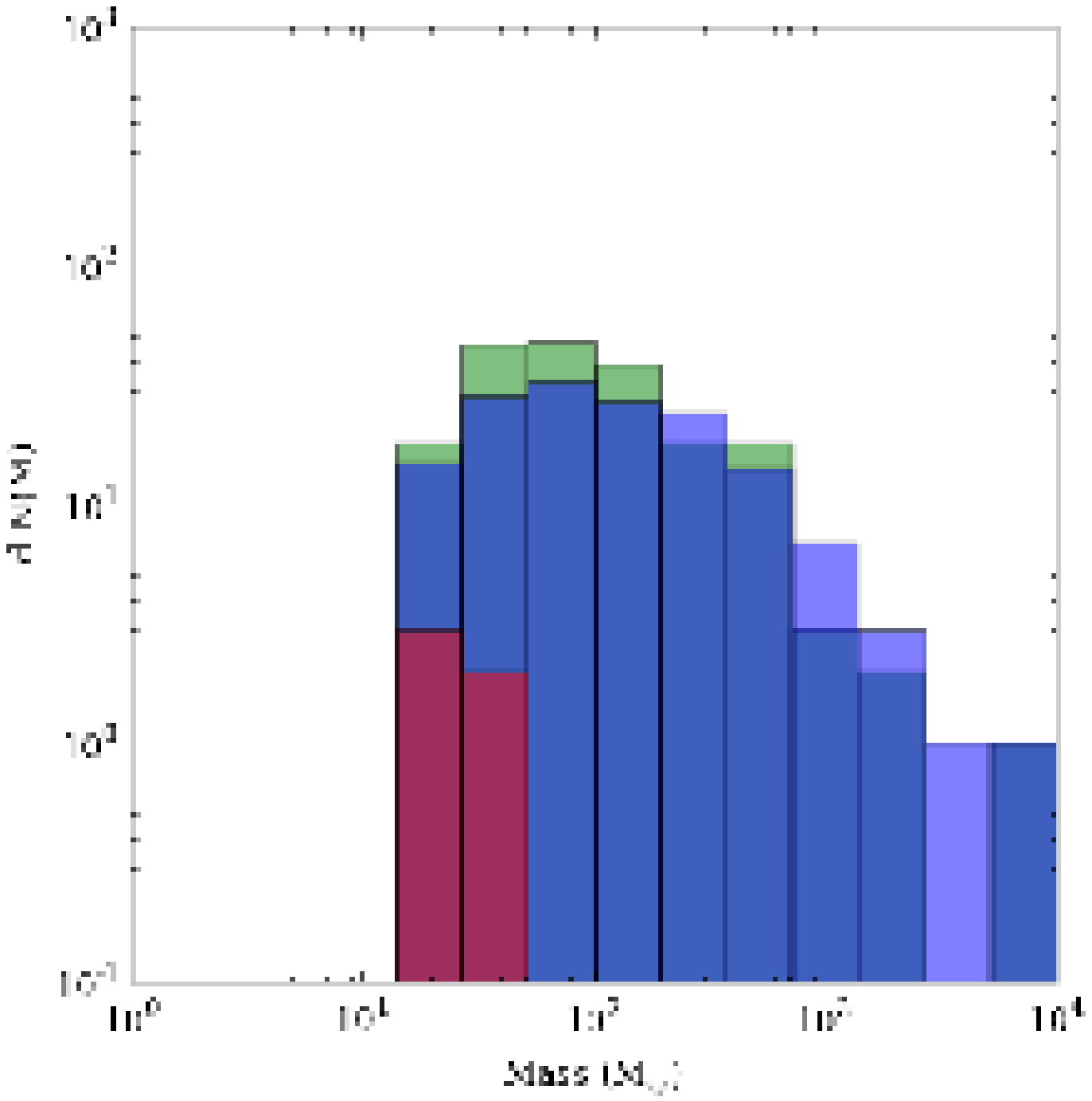}}
     \vspace{.1in}
     \subfloat[Run UZ]{\includegraphics[width=0.39\textwidth]{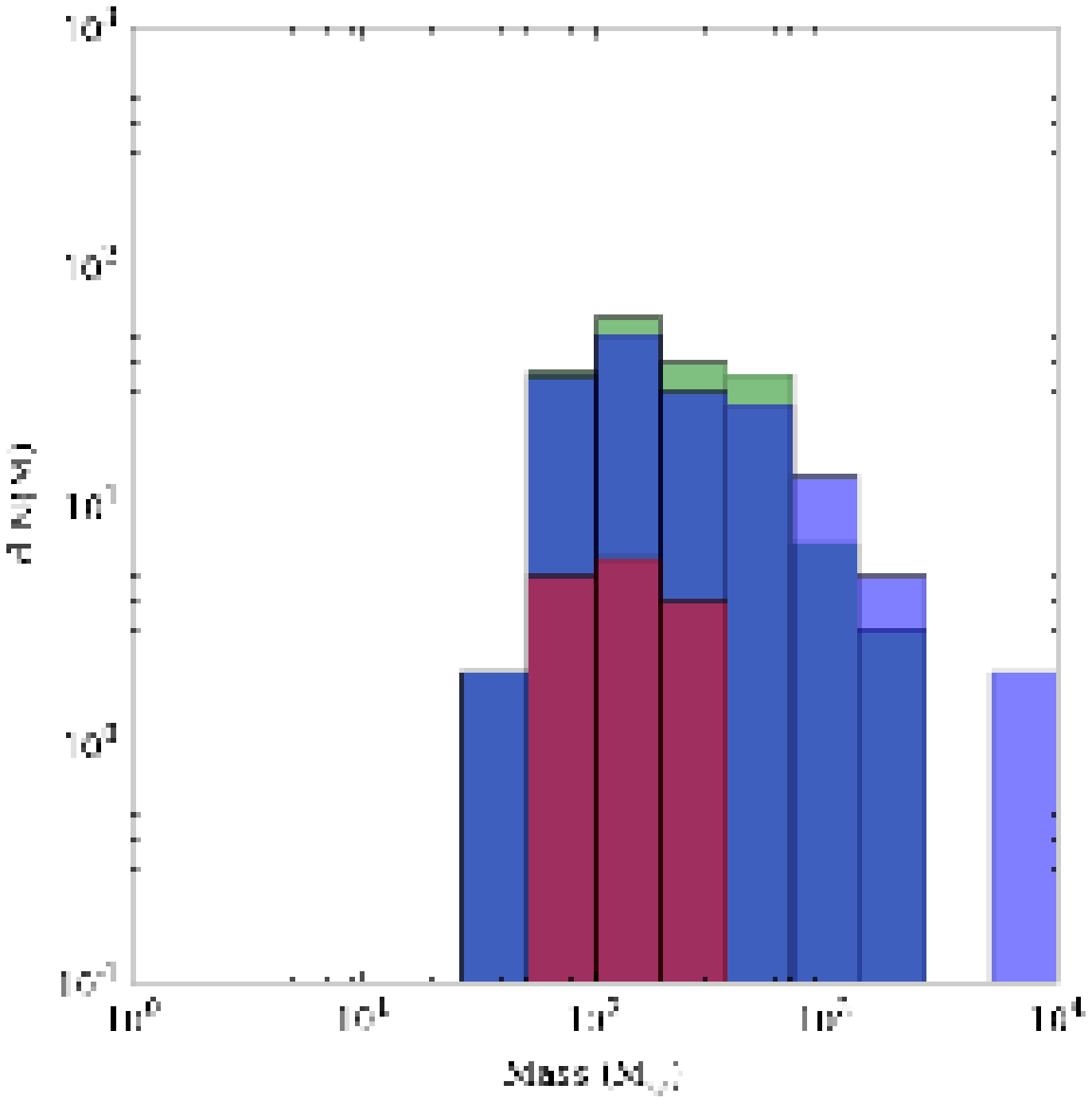}}     
     \vspace{.1in}
     \subfloat[Run UP]{\includegraphics[width=0.39\textwidth]{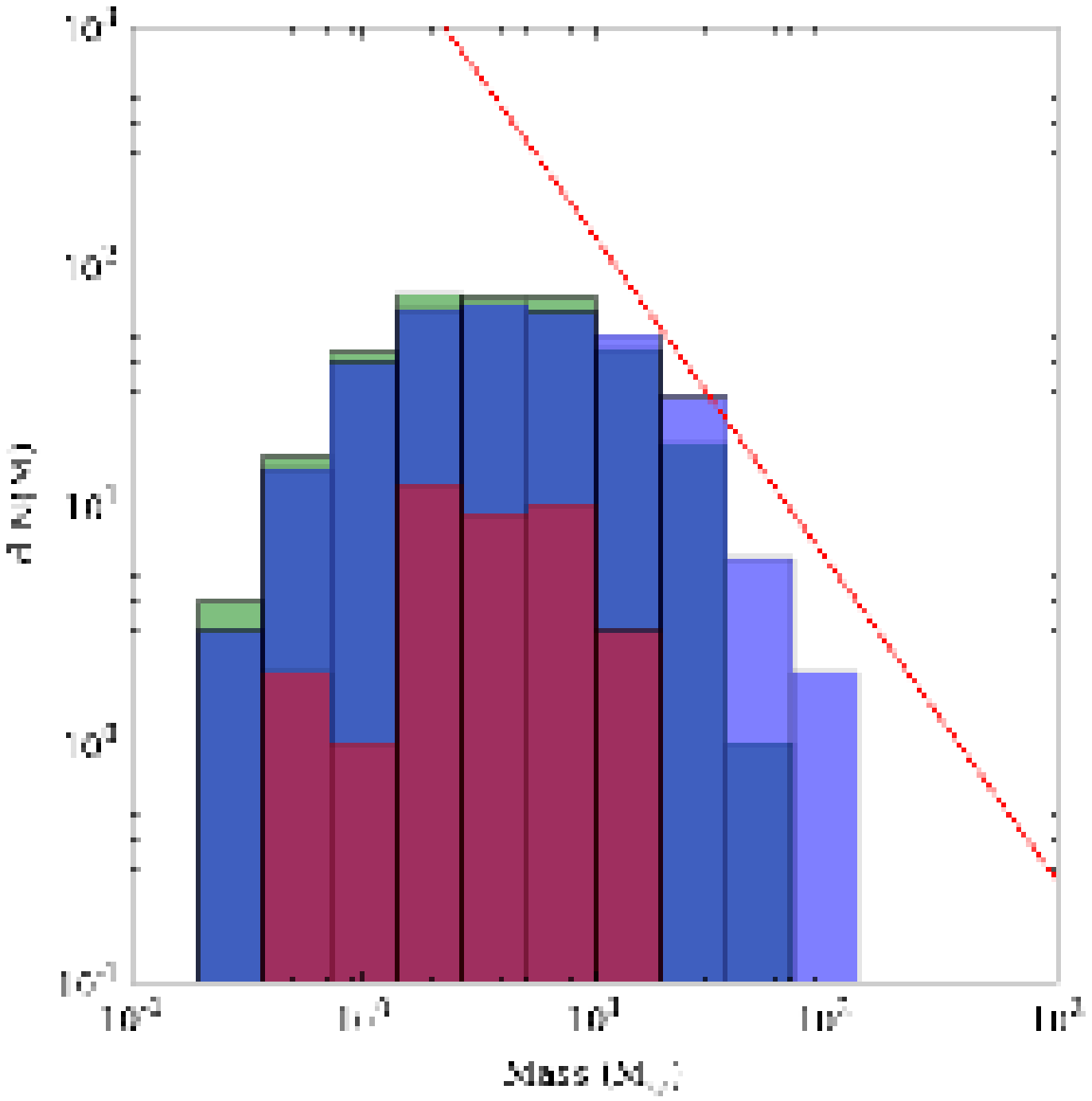}}
     \vspace{.1in}
     \subfloat[Run UQ]{\includegraphics[width=0.39\textwidth]{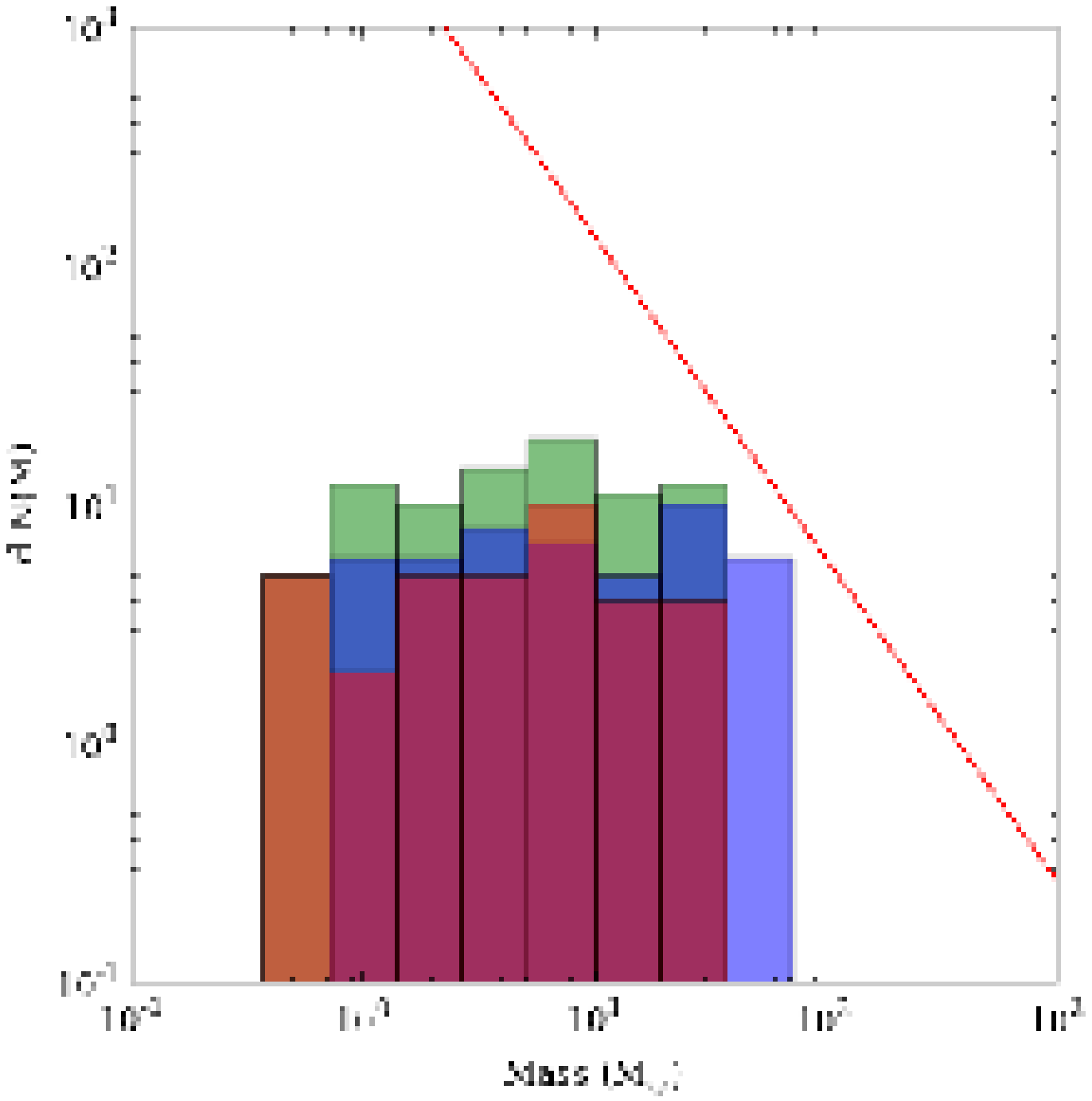}}
     \hspace{.1in}
     \subfloat[Run UF]{\includegraphics[width=0.39\textwidth]{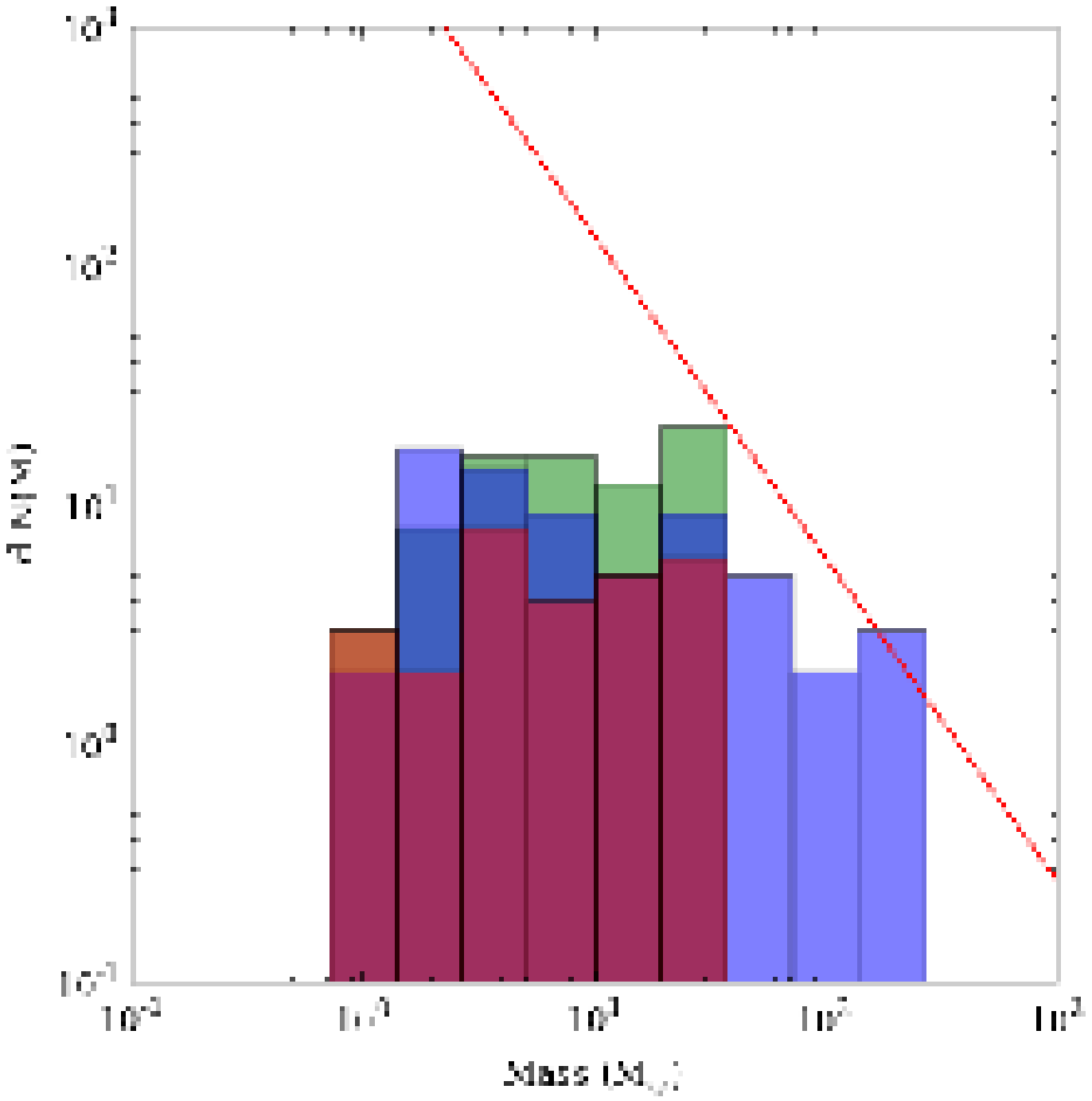}}
    \caption{Comparison of the cluster and stellar mass functions in the ionized (green histograms) and control (blue histograms) Runs UU, UC, UZ, UP, UQ and UF. Red lines in the plots on the bottom row indicate the Salpeter slope for comparison purposes. Triggered objects are depicted as red histograms. Runs UB and UV are not shown because they contain too few objects.}
   \label{fig:compare_massfunc}
\end{figure*}

We show mass functions of sink particles from Runs UU, UC, UZ, UF, UQ and UP in Figure \ref{fig:compare_massfunc}. We do not plot mass functions for Runs UB or UV because they contain so few objects. As in Paper II, each plot shows three semi--transparent histograms representing the mass functions of all objects in the control (blue) and feedback (green) calculations, with the mass function of triggered objects (a subset of the objects formed in the feedback runs) overlaid as red histograms . We note several general features.\\
\indent In all cases, the high--mass end of the histograms is less well--populated in the ionized simulations. This is due to two effects. In all simulations, ionization reduces accretion, particularly on the most massive objects, which are also of course the ionizing sources. Most of them do continue to accrete, but at much reduced rates compared to their counterparts in the control runs, particularly when feedback succeeds in clearing out the central volumes of the clouds, which terminates accretion onto most objects. Growth towards very large masses by accretion is thus suppressed. Additionally, in simulations where sink particles represent small clusters and are permitted to merge with each other, local expulsion of gas and weakening of the gravitational potential reduces the rate of mergers since it retards local collapse. Thus in these simulations, growth to large cluster masses by mergers is also suppressed. In most simulations, the mass functions are therefore shifted towards lower masses, as was observed in Paper II.\\
\indent Of the three mass functions of stellar--mass sinks (Runs UF, UP and UQ) only that of UP resembles a `normal' mass function, being approximately log--normal in shape, with a peak at the rather high mass of $\sim3$M$_{\odot}$. The mass functions in Runs UF and UQ are flat and the suppression of accretion to high masses in Run UF results in a mass function in the feedback run which is flatter even than in the corresponding control simulation. Such flat mass functions have been observed before in simulations of partially--bound clouds \cite{2001ApJ...556..837K,2008MNRAS.386....3C} but we show here that photoionizing feedback, although it does shift the mass function somewhat towards lower masses, does not produce a mass function with a canonical slope. However, we also note that the mass functions in Runs UF and UQ are poorly--sampled so this inference should be treated with some caution.\\
\section{Discussion}
\subsection{Formation of pillars}
The mechanisms by which the prominent elongated pillar structures found in, for example M16 \citep{1996AJ....111.2349H} and Carina \citep{2005AJ....129..888S} are formed is a subject of lively debate. Several mechanisms have been proposed to explain their existence, such as the irradiation of turbulent gas \citep{2010ApJ...723..971G,2012A&A...546A..33T}, shielding of material from point radiation sources by dense clumps \citep{1989ApJ...346..735B}, the amplification of perturbations in the material on which the radiation field impinges \citep{2012A&A...538A..31T}, or the photoevaporation of material from some clumps in to the shadows of others \citep{2010MNRAS.403..714M}. The action of hydrodynamic instabilities in the kind of strongly inhomogeneous gas present in our turbulent cloud is very difficult to isolate. There are however clear instances of what appear to be dense clumps shielding material behind them from radiation, leading to column--like structures when the surrounding low--density material is destroyed or swept up. In Papers I and II we observed the formation of a pronounced pillar by the decapitation and partial photoevaporation of an accretion flow and we see the same phenomenon occurring in some simulations here. We also see a fourth mechanism of pillar formation in at least two simulations, namely Runs UF and UQ; two expanding bubbles meet, squeezing a thin streamer of material between them, one end of which is eventually pinched off, leaving an object resembling a pillar separating the two bubbles. In this case, of course, the pillar does not point towards any of the ionizing sources, but in fact in a perpendicular direction.\\ 
\subsection{Triggering, suppression and redistribution}
In a recent contribution \citep{2012arXiv1208.3364D}, we suggested that the effects of feedback on the stellar component of embedded clusters could be divided into three categories; triggering, suppression and redistribution. Triggering has several possible definitions and it is not clear \emph{a priori} which if any is best. It could mean a temporary increase in the star formation rate, an increase in the star formation efficiency over some timescale, the formation of a larger number of stars (which need not change the SFE) or the involvement of different bodies of gas in the star formation process without necessarily changing the SFR, SFE or the number of stars. The opposites of the first three of these possibilities would constitute suppression of star formation. Lastly, it is possible although perhaps not likely, that the stellar population could be completely or largely unchanged by the action of feedback, but that the stars are found in different places with different velocities. We term this \emph{redistribution}. Each of these three mechanisms may of course be global or local. We showed in Section 3 that all of these process do happen and that their strengths tend to be linked; either feedback does relatively little to the structure of the stellar content of the clouds and none of these happen, or the cluster is strongly affected by feedback and they all occur together.\\
\indent In \cite{2007MNRAS.377..535D} and \cite{2012MNRAS.tmp.2723D} we reported on the effects of \emph{external} irradiation of, respectively, unbound and bound molecular clouds by arbitrarily--placed O--stars. Comparing the simulations presented in Papers I, II, III and here to these earlier calculations, we can draw some general inferences. External ionizing feedback is, if anything, mildly globally positive, in the sense of increasing the star formation efficiency or leaving it unchanged. This is largely due to the rocket effect acting on the skin of the externally--irradiated cloud driving low--density and otherwise non--star--forming gas towards the dense cores of the clouds. Internal feedback, by contrast, is globally negative, since its principal effect is to destroy the dense cores of the clouds. However, in all cases, triggering, suppression and redistribution of star formation can occur on local scales.\\
\indent We showed in Papers I and III that the fraction of gas unbound by ionizing feedback over the $t_{\rm SN}$ timescale is a strong function of the clouds' escape velocities. When considering the outcome of star formation in terms of the observable characteristics of the clusters and stellar populations formed, it is more difficult to draw any general conclusions. While the influence of feedback on the gaseous content of the clouds can be inferred reasonably accurately from their appearance, the changes wrought on the stellar populations and distributions are harder to deduce. While the prominent bubbles in Runs UQ, UF and I dominate the spatial distribution of the stars and are responsible for substantial local triggered star formation, similarly obvious structures in Runs UU, UV, UP and D in fact betoken only small changes to the stellar contents of the clouds relative to the control simulations. Conversely, the bubbles in Run J are much less well--defined, but the mass function and geometry of the embedded cluster in that calculation are both very different thanks to feedback, as discussed in Paper II.\\
\subsection{Changes in cloud structure}
\indent \cite{2012MNRAS.427..625W} studied the influence of cloud structure on the effects of feedback by constructing fractal clouds with fractal dimensions $\mathcal{D}$ in the range 2.0--2.8, all containing a single central ionizing source. The effect of ionizing feedback was positive on short timescales, increasing the star formation rate relative to control runs with no feedback, but became negative at later times by destroying the clouds. The effect of the clouds' initial fractal dimensions was largely to change the geometry of their final states. For $\mathcal{D}<2.4$, the expanding HII regions generated large coherent structures, which was dubbed \emph{shell--dominated} evolution. For $\mathcal{D}>2.6$, they found instead that many small--scale and often pillar--like structures were formed and they refer to this as \emph{pillar--dominated} evolution. They did not attempt to measure how the fractal dimensions of their clouds changed with time, however.\\
\indent We used a simple box--counting method to estimate the fractal dimensions of our clouds, both initially and in their final states. We interpolated the SPH density fields onto 128$^{3}$ grids, partitioned these grids into $s\times s\times s$ subgrids, with $s=\{2,4,8,16, 32, 64,128\}$ and counted the number $n$ of subgrids containing at least one pixel whose density exceeded a threshold $\rho_{0}$. The fractal dimension of the density distribution is then given by d log $n$/d log $s$. The results are somewhat sensitive to the choice of $\rho_{0}$ in that, if a sufficiently small value is chosen, almost all pixels contain gas at or above that density and a fractal dimension close to 3.0 is returned. We experimented with  threshold values parameterized by the cloud man densities $\bar{\rho}$. The results for all clouds were rather similar and we show in Table \ref{tab:fracdim} the results for run UQ. We measured the fractal dimensions of the initial conditions at the onset of ionization, and of the final states of the control and feedback simulations. Results from the other runs are qualitatively similar, although differences between feedback and control runs are of course less marked for the more massive clouds on which feedback has little impact.\\
\begin{table}
\begin{tabular}{l|l|l|l|l|}
Cloud&$\rho_{0}$=0.1$\bar{\rho}$&$\rho_{0}$=1.0$\bar{\rho}$&$\rho_{0}$=10$\bar{\rho}$&$\rho_{0}$=100$\bar{\rho}$\\
\hline
Initial &2.92&2.84&2.69&2.27\\
Final (no fdbck.) &2.94 &2.84 &2.51 &1.83\\
Final (fdbck.) &2.99 &2.75 &2.35 &2.09 \\
\end{tabular}
\caption{Estimated fractal dimensions for the initial state of Run UQ at the onset of ionization, and the final states of the control and feedback simulations, as measured using four different values of $\rho_{0}$ parameterized by the cloud mean densities $\bar{\rho}$.}
\label{tab:fracdim}
\end{table}
Measurements of $\mathcal{D}$ with $\rho_{0}$=0.1$\bar{\rho}$ are all close to 3.0, so setting the threshold so low tells us little about the cloud structure. For Run UQ and all the other runs, $\mathcal{D}$ for the initial conditions measured using $\rho_{0}$=1.0$\bar{\rho}$ or $\rho_{0}$=10$\bar{\rho}$, which seem reasonable values to us, lies in the range 2.7--2.85. We cannot therefore make any inferences on the influence of the initial fractal dimensions of our clouds on the outcome of our simulations. However, it is clear from Table \ref{tab:fracdim} that the fractal dimensions measured at the higher densities evolve in different ways in the feedback and control simulations. In both cases, $\mathcal{D}$ generally decreases as the simulations evolve, indicating that more structure is emerging in all clouds. However, in the runs without feedback, $\mathcal{D}$ decreases faster and further, dropping below 2.0 at the highest densities considered here. This may be indicative of the highest--density gas settling into linear or filamentary structures. In the feedback runs however, even at the highest densities, $\mathcal{D}$ never falls below 2.0, indicating that the gas in these simulations has a more sheet-- or shell--like geometry (corresponding to \cite{2012MNRAS.427..625W}'s shell--dominated models). This seems to confirm the impression given by the images of the clouds in Figure \ref{fig:compare_end} that the main effect of feedback on the clouds' appearance is to generate large--scale shell structures, while disrupting the filaments in which the stars originally formed.
\section{Conclusions}
\indent Our conclusions may be summarized as follows:\\
\indent (i) The presence of even rather dramatic--looking structures such as large bubbles or pillars does not necessarily indicate that the stellar content of an embedded cluster has been strongly influenced by whatever feedback mechanism is responsible for blowing the bubbles. Particularly in partially--unbound clouds, there is a tendency of bubbles to expand into low--density and largely quiescent gas, lessening its impact on the denser material where much of the star formation is taking place.\\
\indent (ii) The effect of \emph{internal ionizing feedback} on star formation in GMCs is overall close to zero or negative, in the sense of decreasing the global star formation efficiencies and rates, regardless of whether clouds are initially bound or not. Local triggering does occur, but in none of the simulations presented here or in Papers I and II is it enough to outweigh the negative feedback due to the destruction of the densest star--forming gas and the disruption of large--scale accretion flows.\\
\indent (iii) Interpretation of age gradients in star--forming regions should be treated with caution, especially if they are used as evidence for triggering. While we do observe age--gradients across the face of some our model clouds (UP, I), they are present in the triggered and spontaneously--formed populations alike and trace the progress of ionization fronts through the clouds, which cut off the supply of cold gas from which sinks accrete as they go. Plotting the ages of sinks with respect to distance from the nearest ionizing star yields little evidence of a correlation in either triggered or spontaneously--formed objects.\\
\indent (iv) The three possible ways in which feedback can influence the structure of embedded clusters -- triggering, suppression and redistribution -- can all occur locally in the same cloud simultaneously and are broadly correlated in that the presence or absence of one implies the presence or absence of the others.\\
\indent (v) The association of triggered stars with pillar--like structures and bubble walls is stronger in initially--unbound clouds than in bound clouds, but this correlation is not perfect in either case.\\
\indent (vi) It is exceedingly difficult to infer which stars in a system have been triggered or redistributed from observing either the stellar or gaseous content of an embedded cluster, or the relative geometries of the two components at a given time. We stressed in \cite{2012arXiv1208.3364D} that, to assess the importance of triggering in a given real system, one must first have a model of how the system would appear in the absence of the feedback mechanism under study. However, even with such a model in place, it is still difficult to see how any but the most general statements concerning triggering can be made.\\
\section{Acknowledgments}
{\bf We thank the anonymous referee for a careful reading of the manuscript and several interesting suggestions which improved the paper.} This research was supported by the DFG cluster of excellence `Origin and Structure of the Universe' (JED, BE).

\bibliography{myrefs}

\label{lastpage}

\end{document}